\documentclass[iop,apj,tighten, twocolappendix, numberedappendix, appendixfloats]{emulateapj}
\pdfoutput=1 
\usepackage{apjfonts} 
\usepackage{amsmath,amstext}
\usepackage[breaklinks,colorlinks,citecolor=blue, linkcolor=magenta]{hyperref} 
\usepackage{enumitem} 

\newcommand{\Mstar}{M_\textnormal{star}}

\shorttitle{Stellar Kinematics and Dynamical Modeling of Low-Mass Galaxies}
\shortauthors{El-Badry et al.}

\begin{document}
\defcitealias{El-Badry_2016}{E16}

\title{When the Jeans don't fit: How stellar feedback drives stellar kinematics and complicates dynamical modeling in 
low-mass galaxies}

\author{Kareem El-Badry\altaffilmark{1,2}, Andrew R. Wetzel\altaffilmark{3,4,5,8,9}, Marla Geha\altaffilmark{2}, 
Eliot Quataert\altaffilmark{1}, Philip F. Hopkins\altaffilmark{3}, \\ Dusan Kere{\v{s}}\altaffilmark{6}, 
T.K. Chan\altaffilmark{6}, and Claude-Andr{\'{e}} Faucher-Gigu{\`{e}}re\altaffilmark{7}}

\affil{$^1$Department of Astronomy, University of California, Berkeley, CA, USA. kelbadry@berkeley.edu}
\affil{$^2$Department of Astronomy, Yale University, New Haven, CT, USA}
\affil{$^3$TAPIR, California Institute of Technology, Pasadena, CA, USA}
\affil{$^4$Carnegie Observatories, Pasadena, CA, USA}
\affil{$^5$Department of Physics, University of California, Davis, CA, USA}
\affil{$^6$Department of Physics, Center for Astrophysics and Space Sciences, University of California at San Diego, 
La Jolla, USA}
\affil{$^7$Department of Physics and Astronomy and CIERA, Northwestern University, Evanston, IL, USA}

\altaffiltext{8}{Moore Prize Fellow}
\altaffiltext{9}{Carnegie Fellow in Theoretical Astrophysics}

\begin{abstract}
In low-mass galaxies, stellar feedback can drive gas outflows that generate non-equilibrium fluctuations in the 
gravitational potential. Using cosmological zoom-in baryonic simulations from the Feedback in Realistic Environments 
(FIRE) project, we investigate how these fluctuations affect stellar kinematics and the reliability of Jeans dynamical 
modeling in low-mass galaxies. We find that stellar velocity dispersion and anisotropy profiles fluctuate significantly 
over the course of galaxies' starburst cycles. We therefore predict an observable correlation between star formation 
rate and stellar kinematics: dwarf galaxies with higher recent star formation rates should have systemically higher 
stellar velocity dispersions. This prediction provides an observational test of the role of stellar feedback in 
regulating both stellar and dark-matter densities in dwarf galaxies. We find that Jeans modeling, which treats galaxies 
as virialized systems in dynamical equilibrium, overestimates a galaxy's dynamical mass during periods of 
post-starburst gas outflow and underestimates it during periods of net inflow. Short-timescale potential fluctuations 
lead to typical errors of $\sim 20\%$ in dynamical mass estimates, even if full 3-dimensional stellar kinematics -- 
including the orbital anisotropy -- are known exactly. When orbital anisotropy is \textit{not} known a priori, typical 
mass errors arising from non-equilibrium fluctuations in the potential are larger than those arising from the 
mass-anisotropy degeneracy. However, Jeans modeling alone \textit{cannot} reliably constrain the orbital anisotropy, 
and problematically, it often favors anisotropy models that do not reflect the true profile. If galaxies completely 
lose their gas and cease forming stars, fluctuations in the potential subside, and Jeans modeling becomes much more 
reliable.
\end{abstract}

\maketitle

\section{Introduction}
Low-mass galaxies $(M_{\rm star} \lesssim 10^{9.5}\,M_{\odot})$ have shallow gravitational potentials that make them 
especially sensitive to stellar feedback-driven gas outflows. These outflows can produce rapid fluctuations in the 
gravitational potential, which can in turn alter the kinematics and distribution of stars and dark matter on short 
timescales \citep[e.g.,][]{Read_2005, Pontzen_2012, DiCintio_2014, Onorbe_2015, Governato_2015, Fry_2015, Chan_2015, 
El-Badry_2016}. Isolated low-mass galaxies are thus ideal laboratories for studying the interplay between stellar 
feedback, gas outflows, and the distribution and dynamics of stars and dark matter.

One of the most widely used techniques for studying galaxies' mass distributions is Jeans dynamical modeling 
\citep{Jeans_1915, Binney_1980, Merritt_1985, Dejonghe_1992}. Jeans modeling aims to extract information about a 
galaxy's underlying gravitational potential from the distribution and kinematics of luminous tracers, which often 
constitute only a small fraction of the total dynamical mass. Jeans modeling is predicated on the assumption of 
dynamical equilibrium (that this, $\Phi({\mathbf x},t) = \Phi({\mathbf x})$, where $\Phi$ is the gravitational 
potential), so that stellar orbits can be assumed to instantaneously trace the gravitational potential.

Jeans modeling traditionally is used to study galaxies with dispersion-supported stellar kinematics and little or no 
gas, while dynamical modeling of gas-rich galaxies generally is based on measurements of gas rotation velocity. 
However, there are several potential advantages to using stellar rather than gas kinematics to probe a galaxy's 
gravitational potential. First, many gas-rich galaxies with high specific star formation rates (sSFR$ = $SFR$ / 
\Mstar$) have irregular gas distributions and lack coherent disks, making accurate measurements of the gas rotation 
velocity infeasible, particularly in the central regions \citep{Cannon_2004, Lelli_2014, Read_2016b}. In addition, 
turbulence can drive gas velocities in the interstellar medium, even where there is a well-defined net rotation speed 
\citep{Fillmore_1986, Rhee_2004, Valenzuela_2007, Oh_2011, Pineda_2016}. In this case, the observed gas rotation speed 
is not an accurate tracer of the circular velocity ($v_{\rm rot}(r) < \sqrt{GM(r)/r}$), so rotation curve fitting will 
systematically underestimate the total enclosed mass.

Stars are (effectively) collisionless tracers of the gravitational potential, so their orbits are not directly coupled 
to the feedback processes that produce non-gravitational support in the gas. It is therefore promising (though 
observationally challenging) to probe the mass distributions of gas-rich galaxies with stellar kinematics, which serve 
as an independent tracer of the gravitational potential \citep{Cinzano_1994, Krajnovic_2005, Adams_2012, Adams_2014}.  

Among the largest sources of uncertainty in stellar Jeans modeling is poor knowledge of the stellar velocity 
anisotropy, $\beta$, which characterizes the relative ``pressure'' between the radial and tangential components of 
stellar orbits. Direct measurement of $\beta$ requires full 3D velocity information. Because observational studies can 
readily measure only a single line-of-sight velocity component, few observational constraints on $\beta$ exist. Rather 
than measuring $\beta$ directly, through proper motions, most studies of nearby galaxies attempt to indirectly 
constrain $\beta$ using dynamical models \citep[e.g.,][]{vanderMarel_1994, Diakogiannis_2014b, Diakogiannis_2014c}. 
Usually, the observed stellar velocities dispersion profile is compared to the dispersion profile predicted by the 
Jeans equations for a particular guess of $\beta(r)$ and $M_{\rm dyn}(r)$ profiles. One then chooses the combination of 
$\beta(r)$ and $M_{\rm dyn}(r)$ profiles that most accurately recover the observed dispersion profile.

Observational works typically assume a parameterized form for the gravitational potential (for example, an NFW profile) 
and either assume $\beta(r) = {\rm const.}$ \citep[e.g.,][]{Geha_2002, Lokas_2005, Koch_2007, Walker_2007, 
Battaglia_2008, Lokas_2009, Walker_2009, Diakogiannis_2014} or assume a theoretically motivated functional form for 
$\beta(r)$ \citep[e.g.,][]{Kleyna_2001, Wilkinson_2004, Mamon_2005, Gilmore_2007, Battaglia_2008, Mamon_2013, 
Diakogiannis_2014, Mashchenko_2015}. A variety of qualitatively different forms of $\beta(r)$ are regularly used in the 
observational literature to model the anisotropies of the same galaxies. It is generally taken for granted that the 
anisotropy and dynamical mass profiles that predict the dispersion profile in closest agreement with the observed 
dispersion profile reflect the galaxy's true $\beta(r)$ and $M_{\rm dyn}(r)$ profiles, but the validity of this 
assumption has not been investigated in detail.

Another major uncertainty in Jeans modeling concerns the assumption of dynamical equilibrium. \citet[][hereafter 
E16]{El-Badry_2016} showed that the stellar kinematics of star-forming gas-rich dwarf galaxies can fluctuate on short 
$(\sim 100\,{\rm Myr})$ timescales, as stellar feedback-driven gas outflows and inflows produce a time-varying 
gravitational potential that transfers energy to stars and dark matter. Such galaxies are rarely in dynamical 
equilibrium. One might worry, then, that any dynamical model that treats galaxies as virialized, equilibrium systems 
could produce biased or incorrect mass estimates for bursty, gas-rich galaxies.

In this work, we explore the role of stellar feedback in driving stellar kinematics in low-mass galaxies, emphasizing 
observable relations that can test the role of star formation and feedback in driving stellar kinematics. We then 
explicitly test the reliability of Jeans dynamical models in low-mass galaxies. We use high-resolution baryonic 
simulations from the Feedback in Realistic Environments (FIRE)\footnote{See the FIRE project website: 
fire.northwestern.edu} project. Because both the underlying mass distribution and 3D kinematics of simulated galaxies 
can be measured exactly, simulations make it possible to measure how robustly dynamical models can recover a galaxy's 
true mass and anisotropy profile and how non-equilibrium fluctuations can bias mass estimates derived from Jeans 
modeling.

This paper is organized as follows. In Section~\ref{sec:simulations}, we describe the FIRE simulations and our galaxy 
sample. In Section~\ref{sec:presenting_profiles}, we present velocity anisotropy and dispersion profiles for our 
simulated galaxies. In Section~\ref{sec:observational_prediction}, we study the relationship between stellar kinematics 
and star formation rate, making the observable prediction that at fixed stellar mass, galaxies with higher sSFR should 
have systemically higher stellar velocity dispersion. In Section~\ref{sec:spherical_jeans_modeling}, we construct 
spherical Jeans models for our galaxies, considering both the case in which $\beta(r)$ is modeled with an unknown 
parameter and the case in which it can be measured directly. In Section~\ref{sec:dwarf_spheroidal}, we compare the 
reliability of Jeans models in gas-rich galaxies to their reliability in gas-poor, quiescent galaxies. Finally, in 
Section~\ref{sec:summary_discussion}, we summarize our results and discuss avenues for future research.

\section{Simulations}
\label{sec:simulations}

We use cosmological zoom-in baryonic simulations from the Feedback in Realistic Environments (FIRE) project 
\citep{Hopkins_2014}. The galaxies which comprise our sample were first presented in \citep{Hopkins_2014} and 
\citet{Chan_2015} and were also studied in \citetalias{El-Badry_2016}. We briefly summarize the simulations here, 
directing the reader to previous works for more details.

Our simulations were run using the \textsc{Gizmo} code \citep{Hopkins_2015}, which employs pressure-entropy based 
smooth particle hydrodynamics \citep[P-SPH;][]{Hopkins_2013} and an improved version of the TreePM gravity solver from 
\textsc{Gadget-3} \citep{Springel_2005}. Initial conditions were generated at $z = 100$ using the MUSIC code 
\citep{Hahn_2011}. All simulations use a flat $\Lambda$CDM cosmology with 
$(\Omega_M, \Omega_{\Lambda}, \Omega_{\rm b}, h)=(0.272, 0.728, 0.0455, 0.702)$.

\textsc{Gizmo} incorporates radiative cooling and heating rates for gas from \textsc{Cloudy} \citep{Ferland_2013} 
across $10 - 10^{10}\,{\rm K}$, with atomic, molecular, and metal-line cooling computed for 11 elements. Ionization and 
heating rates include a redshift-dependent, spatially uniform ultraviolet background computed in 
\citep{FaucherGiguere_2009}. Star formation occurs only in dense, locally self-gravitating molecular clouds with 
density $n > 50\,{\rm cm^{-2}}$ and proceeds with an instantaneous efficiency of 100\% per free-fall time (though 
stellar feedback quickly regulates the local gas density, leading to much lower resultant efficiency; see Orr et al., 
in prep., for more details). 

Each star particle represents a single stellar population with the same mass and metallicity as its progenitor gas 
particle and a \citet{Kroupa_2002} initial mass function. Once stars form, they begin to deposit energy, momentum, and 
metals into nearby gas particles through a variety of feedback processes. Energy, momentum, mass, and metal returns are 
calculated directly from stellar evolution models at each timestep, as computed from \textsc{STARBURST99} 
\citep[v7.0;][]{Leitherer_1999, Leitherer_2005, Leitherer_2010, Leitherer_2014}. We include the effects of stellar 
winds, radiation pressure from massive stars, local photoionization and photoelectric heating, and core-collapse and 
type Ia supernovae, as detailed in \citet{Hopkins_2014}.

\begin{table}[tbp]
\centering
\caption{Parameters of the simulations at $z=0$}
\label{tab:properties}
\begin{tabular}{p{0.8cm} | p{0.8cm}| p{1.1cm} | p{1.3 cm} | p{0.7cm} | p{0.7cm} | p{0.7cm}}

Name & $R_{90 \rm m}$ $(\rm kpc)$  & $\log (M_{\rm star})$ $(M_{\odot})$ & $\log (M_{200 \rm m})$ $(M_{\odot})$ & 
$m_b$ $(M_{\odot})$ & $\epsilon_{\rm gas}$ $(\rm pc)$ & $\epsilon_{\rm star}$ $(\rm pc)$ \\ 
\hline
\texttt{m10}   &  1.40  &  6.35   &   9.92   & 2.6e2 & 3  & 7  \\
\texttt{m10.1} &  3.97  &  7.22   &   10.16  & 2.1e3 & 4  & 7  \\
\texttt{m10.2} &  6.24  &  7.72   &   10.23  & 2.1e3 & 4  & 7  \\
\texttt{m10.6} &  9.02  &  8.46   &   10.60  & 2.1e3 & 10 & 21 \\ 
\texttt{m11}   &  15.45 &  9.32   &   11.17  & 7.1e3 & 7  & 14 \\
\texttt{m11v}  &  14.05 &  9.36   &   11.28  & 5.7e4 & 7  & 14 \\
\texttt{m11.2} &  14.87 &  9.59   &   11.23  & 1.7e4 & 10 & 21 \\
\hline
\end{tabular}
\begin{flushleft}
$R_{90 \rm m}$ is the radius enclosing 90\% of the stellar mass. $M_{\rm star}$ and $M_{200 \rm m}$ are the total mass 
and stellar mass inside $0.1 \, R_{200 \rm m}$ and $R_{200 \rm m}$, respectively, where $R_{200 \rm m}$ is the radius 
within which the matter density is $200 \times$ the mean matter density. $m_b$ is the average baryon particle mass. 
$\epsilon_{\rm gas}$ and $\epsilon_{\rm star}$ are the minimum gravitational softening length for gas and stars, in 
physical units.
\end{flushleft}
\end{table}

We study a sample of 7 low-mass galaxies, which were first presented by \citet{Hopkins_2014} and 
\citet{Chan_2015}.\footnote{We use the same naming convention as \citet{Hopkins_2014} for \texttt{m10}, \texttt{m11}, 
and \texttt{m11v}. \citet{Chan_2015} referred to simulations \texttt{m10.1}, \texttt{m10.2}, \texttt{m10.6}, and 
\texttt{m11.2} as \texttt{m10h1297}, \texttt{m10h1146}, \texttt{m10h573}, and \texttt{m11h383}, respectively.}
Table~\ref{tab:properties} provides a brief summary of their properties at $z=0$; Table~1 of \citetalias{El-Badry_2016} 
presents additional simulation parameters.\footnote{We use the same simulation sample as \citetalias{El-Badry_2016}, 
with two exceptions. First, we study a different galaxy in simulation \texttt{m10.2}. \citetalias{El-Badry_2016} 
studied the galaxy with the largest stellar mass, but we discovered this galaxy to be contaminated with 
lower-resolution dark matter particles. We now study the uncontaminated galaxy, which \citet{Chan_2015} also studied. 
Second, we do not include the Milky Way (MW) mass galaxy \texttt{m12i} in our sample, because spherically-symmetric 
Jeans modeling is not appropriate given its rotation-supported disk.} At $z=0$, these galaxies have stellar masses in 
the range $M_{\rm star} = 10^{6.3 - 9.6} M_{\odot}$ and halo masses $M_{200 \rm m} = 10^{9.9 - 11.3} M_{\odot}$. 
Galaxies from the FIRE project have been shown to reproduce many key observed properties of low-mass galaxies, 
including the $M_{\rm star}-M_{\rm halo}$ relation \citep{Hopkins_2014}, the $\Mstar-$size relation 
\citepalias{El-Badry_2016}, realistic gas outflows \citep{Muratov_2015}, cored density profiles \citep{Chan_2015}, 
dispersion-supported stellar kinematics \citep{Wheeler_2015}, inverted metallicity gradients 
\citepalias{El-Badry_2016}, and the redshift evolution of the $\Mstar-$metallicity relation \citep{Ma_2016}. As 
\citetalias{El-Badry_2016} showed, their global properties, including stellar masses, sizes, and the magnitude of 
feedback-driven potential fluctuations, are well-converged with resolution.

Like observed isolated galaxies in this mass range \citep{Weisz_2012}, our simulated galaxies have bursty star 
formation histories. \citet{Sparre_2015} analyzed the burstiness of the SFHs of some of the galaxies in our sample and 
compared with observed galaxies at this mass, finding that these FIRE simulations reproduce the slope and scatter of 
the observed main sequence of star formation and overall agree with the observed level of burstiness. They did find 
that the most extreme starburst cycles in FIRE are somewhat stronger than inferred for gas-rich dwarf galaxies near the 
MW, so FIRE simulations may overpredict the fraction of isolated galaxies at $M_{\rm star} \lesssim 10^{9.5} M_{\odot}$ 
that are (temporarily) quiescent. Analyzing dwarf galaxies that form around a MW-mass host in a different FIRE 
simulation, \citet{Wetzel_2016} also showed that the FIRE model produces a range of star formation histories that 
agrees well with observed dwarf galaxies around the MW.

To showcase how feedback-driven potential fluctuations affect stellar kinematics and the reliability of Jeans modeling 
in different mass regimes, we present detailed results primarily from two of our galaxies: \texttt{m10} and 
\texttt{m10.6}. These galaxies represent the regimes in which feedback-driven outflows do (\texttt{m10.6}) and do not 
(\texttt{m10}) cause strong fluctuations in the potential. Both galaxies have high total gas fractions 
($f_{\rm gas} = M_{\rm gas}/(M_{\rm gas}+M_{\rm star})\approx 0.75$ at $z=0$), which is typical for observed isolated 
galaxies at these mass scales \citep{Bradford_2015}.

\texttt{m10} has the lowest stellar mass and highest resolution in our sample, with 
$M_{\rm star} = 10^{6.35} M_{\odot}$ and $m_b = 260 M_{\odot}$. Because of its low baryon fraction, gas does not 
constitute a significant fraction of this galaxy's total mass, and thus gas outflows do not cause strong fluctuations 
in the potential. On the other hand, \texttt{m10.6} $(M_{\rm star} = 10^{8.46} M_{\odot};\,m_b = 2100 M_{\odot})$ 
represents the mass regime in which feedback-driven gas outflows are most efficient in driving strong fluctuations in 
the potential. We summarize our results across our full mass range in Section~\ref{sec:mass_scaling}.

\section{Stellar velocity dispersion and anisotropy profiles}
\label{sec:presenting_profiles}

\begin{figure*}
\includegraphics[width=\textwidth]{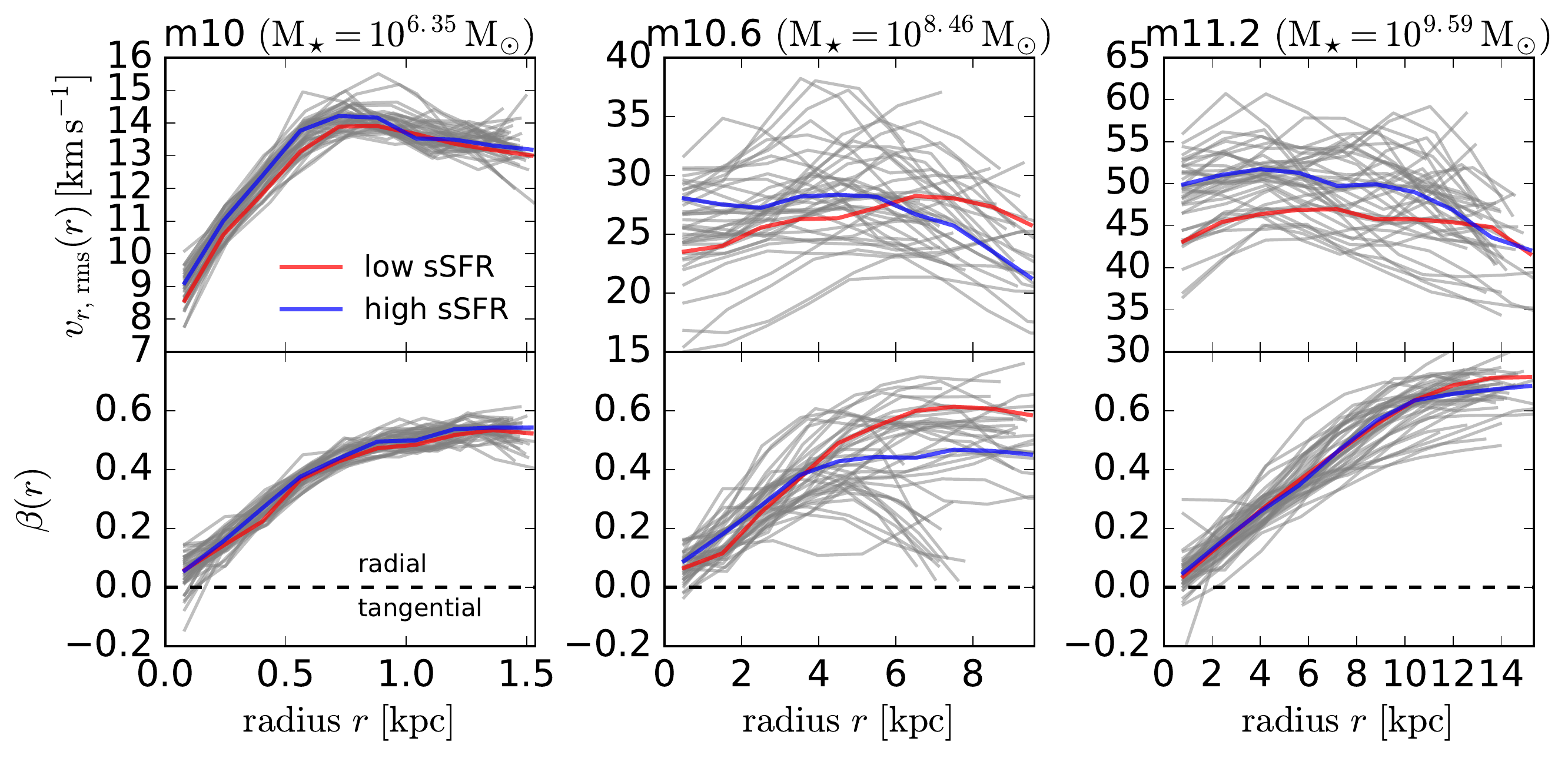}
\caption{Profiles of stellar rms radial velocity, $v_{r,{\rm rms}} \equiv \sqrt{\smash[b]{\overline{v_{r}^{2}}}}$ 
(top), and velocity anisotropy, $\beta$ (bottom), for three galaxies that span the mass range of our simulations. Gray 
lines show the last 40 snapshots since $z \approx 0.2$, with typical snapshot spacing of $50-100\,{\rm Myr}$. Profiles 
are plotted out to $R_{90 \rm m}$, the radius that encloses 90\% of the stellar mass, at each snapshot. Red (blue) 
lines show weighted median profiles of the 10 snapshots with the lowest (highest) sSFR (smoothed over 100 Myr). Scatter 
in $\beta$ and $v_{r,{\rm rms}}$ profiles is largest in \texttt{m10.6}, the galaxy with the largest potential 
fluctuations at late times. Snapshots with low (high) sSFR have systemically lower (higher) $v_{r,{\rm rms}}$ values. While $\beta(r)$ profiles fluctuate significantly, they do not correlate with instantaneous sSFR as strongly.}
\label{fig:anisotropy_age_profiles}
\end{figure*}

The two stellar kinematic ingredients required for dynamical modeling of a spherical system are (1) the rms radial 
velocity, $v_{r, {\rm rms}} \equiv \sqrt{\smash[b]{\overline{{v_r^2}}}}$, where 
$v_{r}=\mathbf{v}\cdot\mathbf{r}/\left|\mathbf{r}\right|$, and (2) the velocity anisotropy, commonly quantified through 
the parameter $\beta$ \citep{Binney_2008}, which is defined as
\begin{equation}
\label{eqn:beta}
\beta = 1 - \frac{\overline{v_{\theta}^{2}} + \overline{ v_{\phi}^{2}} }{2\overline{v_{r}^{2}}}.
\end{equation}
In this parameterization, anisotropy ranges between $\beta = 1$ for completely radial orbits and $\beta=-\infty$ for 
completely tangential orbits, with $\beta = 0$ corresponding to isotropy, (that is, equality between radial and 
tangential components). In this section, we present the (time-dependent) profiles of $v_{r, {\rm rms}}(r)$ and 
$\beta(r)$ in our simulated galaxies. Note that, unlike the line-of-sight velocities that are typically measured 
observationally, $v_{r, {\rm rms}}$ and $\beta$ are 3D quantities.

Figure~\ref{fig:anisotropy_age_profiles} shows the radial dependence of the stellar rms radial velocity 
$v_{r,{\rm rms}}$ (top), and anisotropy $\beta$ (bottom) in three galaxies spanning the mass range of our sample. We 
plot profiles for each of the last 40 simulation snapshots since $z \approx 0.2$ to showcase short-timescale 
variations. Blue and red lines show the mass-weighted median profiles for the 10 snapshots with the highest and lowest 
specific star formation rate, respectively. 

Both the rms radial velocity and anisotropy profiles in \texttt{m10} are fairly stable across these 40 snapshots, 
reflecting the relatively calm evolution of this galaxy at $z \sim 0$. As \citetalias{El-Badry_2016} showed, the baryon 
fraction of \texttt{m10} is so low that feedback-driven gas outflows do not displace enough mass to significantly 
change the total potential, so the galaxy's stellar kinematics do not fluctuate significantly at late times.

In contrast, both the shape and normalization of the $v_{r, {\rm rms}}(r)$ and $\beta(r)$ profiles fluctuate 
dramatically in \texttt{m10.6}. At fixed radius, the rms radial velocity changes by nearly a factor of two over the 
course of the starburst cycle. The anisotropy profile is more stable at small radii, but at large radii, it changes 
between isotropy $(\beta \approx 0)$ and highly radial orbits $(\beta \approx 0.7)$ on timescales of only a few 100 
Myr. Fluctuations in $v_{r,{\rm rms}}(r)$ and $\beta(r)$ are qualitatively similar in \texttt{m11.2} and \texttt{m10.6} 
but are somewhat weaker in \texttt{m11.2}, consistent with the mass scaling of potential fluctuations in 
\citetalias{El-Badry_2016}.

In all three galaxies, the median rms radial velocity is higher at fixed radius during episodes of higher sSFR. This is 
because, to first order, $v_{r,{\rm rms}}$ in a dispersion-supported system traces the depth of the gravitational 
potential.\footnote{In equilibrium, $v_{r,{\rm rms}} = \sigma_r \equiv(\overline{v_{r}^{2}}-\overline{v_{r}}^{2})^{1/2}$. Note, however, that it is the the \textit{total} dispersion 
$(\sigma_r^2 + \sigma_{\phi}^2 + \sigma_{\theta}^2)^{1/2}$ that traces the depth of the potential, so the relationship 
between $v_{r,{\rm rms}}$ and potential depth also depends on $\beta(r)$. This is why $v_{r,{\rm rms}}$ is not 
necessarily highest at $r=0$, where the potential is deepest.} The potential is deepest when cold gas accumulates in 
the galactic center, which is also when the sSFR is highest. Similarly, the sSFR falls when gas is driven into the 
outskirts of the galaxy and becomes rarefied; this gas displacement also shallows the gravitational potential. The 
$v_{r, {\rm rms}}$ profiles are steeper during snapshots with high sSFR, a consequence of the potential rising more 
steeply when gas is concentrated in the center. 

While Figure~\ref{fig:anisotropy_age_profiles} also shows significant fluctuations in $\beta(r)$, these fluctuations 
are not maximal during the highest- and lowest-sSFR snapshots. As we will show in 
Section~\ref{sec:observational_prediction}, the time-evolution of $\beta$ \textit{is} closely related to the star 
formation history, but there is a significant time-offset between changes in sSFR and changes in $\beta$, such that the 
relation between $\beta$ and sSFR is not evident in Figure~\ref{fig:anisotropy_age_profiles}.

\begin{figure}
\includegraphics[width=\columnwidth]{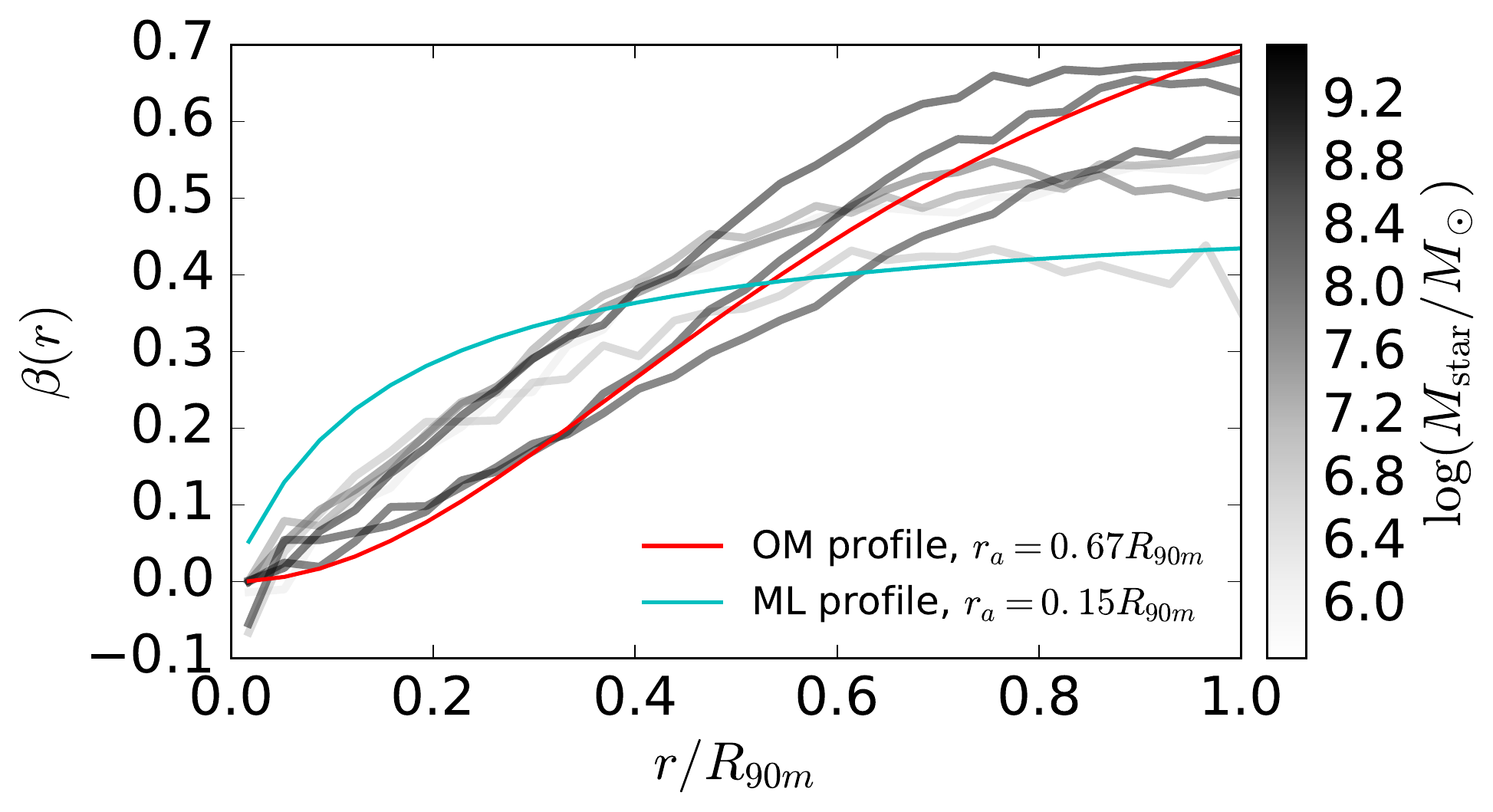}
\caption{Time-averaged profiles of stellar velocity anisotropy, $\beta$, for our 7 low-mass galaxies. Gray curves show 
the mass-weighted median from the last 40 simulation snapshots since $z \approx 0.2$. All profiles have a similar 
shape, with $\beta \sim 0$ near the galactic center and increasingly radial orbits at large radii. We find no clear 
trend in profile shape with stellar mass. Red and cyan curves show Osipkov-Merritt (OM) and Mamon-{\L}okas (ML) 
profiles (Equations~\ref{eqn:beta_OM} and \ref{eqn:beta_ML}) as fit to the median profile of all galaxies.}
\label{fig:median_anisotropy_profiles}
\end{figure}

Despite their strong fluctuations in stellar kinematics, the shape of galaxies' $\beta(r)$ profiles are remarkably 
similar when averaged over many snapshots. None of our low-mass galaxies are tangentially biased ($\beta < 0$), and 
except for a few non-equilibrium snapshots of \texttt{m10.6}, $\beta$ increases monotonically with radius. 
Figure~\ref{fig:median_anisotropy_profiles} shows the median radial anisotropy profiles of all 7 low-mass galaxies in 
our simulations, which we calculate by taking the median of $\beta$ values from the last 40 snapshots in each radial 
bin, where $\beta$ values from individual snapshots are weighted by the total stellar mass in the correspond radius bin 
in that snapshot. Consistent with other simulations of dispersion-supported galaxies \citep{Dubinski_1998, Saiz_2004, 
Dekel_2005, Mamon_2013, Campbell_2016}, the anisotropy profiles of all galaxies in 
Figure~\ref{fig:median_anisotropy_profiles} are approximately isotropic $(\beta = 0)$ near the center and become 
increasingly radially biased at large radii.

$\beta(r)$ profiles that are monotonically increasing are a natural consequence of feedback-driven ``breathing modes'', which continually revirialize our low-mass galaxies and place stars on predominantly radial orbits. \citetalias{El-Badry_2016} showed that star formation in these galaxies occurs almost exclusively in the galactic center ($> 90\%$ of stars form within $r \lesssim 0.4 R_{90 m}$, where $R_{90 m}$ is the radius enclosing $90\%$ of the stellar mass at $z = 0$), so stars at large radii must have migrated outward on radial orbits. Except when close to apocenter, these stars will have $v_{r} > v_{\rm tan}$, where $v_{\rm tan}$ is the tangential velocity.
On the other hand, the stars near the galactic center represent a mix of stars that formed there (and inherited the mixed radial + tangential kinematics of their gas clouds) and stars that are passing through the center on radial orbits. Even stars on radial orbits have their periapsis, where necessarily $v_{\rm tan} > v_r$, at small radius. This preferentially drives $\beta$ to be small at small radii: for highly elongated radial orbits, orbital apoapses are distributed across a range of radii, while periapses are all near the center. Both dissipationless \citep{vanAlbada_1982, Londrillo_1991, Dubinski_1998, Hozumi_2000} and hydrodynamic \citep{Dekel_2005} simulations of galaxy formation find that a monotonically increasing $\beta(r)$ profile, with $\beta = 0$ at the center and $\beta > 0.5$ at large radii, arises naturally during violent relaxation from a variety of initial conditions.

We next examine how well common functional forms of $\beta(r)$, which frequently are used to model $\beta(r)$ in observational Jeans modeling studies, are able to parametrize our simulated galaxies.
We consider two such functional forms. The first is the Osipkov-Merritt (hereafter ``OM'') profile \citep{Osipkov_1979, Merritt_1985}, given by
\begin{equation}
\label{eqn:beta_OM}
\beta_{{\rm OM}}(r)=\frac{r^{2}}{r^{2}+r_a^{2}},
\end{equation}
where $r_a$ is a scale radius, and $\beta \to 0$ for $r \ll r_a$, while $\beta \to 1$ for $r \gg r_a$. The OM profile is of theoretical interest because it gives rise to a mathematically convenient family of spherically-symmetric distribution functions, and because it has a similar form to the anisotropy profiles produced by simulations of spherical collapse \citep{vanAlbada_1982,Londrillo_1991,Hozumi_2000}. The second functional form that we consider is that of \citet[][hereafter the ``ML'' profile]{Mamon_2005}, given by 
\begin{equation}
\label{eqn:beta_ML}
\beta_{{\rm ML}}(r)=\frac{1}{2}\frac{r}{r+r_a}.
\end{equation}
Like the OM profile, the ML profile goes to $\beta = 0$ at $r \ll r_a$, but at $r \gg r_a$, it asymptotically approaches $\beta = 1/2$, not $\beta = 1$. At small radii, the ML profile rises more steeply than the OM model. The ML profile provides a good fit to the anisotropy profiles of dark matter halos in some cosmological simulations \citep{Mamon_2005} and of stars in idealized simulations of major mergers \citep{Dekel_2005}.

Figure~\ref{fig:median_anisotropy_profiles} compares the OM and ML profiles with the time-averaged $\beta(r)$ profiles of our simulated galaxies. We choose $r_a$ values to best-fit the median $\beta(r)$ profile of all 7 galaxies, for illustrative purposes only. The OM profile provides a reasonably good fit for all of our low-mass galaxies, though the anisotropy profiles from the simulations generally rise more steeply than the OM profile at small radii and approach lower $\beta$ values at large radii ($\beta \approx 0.6$ rather than $\beta \to 1$). The ML profile is a poorer fit at small radii, where it rises too steeply. At large radii, the ML profile is a good fit for profiles that plateau at $\beta \leq 0.5$, but problematically, it is unable to accommodate anisotropies greater than $\beta = 0.5$.

Of course, we have little a priori reason to expect that the OM or ML models should provide a particularly good fit to the anisotropy profiles of our simulated galaxies. While, qualitatively, an anisotropy profile that increases from isotropy at $r = 0$ to radial anisotropy at large radii is a natural consequence of violent relaxation, the precise form of $\beta(r)$ depends on the details of a galaxy's formation history, the kinematics of star-forming gas clouds, and the strength of feedback-driven potential fluctuations.

Since different galaxies in our sample have different star formation and evolutionary histories, one might also expect them to have a variety of anisotropy profile forms. The similarity between the median anisotropy profiles across our sample in Figure~\ref{fig:median_anisotropy_profiles} is thus somewhat surprising. It suggests that the semi-periodic potential fluctuations that the galaxies undergo may wash out the kinematic ``memory'' of the galaxies' formation histories, driving them towards a universal anisotropy profile.

In summary, feedback-driven potential fluctuations cause strong fluctuations in the normalization of $v_{r, {\rm rms}}$ and $\beta$ profiles, as well as weaker fluctuations in the profiles' radial shapes.
All of our galaxies' time-averaged $\beta$ profiles are nearly self-similar and are reasonably well fit by an OM profile.

\section{Correlating sSFR with stellar kinematics}
\label{sec:observational_prediction}

In this section, we explore in detail the time (co)evolution of sSFR and stellar kinematics. We then use the correlation between star formation activity and stellar kinematics predicted by our simulations to formulate an observational test of the role of stellar feedback and feedback-driven potential fluctuations in the evolution of low-mass galaxies.
Because stars and dark matter respond kinematically to potential fluctuations in very similar ways \citepalias{El-Badry_2016}, our predictions also represent an observational test of the role of stellar feedback in regulating the  inner dark matter density profile in dwarf galaxies.

\subsection{Evolution of sSFR and stellar kinematics}

\begin{figure}
\includegraphics[width=\columnwidth]{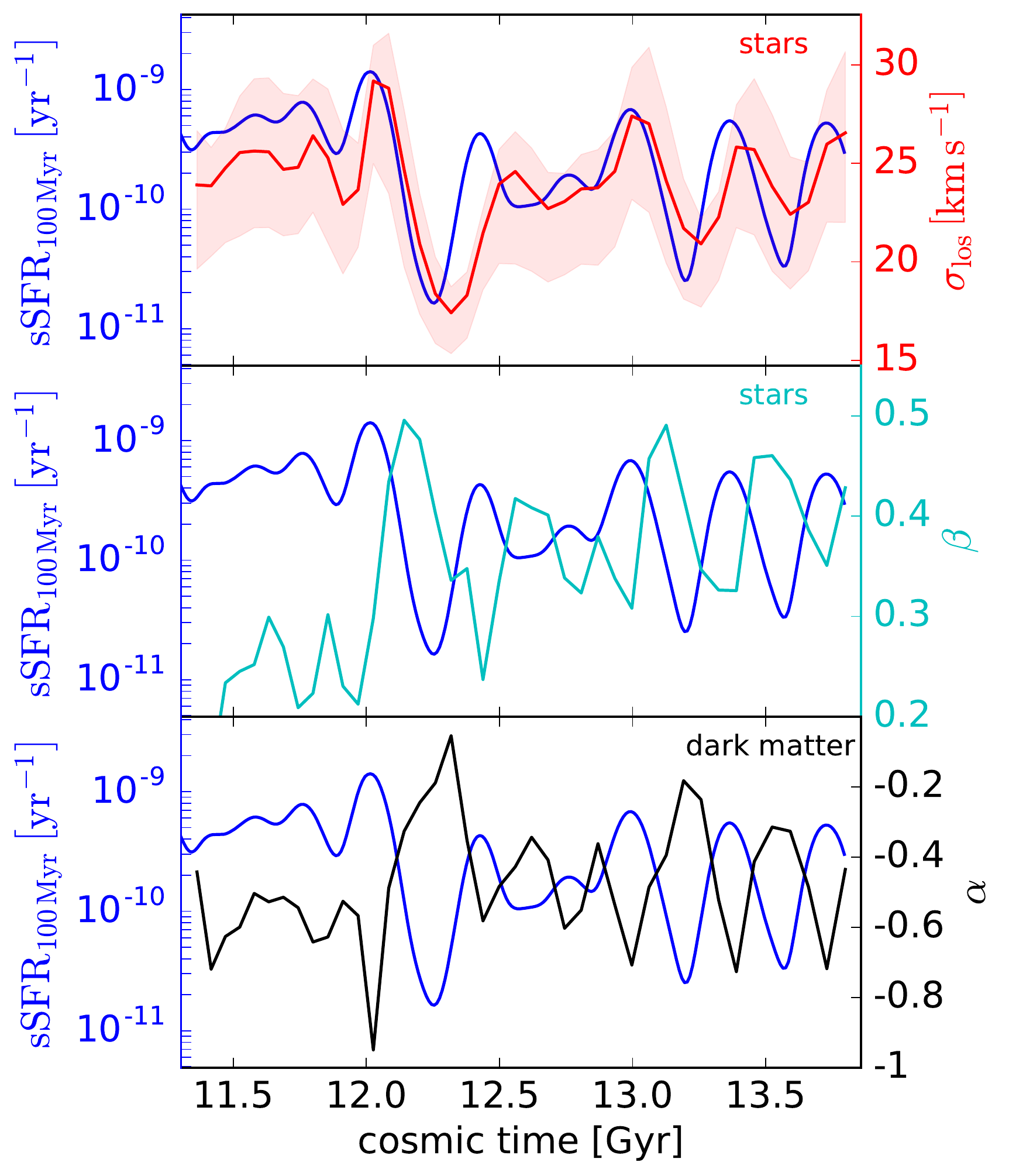}
\caption{
\textbf{Top:} Relation between specific star formation rate (sSFR) and line-of-sight stellar velocity dispersion $\sigma_{\rm los}$ as a function of cosmic time in \texttt{m10.6}. Blue curve shows sSFR, smoothed over 100 Myr, as labeled on the left y-axis, while red curve shows $\sigma_{\rm los}$, as labeled on the right y-axis. We ``observe'' each snapshot along 1000 lines-of-sight distributed uniformly on the unit sphere: red curve shows the median across these viewing angles and shaded region shows 68\% scatter. sSFR and $\sigma_{\rm los}$ are strongly correlated, with a time-offset of $\approx 50$ Myr.
\textbf{Middle:} Same as top panel, but for the velocity anisotropy $\beta$. Anisotropy is also closely related to sSFR. Peaks in $\beta(t)$ correspond to post-starburst periods in which sSFR is falling rapidly. 
\textbf{Bottom}: central slope $\alpha$ of the dark matter density profile $(\rho_{\rm DM}\propto r^{\alpha})$. $\alpha$ is anticorrelated with sSFR and $\sigma_{\rm los}$. The predicted coevolution of sSFR and $\sigma_{\rm los}$ provides an \textit{observable test} of the effect of stellar feedback on stellar kinematics, and by extension, the dark-matter density profile, in low-mass galaxies.}
\label{fig:m10.6_sfr_sigma_time}
\end{figure}

Figure~\ref{fig:m10.6_sfr_sigma_time} shows the time-evolution of sSFR and the line-of-sight stellar velocity dispersion $\sigma_{\rm los}$ in \texttt{m10.6} over the last $\approx 3.5$ Gyr since $z = 0.2$.
Because this galaxy is not perfectly spherically symmetric, $\sigma_{\rm los}$ varies somewhat with viewing angle, with typical differences of 50\% between the most extreme ``edge-on'' and ``face-on'' angles. To quantify this variation, we ``observe'' each snapshot along 1000 different lines of sight distributed uniformly on the unit sphere, calculating $\sigma_{\rm los}$ for each line of sight. We then calculate the median and 68\% scatter across these 1000 $\sigma_{\rm los}$ values for each snapshot.

sSFR and $\sigma_{\rm los}$ track each other remarkably well across all viewing angles. However, $\sigma_{\rm los}(t)$ is temporally offset from sSFR$(t)$ by $\approx 50$ Myr, a result of temporal delay between high sSFR causing a change in gas kinematics and the response of the stars to the changing potential.
We interpret this strong correlation and time delay as follows.
To first order, $\sigma_{\rm los}$ traces the depth of the gravitational potential, which is deepest when significant gas has accumulated in the galactic center, when in turn high gas densities drive high sSFR. As stellar feedback begins heating and rarefying the gas, driving galactic winds, sSFR starts to fall almost instantaneously. However, $\sigma_{\rm los}$ does not fall until \textit{after} significant gas mass is driven out, shallowing the overall gravitational potential, and this takes roughly a dynamical time. Similarly, during the post-starburst cooling phase, stars do not respond kinematically until gas accumulates in the galactic center and the potential contracts, again over approximately a dynamical time. Roughly consistent with the offset in Figure~\ref{fig:m10.6_sfr_sigma_time}, $t_{\rm dyn}$ in \texttt{m10.6} is $< 100\,{\rm Myr}$ in the central few kpc, where star formation occurs.

The middle panel of Figure~\ref{fig:m10.6_sfr_sigma_time} compares the time-evolution of the stellar velocity anisotropy, $\beta$, to that of the sSFR. Like the velocity dispersion, $\beta$ undergoes semi-periodic fluctuations. $\beta$ is highest during post-starburst outflow periods, when the stellar distribution is expanding, and is lowest when gas accumulates in the center and the potential contracts. However, there is a longer time-delay of $\approx 100$ Myr between peaks in sSFR$(t)$ and peaks in $\beta(t)$: the anisotropy does not rise until stars have begun to migrate outwards, when the sSFR has already started to decline.

Finally, the bottom panel of of Figure~\ref{fig:m10.6_sfr_sigma_time} shows the  time-evolution of the central slope $\alpha$ of the dark matter density profile $(\rho_{\rm DM}\propto r^{\alpha})$. We define $\alpha$ as the power law that best fits $\rho_{\rm DM}$ in the interval $r=(1\% –- 2\%) R_{200 \rm m}.$ Here, $\alpha \sim 0$ represents a flat central density profile (a ``core''), while $\alpha \sim −1$ represents a steep Navarro-Frenk-White-like profile (a ``cusp''). See \citet{Chan_2015} for further discussion of $\alpha$. $\alpha$ is anticorrelated with sSFR and $\sigma_{\rm los}$: the dark matter density profile is cuspy when gas is accumulated in the galactic center, leading to high sSFR and $\sigma_{\rm los}$, and is flatter during post-starburst outflow periods, when sSFR and $\sigma_{\rm los}$ also fall. 

The similar time evolution of sSFR, stellar kinematics, and the slope of the dark matter density profile shown in Figure ~\ref{fig:m10.6_sfr_sigma_time} demonstrates the fundamental relation between dark matter core creation and fluctuations in stellar kinematics: both processes are driven by fluctuations in the gravitational potential following bursts of star formation. This means that the relationship between sSFR and stellar kinematics predicted by our model can serve as an observational test for feedback-driven coring scenarios. While galaxies' orbital anisotropies are difficult to constrain observationally, $\sigma_{\rm los}$ can be measured straightforwardly from stellar absorption line widths. We thus quantify the correlation between sSFR and $\sigma_{\rm los}$ predicted by our model, to provide testable predictions for observations.

\subsection{Testable predictions}

\begin{figure}
\includegraphics[width=\columnwidth]{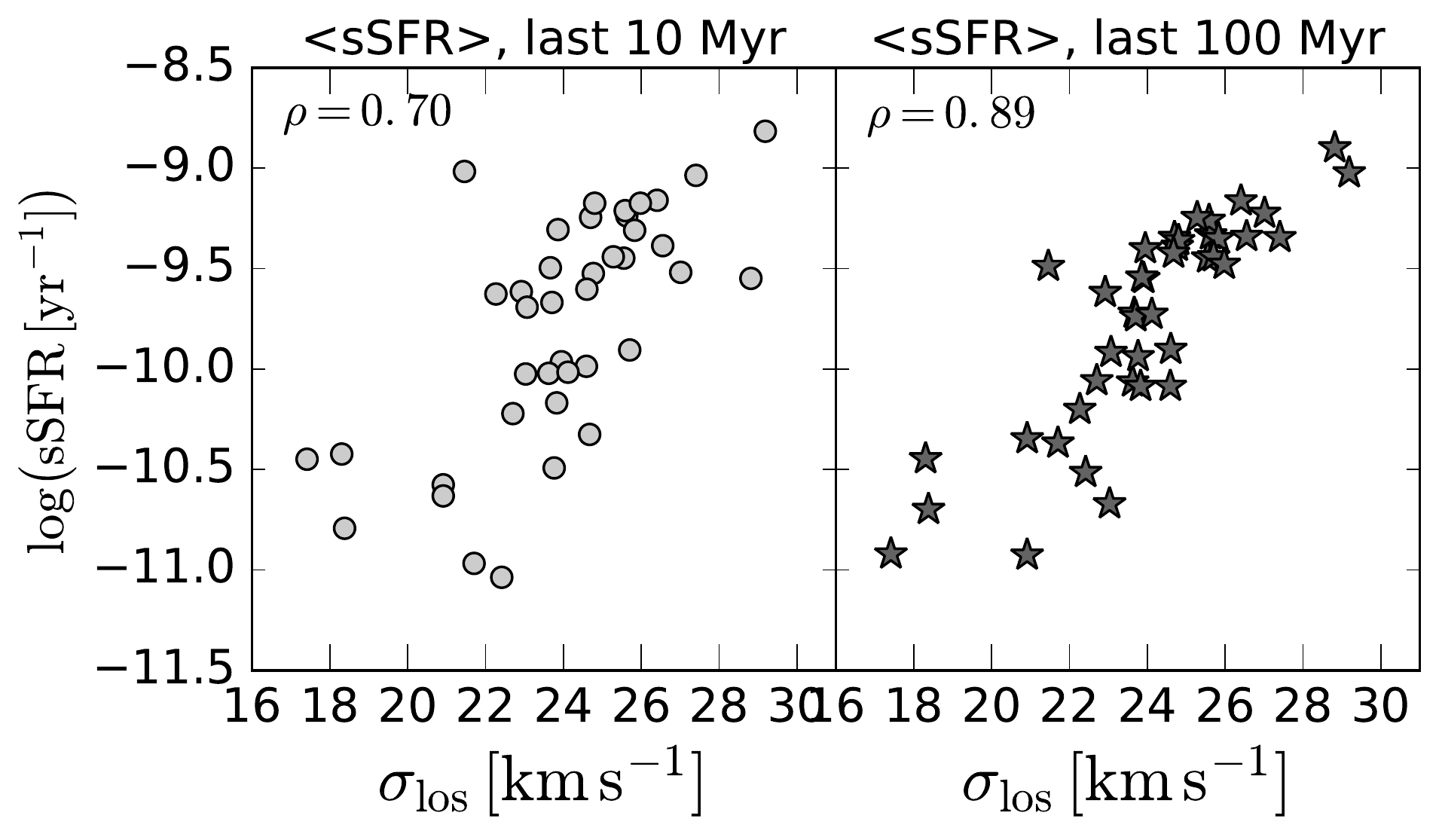}
\caption{
Stellar line-of-sight velocity dispersion, $\sigma_{\rm los}$, versus sSFR for the last 40 snapshots of \texttt{m10.6}. For each snapshot, we calculate $\sigma_{\rm los}$ along 1000 lines of sight as in Figure~\ref{fig:m10.6_sfr_sigma_time} and plot the median.
We show the correlation when sSFR is averaged over the last 10 Myr (left) and the last 100 Myr (right). Averaging sSFR over the last 100 Myr yields a tighter correlation. This is because there is a $\sim 50$ Myr offset between changes in sSFR and subsequent changes in stellar kinematics (see Figure~\ref{fig:m10.6_sfr_sigma_time}).
Each panel indicates the Spearman correlation coefficient, $\rho$.}
\label{fig:m10.6_sfr_vs_sigma_offset}
\end{figure}

Figure~\ref{fig:m10.6_sfr_vs_sigma_offset} shows sSFR versus $\sigma_{\rm los}$ for the last 40 snapshots of \texttt{m10.6}.
As before, we calculate $\sigma_{\rm los}$ along 1000 different lines of sight distributed uniformly on the unit sphere. Points show the median over these sightlines; we omit the scatter between different lines of sight for clarity.
We show sSFR averaged over the last 10 and 100 Myr, chosen to approximate respectively the response timescales of H$\alpha$ and ultraviolet continuum emission, two tracers of star formation commonly used in observations of nearby galaxies \citep[e.g.,][]{Kennicutt_1998, Weisz_2012}.
We show the Spearman correlation coefficient, $\rho$, which quantifies the rank correlation between sSFR and $\sigma_{\rm los}$, in each panel.

The left panel, which shows the relation between $\sigma_{\rm los}$ and sSFR averaged over the last 10 Myr, reveals a moderate correlation, with $\rho = 0.7$. However, the right panels show a significantly stronger correlation, $\rho = 0.89$, between $\sigma_{\rm los}$ and sSFR averaged over the last 100 Myr. We experimented with averaging sSFR over timescales between 0 and 200 Myr and find $\sim 100$ Myr to provide the tightest relation for this galaxy. This correlation reinforces a scenario in which the evolution of $\sigma_{\rm los}$ follows feedback-driven potential fluctuations that are strongest approximately 50 Myr after bursts of star formation. 

The correlation between sSFR and stellar kinematics is stronger when sSFR is averaged over 100 Myr than when it is averaged over 10 Myr, for two reasons. First, the instantaneous time-lag between sSFR and stellar kinematics is roughly 50 Myr (see Figure~\ref{fig:m10.6_sfr_sigma_time}), and averaging sSFR over the 100 Myr period preceding a snapshot is equivalent to averaging over a 100 Myr wide time bin centered 50 Myr before the snapshot. Second, the large-scale outflows that displace enough mass to significantly shallow the potential typically last more than 100 Myr. The 10 Myr averaged sSFR includes some minor maxima and minima due to bursts of localized star formation that do not drive galactic-scale outflows \citepalias[see][]{El-Badry_2016} and thus do not significantly alter stellar kinematics.

The other low-mass galaxies in our simulations produce the tightest correlation between sSFR and $\sigma_{\rm los}$ when sSFR is averaged over timescales ranging from 80 to 150 Myr. We find no clear scaling between this timescale and mass, likely because all the galaxies in our simulations except \texttt{m10} have similar dynamical times \citepalias{El-Badry_2016}.
All the galaxies in our sample except \texttt{m10} produce a positive correlation between 100 Myr averaged sSFR and $\sigma_{\rm los}$ with $\rho > 0.5$. We find no correlation between sSFR and $\sigma_{\rm los}$ in \texttt{m10}, because gas outflows in this galaxy do not displace enough mass to drive significant potential fluctuations.

This predicted correlation between sSFR and $\sigma_{\rm los}$ provides a clear observational test to determine whether real low-mass galaxies undergo feedback-driven potential fluctuations as in our simulations. It is of course not possible to directly trace the time-evolution of individual galaxies' sSFRs and kinematics as in Figure~\ref{fig:m10.6_sfr_sigma_time}. On the other hand, if the late-time evolution of individual galaxies over timescales longer than a few dynamical times is an ergodic process (which is a good approximation in our model), then a statistical survey of many galaxies with similar masses would stochastically sample galaxies at different phases in their burst cycles, and thus be directly comparable to the results of Figure~\ref{fig:m10.6_sfr_vs_sigma_offset}. This implies that, in a representative survey of low-mass galaxies with similar masses and a variety of SFRs, there should be a clear correlation between sSFR and stellar kinematics: galaxies that underwent a starburst in the last $\sim 100$ Myr should exhibit systemically higher $\sigma_{\rm los}$.

This prediction was recently tested by \citet{Cicone_2016}, who investigated the scaling of observed galaxies' stellar absorption line width with sSFR using stacked SDSS spectra from $\sim 160,000$ star-forming galaxies. In good agreement with our model, they found that, for galaxies with $M_{\rm star} \leq 10^{9.5}\,M_{\odot}$, starburst galaxies have systemically different stellar kinematics from galaxies with lower sSFR: absorption line widths (which trace $\sigma_{\rm los}$) increase with sSFR at fixed $M_{\rm star}$. Also consistent with our model, the correlation between $\sigma_{\rm los}$ and sSFR vanishes at higher $M_{\rm star}$. This occurs in our simulations as well \citepalias[see][]{El-Badry_2016}, because galaxies with $M_{\rm star} \gtrsim 10^{9.5}\,M_{\odot}$ have deeper potentials and thus do not undergo the potential fluctuations that regulate short-timescale evolution at lower masses.

More generally, the relationship between sSFR and stellar kinematics predicted by Figure~\ref{fig:m10.6_sfr_vs_sigma_offset} represents a compelling test of the role of stellar feedback in the overall evolution of low-mass galaxies. In particular, a myriad of studies \citep{Navarro_1996, Read_2005, Pontzen_2012, DiCintio_2014, Onorbe_2015, Chan_2015, Tollet_2016, Read_2016b, Wetzel_2016} have suggested that feedback-driven potential fluctuations are responsible for reducing the central densities of dark-matter in low-mass galaxies, flattening the inner density profile and thus providing a ``baryonic solution'' to the ``core-cusp'' and related ``too big to fail'' problems \citep{Flores_1994, Moore_1994, BoylanKolchin_2011, Jiang_2015}.

However, significant uncertainties persist in the nature of such baryonic solutions. The processes through which stellar feedback couples to gas in low-mass galaxies remain imperfectly understood, so it remains an open question whether stellar feedback can reconcile the density profiles of low-mass galaxies predicted by $\Lambda$CDM simulations with observations. Baryonic solutions are difficult to test in large part because of the observational challenges of robustly measuring the inner dark-matter density profiles of low-mass galaxies. However, as demonstrated in Figure ~\ref{fig:m10.6_sfr_sigma_time}, \textit{if stellar feedback-driven potential fluctuations heat the orbits of dark matter in dwarf galaxies, thus flatting the inner density profile and forming a ``core,'' then such feedback also must significantly alter the kinematics of stars}. The predicted observable correlation between sSFR and stellar kinematics thus serves as a clear test for any feedback-driven baryonic solution to discrepancies between $\Lambda$CDM predictions and the observed inner density profiles of dwarf galaxies.

\section{Spherical Jeans Modeling}
\label{sec:spherical_jeans_modeling}

\begin{figure}
\includegraphics[width=\columnwidth]{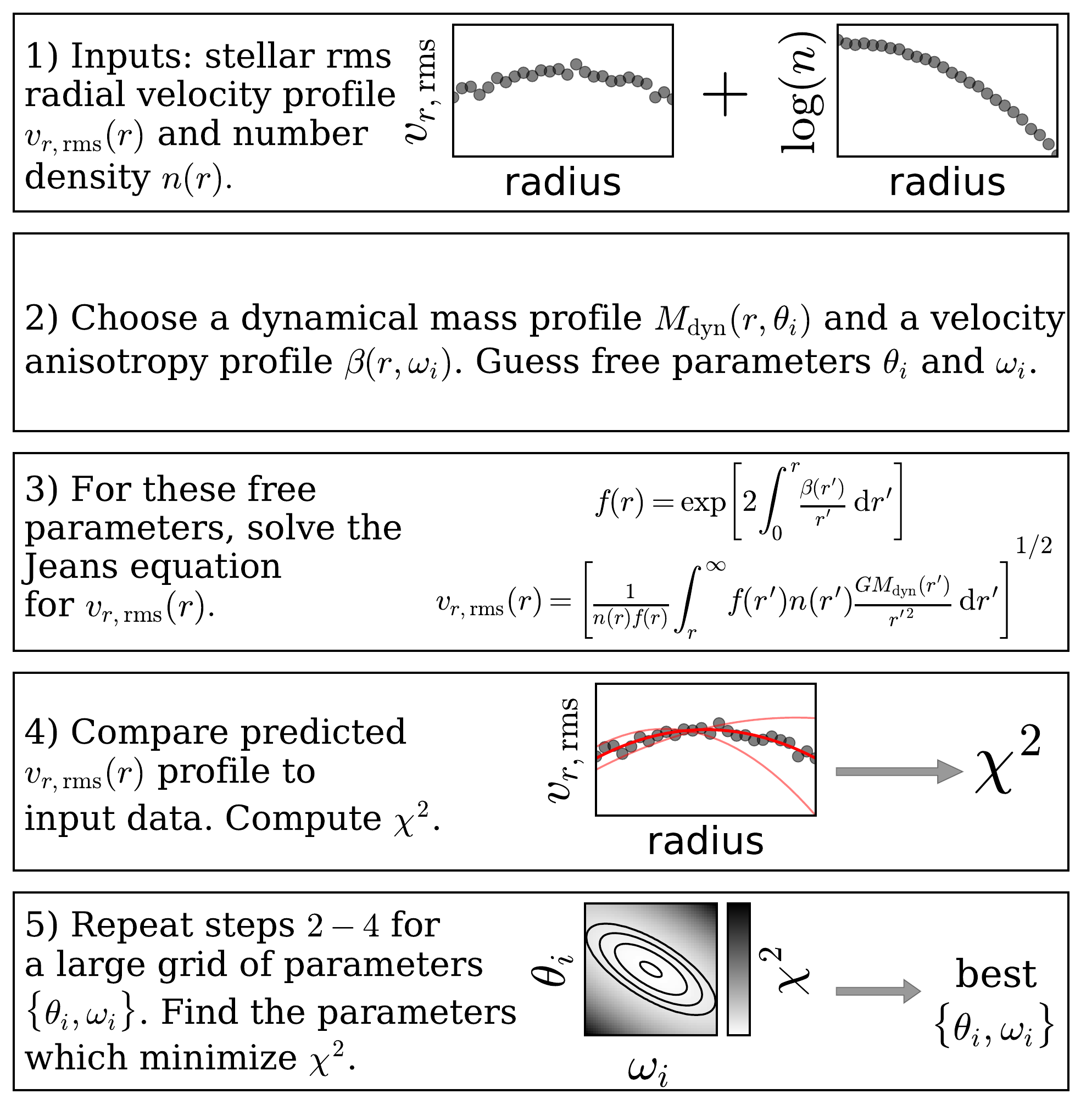}
\caption{Schematic overview of the Jeans modeling procedure. For details, see Section~\ref{sec:modeling_methods} and Appendix~\ref{sec:Jeans_Model_Explanation}.}
\label{fig:schematic}
\end{figure}

Having explored the effects of feedback-driven potential fluctuations on stellar kinematics, we now turn to Jeans dynamical modeling. Our goal is to assess the accuracy with which Jeans modeling can recover the underlying mass profiles of low-mass galaxies. In particular, we seek to test the consequences of bursty star formation and the resulting time-dependent gravitational potential on the accuracy of dynamical mass estimates.
We first describe our framework for building dynamical models of galaxies. Figure~\ref{fig:schematic} provides an overview of the entire procedure. We then present a more thorough description of our modeling approach in the context of other works, with further details in Appendix~\ref{sec:Jeans_Model_Explanation}.

To make our assessment of the effects of feedback-driven potential fluctuations as clear-cut as possible, we do not attempt to account for observational effects. That is, we assume that the rms radial velocity profile $v_{r,{\rm rms}}(r)$ and stellar 3D number density profile $n(r)$ are known exactly.
This is a major simplification over observational studies, for which kinematics are only available along a single line-of-sight and must be de-projected with assumptions about galaxies' 3D shape \citep[e.g.,][]{Binney_1982}.\footnote{
Typically, observational works measure the projected line-of-sight velocity dispersion profile $\sigma_{p}(R)$ and the surface brightness profile $I(R)$ and convert these into a number density $n(r)$ and a radial velocity dispersion $\sigma_r(r)$ profile through an Abel integral transform under the assumption of spherical symmetry. This requires assumptions about the stellar mass-to-light ratio $\Upsilon_{\rm star}(r)$, and the overwhelming majority of observational Jeans modeling studies assume $\Upsilon_{\rm star}(r) = {\rm const}$ for simplicity. However, this choice is not well justified, because many galaxies studied with Jeans modeling have significant radial gradients in their stellar populations \citep[see][and references therein]{Schroyen_2013}.}
Our tests below are thus intentionally idealized, allowing us to disentangle cleanly the effects of incomplete knowledge of stellar kinematics from those of non-equilibrium fluctuations in the gravitational potential. Fully accounting for observational errors will exacerbate the errors in our dynamical mass estimates.

In the first part of our analysis, we assume that the stellar anisotropy profile $\beta(r)$ is known exactly. Then, in Section~\ref{sec:different_beta_models}, we consider the effects of uncertainty in $\beta(r)$ on the accuracy of dynamical modeling.

Note that none of our simulated low-mass galaxies have rotation-supported stellar kinematics; they all have $\overline{v_{ \phi}} / v_{\rm rms}$ < 0.2. Our galaxies are not completely spherically symmetric; at $z=0$ they have typical axis ratios in the range $c/a = 0.4 - 0.7$. However, Jeans modeling errors caused by departure from spherical symmetry are smaller than errors arising from incomplete knowledge of $\beta$ or from non-equilibrium fluctuations: the least spherical galaxy in our sample (\texttt{m10}, with $c/a\approx 0.4$) has the smallest typical mass modeling errors ($\sim 3\%$). Spherical Jeans modeling is most commonly used in observational works to study dSph and dE galaxies, which have a wide variety of axis ratios in the range $c/a \approx 0.2-0.9$ \citep{Wheeler_2015}.

\subsection{Modeling methods}
\label{sec:modeling_methods}

For a spherically symmetric system with no net streaming motions, the kinematics and spatial distribution of tracer particles are related to the dynamical mass profile $M_{\rm dyn}(r)$ by the Jeans equation \citep[e.g.,][]{Binney_1980}: 
\begin{equation}
\label{eqn:Jeans_eq}
\frac{{\rm d}[n(r) \overline{v_{r}^{2}}(r)]}{{\rm d}r} + 2 \frac{\beta(r)}{r} n(r) \overline{v_{r}^{2}}(r) = -n(r) \frac{G M_{\rm dyn}(r)}{r^{2}}
\end{equation}
Here, $n(r)$ is the number density of tracer (star) particles, $\overline{v_r^2}(r)$ is the mean of the squared radial velocity of stars at radius $r$, $\beta(r)$ is the anisotropy parameter given by Equation~\ref{eqn:beta}, and $M_{\rm dyn}(r)$ is the total mass of stars, gas, and dark matter enclosed in a sphere of radius $r$. In the equilibrium case, $\overline{v_r} = 0$, so many authors write $\sigma_{r}^{2}\equiv \overline{v_{r}^{2}} - \overline{v_{r}}^{2}$ in place of $\overline{v_{r}^{2}}$. The goal of Jeans modeling is to calculate the form of $M_{\rm dyn}(r)$ from $n(r)$, $\overline{v_r^2}(r)$, and $\beta(r)$.

Equation~\ref{eqn:Jeans_eq} in principle can be solved explicitly for $M_{\rm dyn}(r)$, yielding a direct, nonparametric estimate of the total mass profile. However, this approach has several problems. First, because typical $\overline{v_r^2}(r)$ and $n(r)$ profiles are noisy and their derivatives are even noisier, this approach generally requires the data to be smoothed \citep[e.g.,][]{Gebhardt_1995} or fit with some smooth function \citep[e.g.,][]{Kleyna_2004, Douglas_2007, Gilmore_2007, Napolitano_2009, Napolitano_2011}. In this case, the shape of the inferred density profile can depend nontrivially on the choice of smoothing procedure \citep{Gebhardt_1994, Gebhardt_1995} or on the choice of functions used to fit $\overline{v_r^2}(r)$, $n(r)$, and $\beta(r)$. Second, if the anisotropy is not known a priori (as is usually the case in observational studies), this approach does not provide any method for constraining $\beta(r)$. Finally, if a system is not in dynamical equilibrium, solving Equation~\ref{eqn:Jeans_eq} directly for $M_{\rm dyn}(r)$ can lead to a predicted mass profile that is not monotonically increasing, requiring unphysical negative densities.

For these reasons, we instead obtain $M_{\rm dyn}(r)$ from Equation~\ref{eqn:Jeans_eq} using a parameterized fit, which is the approach most commonly used in the observational literature \citep[e.g.,][]{Fitchett_1988, Fischer_1992, Hui_1995, Geha_2002, Lokas_2002, Lokas_2005, Walker_2009, Lokas_2010b, Bonnivard_2015, Chen_2016}. We first assume a functional form of $M_{\rm dyn} = M_{\rm dyn}(r,\theta_i)$, parameterized by $N$ free parameters $\{\theta_{i}\}$. In the general case in which we have no a priori knowledge of $\beta(r)$, we also assume a functional form for the anisotropy $\beta = \beta(r, \omega_i)$, parameterized by an additional $W$ free parameters $\{\omega_i\}$.

Then, for an $(N+W)$ dimensional grid of different values of $\{\theta_{i},\omega_i\}$, we integrate Equation~\ref{eqn:Jeans_eq} to solve for $\overline{v_{r}^{2}}(r)$, producing a predicted $\overline{v_{r}^{2}}(r)$ profile for every point $\{\theta_{i},\omega_i\}$ in the grid. Finally, we compare the predicted $\overline{v_{r}^{2}}(r)$ profiles for each $\{\theta_{i}, \omega_i\}$ to the true, measured $\overline{v_{r}^{2}}(r)$ data points. We choose the set of parameters $\{\theta_{i},\omega_i\}$ that minimizes the $\chi^2$ statistic for this comparison as the ``best-fit'' parameters.

This approach requires an a priori assumption about the form of $\rho_{\rm dyn}(r)$ and $M_{\rm dyn}(r)$. A simple solution \citep[e.g.,][]{Gebhardt_2002, Romanowsky_2003, Klimentowski_2007, Walker_2007, Geha_2010, Lokas_2010b} is to assume that mass follows light, that is, $\rho_{{\rm dyn}}\left(r\right) = \Upsilon\times n \left(r\right)$ and $M_{\rm dyn} \left(r \right) = 4\pi\Upsilon\int_{0}^{r}r'^{2}n(r')\,{\rm d}r'$, in which case the mass-to-light ratio $\Upsilon$ is the only model parameter $\{\theta_{i}\}$. This approach is well suited to systems in which the visible tracer dominates the mass, such as globular clusters. However, the density of stellar tracers in our galaxies falls off more steeply at large radius than the total density profile, which is dominated by dark matter. We thus opt for a functional form of $\rho_{\rm dyn}(r)$ that has a different shape than the stellar density $n(r)$.

Following \citet{Adams_2014}, we experimented with two forms of $\rho_{\rm dyn}(r)$ that are common in the literature: a three-parameter ``generalized NFW'' profile \citep[gNFW;][]{Zhao_1996, Wyithe_2001}, which leaves the central density slope as a free parameter but forces the NFW $\rho \propto r^{-3}$ behavior at large radius, and a two-parameter ``Burkert'' profile  \citep{Burkert_1995}, which forces a constant-density core at small radius but also scales as $\rho \propto r^{-3}$ at large radius.
We present a detailed comparison of our results using the two profiles in Appendix~\ref{sec:Jeans_Model_Explanation}, finding that the two-parameter Burkert profile constrains the dynamical masses of all our galaxies as well as or better than the more complicated gNFW profile.
Thus, throughout our analysis we exclusively use the Burkert model, as given by
\begin{equation}
\label{eqn:Burkert_rho}
\rho_{{\rm Burkert}}\left(r\right)=\frac{\rho_{b}}{\left(1+r/r_{b}\right)\left(1+(r/r_{b})^{2}\right)},
\end{equation}
where $r_b$ and $\rho_b$ are free parameters. The Burkert profile forces a core in the density profile: $\rho_{\rm Burkert}(r)$ is constant for $r \ll r_b$ and goes as $\rho \propto r^{-3}$ at $r \gg r_b$.
Equation~\ref{eqn:Burkert_Mr} gives the corresponding form of $M_{\rm dyn}(r)$.

For the case in which the anisotropy is not known a priori, we also require an assumption about the form of $\beta(r)$. The most common approach in the observational literature \citep[e.g.,][]{Geha_2002, Lokas_2005, Walker_2007, Lokas_2009, Walker_2009} is to assume constant anisotropy, that is, $\beta(r) = \beta_0$. In this case, $\beta_0$ is the only model parameter $\{\omega_i\}$. Another popular choice \citep[e.g.,][]{Kleyna_2001, Mamon_2005, Battaglia_2008, Serra_2010, Chen_2016} is the OM profile given in Equation~\ref{eqn:beta_OM}; in this case, one fits for the OM ``anisotropy radius'': $\{\omega_i\}=r_a$. In Section~\ref{sec:different_beta_models}, we investigate how using different models for $\beta(r)$ affects the accuracy of our dynamical mass estimates.  

Armed with a functional form of $M_{\rm dyn}(r)$, which depends on two free parameters $\{\theta_{i}\}=\{r_b, \rho_b\}$, and either a functional form for $\beta(r)$, which depends on additional free parameters $\{\omega_{i}\}$, or a $\beta(r)$ profile measured directly from the simulation, we can compute the $\overline{v_{r}^2}(r)$ profile predicted for a spherical system in equilibrium with the known $n(r)$ profile. Equation~\ref{eqn:Jeans_eq} can be solved for $\overline{v_{r}^2}(r)$ \citep{vanderMarel_1994, Mamon_2005, Read_2016} by introducing an integrating factor
\begin{equation}
\label{eqn:int_factor}
f(r)=\exp\left[2\int_{0}^{r}\frac{\beta(r')}{r'}{\rm d}r'\right].
\end{equation}
Using the boundary condition that $\overline{v_{r}^{2}}$ goes to 0 as $r\to\infty$, we integrate Equation~\ref{eqn:Jeans_eq} and solve for $n\overline{v_{r}^{2}}$, yielding
\begin{equation}
\label{eqn:vr_square}
\overline{v_{r}^{2}}(r)=\frac{1}{f(r)n(r)}\int_{r}^{\infty}f(r')n(r')\frac{GM_{{\rm dyn}}(r')}{r'^{2}}\,{\rm d}r'.
\end{equation}

In practice, we implement this Jeans modeling procedure as follows. We first calculate $n$ in 50 spherical shells spaced linearly between $r_{\rm min} = 0$ and $r_{\rm max} = 2 \, R_{90 \rm m}$. For a given set of model parameters $\left\{r_{b},\rho_{b},\omega_i\right\}$ describing the mass distribution and anisotropy, we evaluate Equation~\ref{eqn:vr_square} numerically to produce an array of predicted rms radial velocities $v_{r,{\rm rms}}(r_i)\equiv \sqrt{\smash[b]{\overline{v_{r}^{2}}}}(r_{i})$ in 25 spherical bins of radius $r_i$, where $r_i$ are spaced linearly between $r=0$ and $r=R_{90 \rm m}$. We verified that our results are not sensitive to binning: increasing or decreasing our bin size by a factor of two typically changes the $v_{r,{\rm rms}}(r_i)$ profile predicted by Equation~\ref{eqn:vr_square} by less than one percent. We discuss the effects of varying binning and the radial range over which $v_{r, \rm rms}$ is sampled in Appendix~\ref{sec:bin_spacing}. We find that dynamical masses inferred from Jeans modeling become significantly less accurate beyond the maximum radius where $v_{r, \rm rms}$ can be measured.

In carrying out the numerical integration to evaluate Equation~\ref{eqn:vr_square}, we set $n(r) = 0$ for $r>r_{\rm max}$. This allows us to use $n(r_i)$ values calculated directly from the simulation in discrete radius bins, without having to fit analytic profiles to $n(r)$ in order to evaluate the integral as $r'\to\infty$. Fitting profiles to $n(r)$ is common in the literature \citep[e.g.,][]{Geha_2002, Lokas_2002, Lokas_2005, Walker_2009, Lokas_2010, Bonnivard_2015, Chen_2016}, but we wished to avoid it because we found that the choice of profile can sometimes affect the predicted $v_{r,{\rm rms}}(r_i)$ profile when $n(r)$ is not smooth. We have also verified that the $v_{r,{\rm rms}}(r_i)$ values predicted by Equation~\ref{eqn:vr_square} are not sensitive to the choice of $r_{\rm max}$ as long as $r_{\rm max} > R_{90 \rm m}$. This is because $n(r)$ falls off quickly at large radius while $\beta(r)$ is approximately constant, so the integrand goes to 0 for $r \gg R_{90 \rm m}$ irrespective of the behavior of $f(r)$. Observational studies can typically measure galaxy surface brightness profiles out to significantly larger radii than they can measure stellar kinematics (which require spectroscopy) \citep{Bundy_2015}, so the form of $n(r)$ at large radius is not a dominant source of uncertainty.

Once we have the $v_{r,{\rm rms}}(r_i)$ profile predicted for a particular set of $\{r_{b},\rho_{b}, \omega_i\}$, we calculate the corresponding $\chi^2$ statistic:\footnote{
The $\chi^2$ statistic is conventionally defined with an uncertainty term $\sigma_i^2$ in the denominator. We do not attempt to account for observational errors, and thus use the model value, $v_{r,{\rm rms}}(r_i)$, instead. This only affects the normalization of $\chi^2$ values, which is arbitrary. The utility of the $\chi^2$ statistic lies in comparing the \textit{relative} goodness of fit of $v_{r, {\rm rms}}$ profiles predicted by different models. We also experimented with using ``Poisson uncertainties'' scaled as $\sigma_i \propto N_i^{-1/2}$ where $N_i$ is the number of star particles in each radial bin, but found no significant differences in our results.}
\begin{equation}
\label{eqn:chi_square}
\chi^{2}=\sum_{i}\frac{\left(v_{r,{\rm rms}}(r_{i})-v_{r,{\rm rms},i}\right)^{2}}{v_{r,{\rm rms}}(r_{i})},
\end{equation}
where $v_{r,{\rm rms},i}$ are the \textit{true} values of $v_{r,{\rm rms}}$ in spherical shells of radius $r_i$.

We carry out the $\chi^2$ minimization using an initial brute force step followed by an optimization step to improve precision. We begin by laying down a coarse grid spanning all plausible regions of $\{r_b, \rho_b,\omega_i\}$ parameter space. We calculate a $v_{r,{\rm rms}}(r_i)$ array and corresponding $\chi^2$ value for each point in the grid and find the gridpoint ${\mathbf p_0}= \{r_{b}, \rho_{b},\omega_{i}\}$ at which the $\chi^2$ value is minimized. We then use ${\mathbf p_0}$ as the starting point for an optimization algorithm to begin searching for the global minimum, which is always in the vicinity of ${\mathbf p_0}$. This details of this minimization procedure are explained in Appendix ~\ref{sec:chi_square_minization}.

It is important to ensure that the $\chi^2$ minimization procedure converges on the true global minimum of $\chi^2(r_b, \rho_b,\omega_i)$ rather than on a local minimum in a different region of parameter space. We have tested our minimization method extensively for snapshots in a variety of dynamical states; see Appendix ~\ref{sec:chi_square_minization}. We find that for the two-parameter Burkert profile, the $\chi^2$ function is sufficiently smooth that there is little danger of converging on a local minimum, as the function contains only a single, global, minimum. Even when the $\chi^2$ function depends on additional free parameters -- that is, for parameterized models of $\beta(r)$ or more complicated $M_{\rm dyn}(r)$ profiles -- we always converge on the true $\chi^2$ minimum. There is, however, significant degeneracy between the profiles' core size $r_b$ and normalization $\rho_b$. The $v_{r,{\rm rms}}$ profiles predicted by Equation~\ref{eqn:vr_square} depend explicitly only on the total enclosed mass $M_{\rm dyn}(r)$, so similar $v_{r,{\rm rms}}$ profiles result from mass profiles with small $r_b$ and high $\rho_b$ and profiles with large $r_b$ and low $\rho_b$, as the two families of profiles have similar total enclosed masses. We find additional degeneracies between $\rho_b$ and anisotropy, which we describe in Appendix~\ref{sec:mass_anisotropy_degeneracy}.

We have checked that, like the predicted $v_{r, {\rm rms}}$ profiles, our dynamical mass estimates are not sensitive to changes in the number of radial bins in which we calculate $v_{r,{\rm rms}}$. This remains true as long as bins are linearly spaced. Switching to logarithmically spaced bins can nontrivially change the best-fit $M_{\rm dyn}(r)$ profile returned by $\chi^2$ minimization. This can be understood as follows. Because all bins are weighted equally in Equation~\ref{eqn:chi_square}, if bins are linearly spaced, our procedure returns the $\{r_b,\rho_b,\omega_i\}$ set that produces the $v_{r,{\rm rms}}$ profile in closest \textit{overall} agreement with the true profile. If, on the other hand, bins are logarithmically spaced, $v_{r,{\rm rms}}$ will be most frequently sampled at small radius, and $\chi^2$ minimization will return the $\{r_b,\rho_b,\omega_i\}$ set that produces the $v_{r,{\rm rms}}$ profile in closest agreement with the true $v_{r,{\rm rms}}$ profile \textit{at small radius}. We opt to use linearly rather than logarithmically spaced bins, because this is the standard choice in the observational literature \citep[e.g.,][]{Geha_2002, Lokas_2009, Walker_2009}, and because logarithmic bins become undersampled and dominated by Poisson noise at small radius. We compare results using different radial binning schemes in Appendix~\ref{sec:bin_spacing}.

\subsection{Modeling results for known \texorpdfstring{$\beta(r)$}{Lg}}
\label{sec:jeans_modeling_results}

\begin{figure}
\includegraphics[width=\columnwidth]{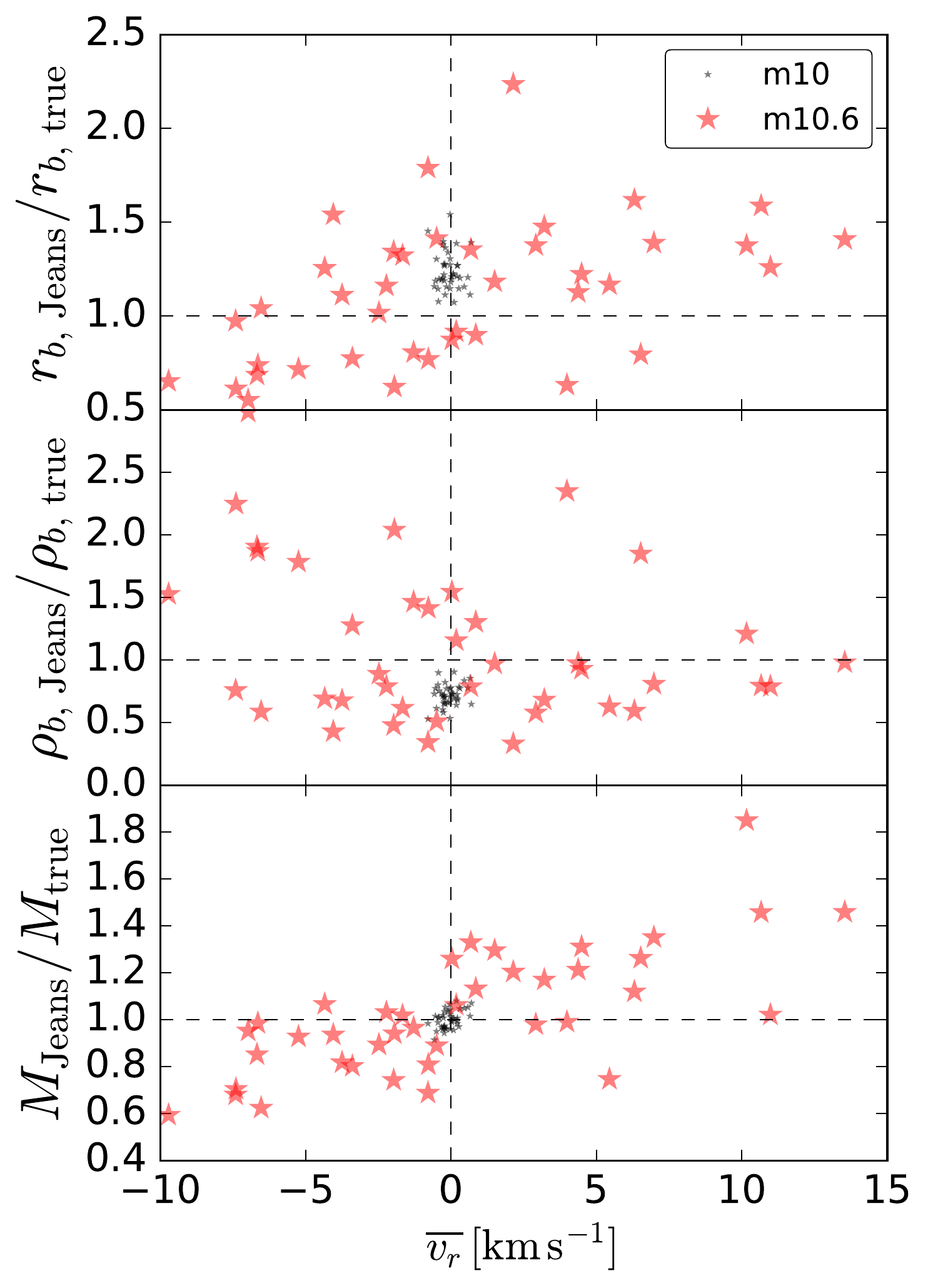}
\caption{
Jeans modeling of \texttt{m10} and \texttt{m10.6}. Each panel shows the ratio of a dynamical quantity predicted by Jeans modeling to the true value of that quantity. \textbf{Top}: $r_b$, the Burkert profile scale radius. \textbf{Middle}: $\rho_b$, the Burkert characteristic density. \textbf{Bottom}: total dynamical mass enclosed within $R_{90 \rm m}$. Individual points correspond to the last 40 simulation snapshots since $z \sim 0.2$. $x$-axis shows  $\overline{v_r}$, the mean radial velocity of stars within $R_{90 \rm m}$, which approximately traces the dynamical state of the system: $\overline{v_r} > 0$ corresponds to outflow, $\overline{v_r} < 0$ to inflow, and $\overline{v_r} = 0$ to (instantaneous) equilibrium. Although Jeans modeling accurately constrains the total mass of \texttt{m10}, it does not recover the true best-fit values of $r_b$ and $\rho_b$. Dynamical mass errors are much larger in \texttt{m10.6}, a consequence of the feedback-driven potential fluctuations the galaxy undergoes, and clearly correlate with $\overline{v_r}$.
}
\label{fig:two_gals_jeans_vs_real}
\end{figure}

Before investigating the effects of using different models for $\beta(r)$ in Jeans modeling, we first consider the ideal case in which anisotropy is known exactly, that is, we measure $\beta(r)$ directly from the simulation. While measuring $\beta(r)$ directly remains infeasible for nearby low-mass galaxies, it is feasible for globular clusters and is expected to become possible for dwarf galaxies within $\sim 100$ kpc in the next decade \citep[e.g.,][]{Kallivayalil_2015}.

Figure~\ref{fig:two_gals_jeans_vs_real} shows the results of Jeans modeling with known $\beta(r)$ for two of our galaxies. In each panel, we plot (for the last 40 snapshots since $z \approx 0.2$) the ratio of the parameters of the best-fit Burkert profile found through Jeans modeling to those found by fitting the known density profile $\rho_{\rm dyn}(r)$ directly. 
The top and middle panels show best-fit $r_b$ and $\rho_b$ parameter values, while the bottom panel shows the total mass enclosed within $R_{90 \rm m}$ by the Burkert profile with those $r_b$ and $\rho_b$ values. Parameters in all panels are plotted against $\overline{v_r}$,  the mean radial velocity of stars. To first order, $\overline{v_r}$ is an indicator of the dynamical state of the galaxy. When $\overline{v_r} > 0$, stars are moving outward in a shallowed potential following a gas outflow. When $\overline{v_r} < 0$, stars are migrating back inwards as gas re-accumulates in the galactic center and the potential contracts. Finally, when $\overline{v_r} = 0$, the galaxy either is transitioning between these two regimes or is undergoing more passive evolution.

Estimates of the total enclosed mass are both more accurate and better converged across snapshots in \texttt{m10} than in \texttt{m10.6}. In all of the last 40  snapshots of \texttt{m10}, Jeans modeling recovers the true total mass within $R_{90 \rm m}$ with errors of less than 10\%. In contrast, the fractional error in the recovered dynamical mass is $\sim 50\%$ in many snapshots of \texttt{m10.6}, with the most extreme error overestimating the total mass by nearly a factor of two. This is a direct consequence of the larger potential fluctuations in \texttt{m10.6}, which prevent the galaxy from reaching dynamical equilibrium. \texttt{m10} undergoes only very weak potential fluctuations, as evidenced its low $|\overline{v_r}|$ values. 

Although Jeans modeling reliably recovers the total enclosed mass of \texttt{m10}, the agreement between the best-fit $r_b$ and $\rho_b$ parameters found through Jeans modeling and those found with direct fitting is not very good. In all snapshots, Jeans modeling predicts a Burkert profile with a larger core and lower central density than the true best-fit Burkert profile. Through Equation~\ref{eqn:vr_square}, Jeans modeling depends explicitly only on the total enclosed mass $M_{\rm dyn}(r)$. Because models with large cores and low central densities produce similar integrated mass profiles to models with small cores and high central densities, Jeans modeling is unable to reliably distinguish between the two. Of course, Equation~\ref{eqn:vr_square} is not completely blind to the shape of the density profile, because $\rho_{\rm dyn}(r)$ is related to $M_{\rm dyn}(r)$ through a derivative, but, as can be seen from Equation~\ref{eqn:Jeans_eq}, $\rho_{\rm dyn}(r)$ depends on the \textit{second} derivatives of $n(r)$ and $v_{r,{\rm rms}}(r)$, and the slope of $\rho_{\rm dyn}(r)$ depends on their \textit{third} derivatives. These can differ significantly even between $n(r)$ and $v_{r,{\rm rms}}(r)$ profiles that are qualitatively similar. Thus, even small deviations from dynamical equilibrium or spherical symmetry can introduce nontrivial errors in the shape of galaxies' recovered density profiles, even if their total dynamical masses are well-constrained. See \citet{Wolf_2010} and \citet{Battaglia_2013} for further discussion of the sensitivity of Jeans modeling to different dynamical quantities.

The relation between $\overline{v_r}$ and $M_{\rm Jeans}/M_{\rm true}$, which can be seen for \texttt{m10.6} in the bottom panel of  Figure~\ref{fig:two_gals_jeans_vs_real}, demonstrates the effect of non-equilibrium fluctuations on the accuracy of Jeans modeling. 
Snapshots with $\overline{v_r} > 0$ (outflow) consistently predict dynamical masses that are too large, while in snapshots with $\overline{v_r} < 0$ (inflow), Jeans modeling almost always underestimates the total enclosed mass. The best-fit Burkert $r_b$ and $\rho_b$ values also reflect this to some extent: in snapshots with $\overline{v_r} > 0$, Jeans modeling typically predicts $r_b$ and $\rho_b$ values that are larger and smaller, respectively, than the true values. The opposite is true for snapshots with $\overline{v_r} < 0$.  Mass errors are all smaller ($< 10\%$) in \texttt{m10}, because $\overline{v_r}$ is only a few percent of $\sigma_r$, and the galaxy is never far from dynamical equilibrium. However, there is still a weak correlation evident between $\overline{v_r}$ and $M_{\rm Jeans}/M_{\rm true}$ in \texttt{m10}.

\begin{figure}
\includegraphics[width=\columnwidth]{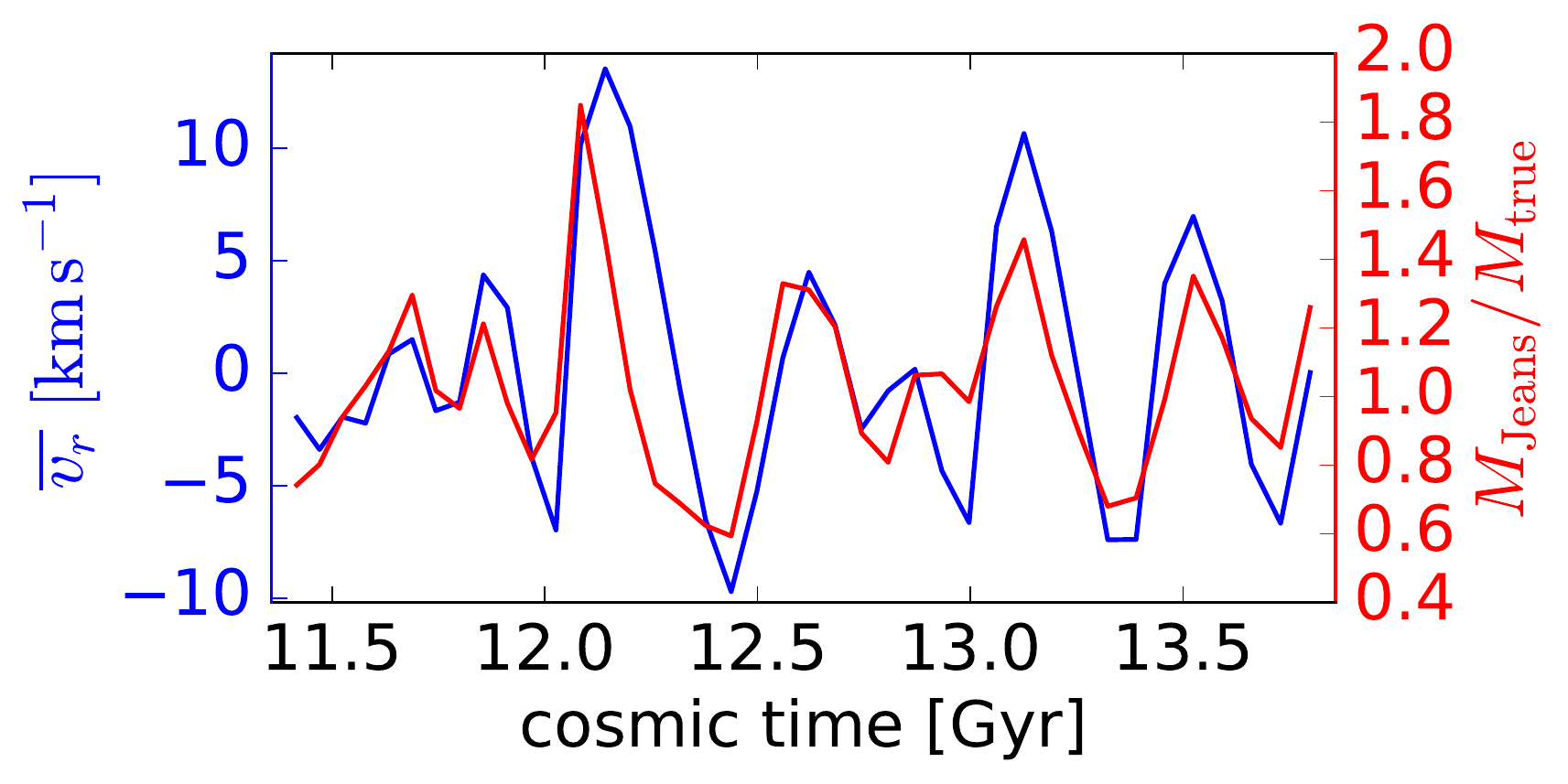}
\caption{
Time evolution of the accuracy of dynamical mass estimates from Jeans modeling in \texttt{m10.6}.
\textbf{Left}: mean stellar radial velocity $\overline{v_r}$, which traces the dynamical state of the galaxy.
\textbf{Right}: Ratio of $M_{\rm Jeans}$, the total dynamical mass inside $R_{90 \rm m}$ as calculated from Jeans modeling, to the true enclosed dynamical mass. Jeans modeling generally overpredicts the dynamical mass during outflow periods ($\overline{v_r} > 0$) and underpredicts it during inflow $(\overline{v_r} < 0$).}
\label{fig:jeans_error_vs_vr_time}
\end{figure}

We further investigate the relationship between non-equilibrium fluctuations in the stellar distribution and Jeans modeling mass errors in Figure~\ref{fig:jeans_error_vs_vr_time}, which shows side-by-side the time evolution of $\overline{v_r}$ and the error in the dynamical mass predicted by Jeans modeling in \texttt{m10.6} since $z=0.2$. Overall, the enclosed mass error and the mean radial velocity trace each other remarkably closely. During each inflow/outflow episode, Jeans modeling predictably first overestimates and then underestimates the dynamical mass. The fractional error in the Jeans mass estimate scales approximately with the magnitude of the $\overline{v_{r}}$ fluctuation.
Snapshots with $\overline{v_r} = 0$ tend to have $M_{\rm Jeans}/M_{\rm true} \approx 1$. In a time-averaged sense, potential fluctuations thus do not systemically \textit{bias} dynamical mass estimates, but rather, they drive significant (of order unity) scatter.

This relationship between $\overline{v_r}$ and error in Jeans modeling mass estimates can be understood as follows. When $\overline{v_r} > 0$, the potential has just become shallower because of a gas outflow. As measured by $v_{r,{\rm rms}}$, the stellar velocity dispersion is then too high for the newly shallowed potential to sustain the current stellar distribution, so stars begin to move outward. However, Jeans modeling necessarily assumes that the galaxy is in equilibrium, and thus that the current (high) $v_{r,{\rm rms}}$ values require a deeper potential. On the other hand, when $\overline{v_r} < 0$, the potential has just contracted again because of renewed gas accumulation in the galactic center. In this case, the $v_{r,{\rm rms}}$ values have not yet had time to respond to the recently deepened potential and are thus lower than they would be in equilibrium. Jeans modeling therefore interprets the current (low) $v_{r,{\rm rms}}$ values as indicative of a shallow gravitational potential and consequently underestimates the enclosed mass.

\subsection{Scaling with mass}
\label{sec:mass_scaling}

\begin{figure}
\includegraphics[width=\columnwidth]{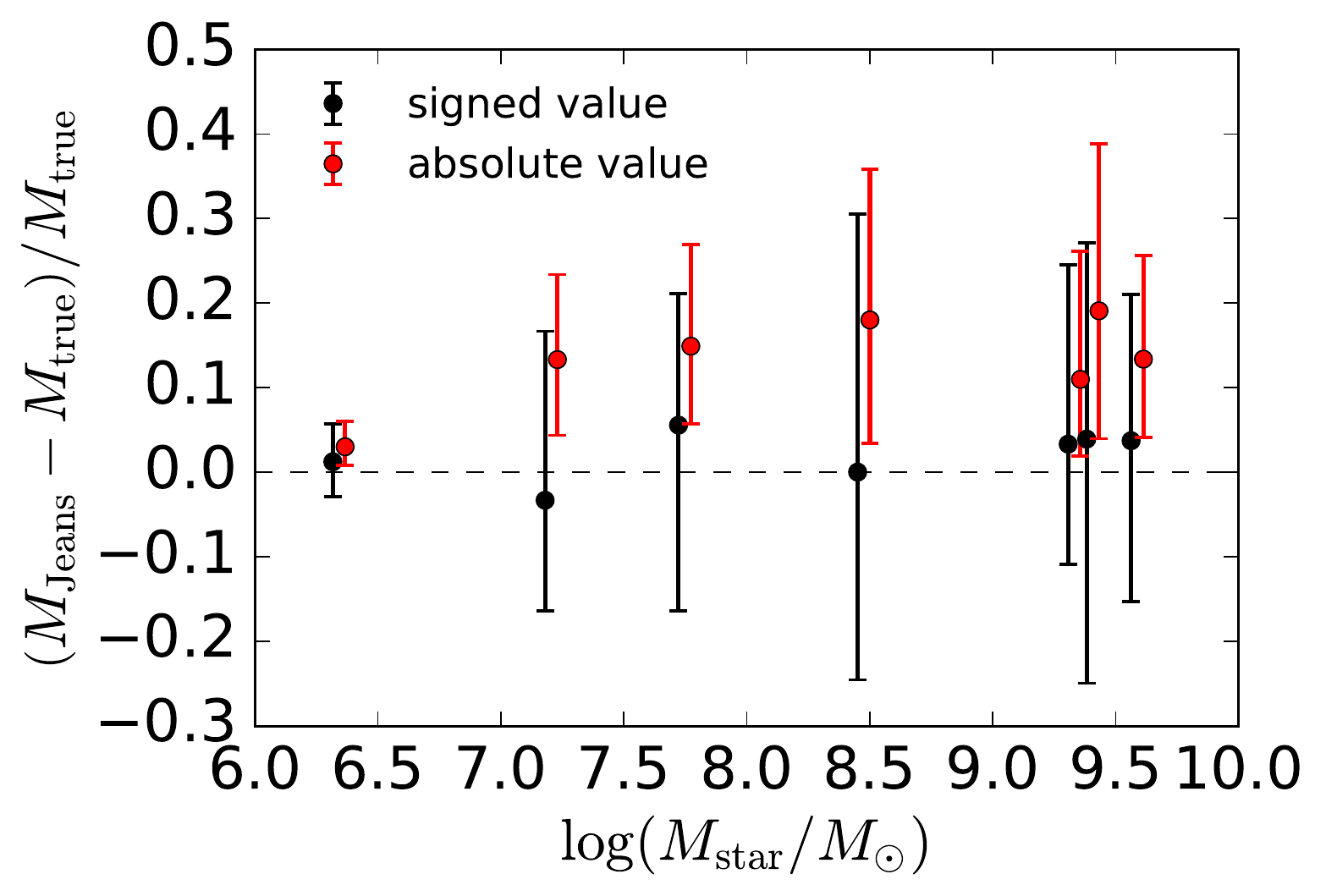}
\caption{
Fractional error in $M(< R_{90 \rm m})$ from Jeans modeling as a function of $M_{\rm star}(z=0)$, for all 7 galaxies in our sample. Points show the mean across the last 40 simulation snapshots since $z \approx 0.2$; error bars show the 68\% scatter across these snapshots. Black and red points show the net and absolute fractional error, respectively (with a small horizontal offset for clarity). In the mass regime where stellar feedback produces the strongest potential fluctuations, the typical errors in total mass are $\sim 20\%$.}
\label{fig:all_err_vs_mass}
\end{figure}

We now investigate how the accuracy of dynamical mass estimates from Jeans modeling scales with galaxy mass across all 7 of our simulated low-mass galaxies. To capture the full range of dynamical states over the course of several starburst cycles, we construct Jeans models for the last 40 snapshots of each simulation.

Figure~\ref{fig:all_err_vs_mass} shows the fractional error in Jeans dynamical mass estimates for all galaxies as a function of stellar mass at $z = 0$. For each snapshot, we calculate $M_{\rm dyn}(< R_{90 \rm m})$ for the Burkert profile recovered by Jeans modeling and compute both the net (signed) and absolute (unsigned) fractional difference between this and the true mass enclosed within $R_{90 \rm m}$, that is, ${\rm signed} = (M_{\rm Jeans} - M_{\rm true})/M_{\rm true}$, and ${\rm absolute} = \left|M_{\rm Jeans} - M_{\rm true}\right|/M_{\rm true}$. We then plot the median value and 68\% scatter in both the signed (black) and unsigned (red) mass error for the last 40 snapshots. The black points thus show whether Jeans modeling is recovering masses that are systemically too large or too small, while the red points measure the typical absolute error.

Both the net and absolute error are smallest in \texttt{m10}, with typical absolute mass errors of less than 5\%. The absolute error in dynamical mass estimates increases with stellar mass until $M_{\rm star} \sim 10^{8.5}\,M_{\odot}$. The absolute mass error flattens off around $10^9 M_{\odot}$ and begin to decline at higher masses, though there are not enough galaxies in our sample to robustly determine the exact mass scaling. That said, given that the errors in Jeans modeling are caused by feedback-driven potential fluctuations, we expect the mass-scaling to reflect the mass-scaling of these potential fluctuations, which has been investigated in detail by a number of works using a variety of feedback prescriptions \citep[e.g.,][]{DiCintio_2014, Chan_2015, Tollet_2016}, with a consistent conclusion that fluctuations are strongest at $M_{\rm star} \sim 10^9 M_{\odot}$ and $M_{200 \rm m} \sim 10^{11} M_{\odot}$.
Because stellar radial migration, dark matter core creation, and errors in Jeans modeling are all driven by the same potential fluctuations, we expect them all to follow similar mass scalings.

\subsection{Jeans modeling with unknown \texorpdfstring{$\beta(r)$}{Lg}}
\label{sec:different_beta_models}

Thus far, we assumed that the velocity anisotropy, $\beta(r)$, is known exactly. We now consider the effects of ignorance of $\beta(r)$ on the accuracy of Jeans mass estimates. We test several common treatments of $\beta(r)$ used in Jeans modeling.
\begin{enumerate}
\item Isotropy: $\beta(r) = 0$.
\item Constant anisotropy: $\beta(r) = \beta_0$.
\item OM anisotropy profile:  $\beta(r) = \beta_{\rm OM}(r, r_a)$, where $\beta_{\rm OM}$ is given in Equation~\ref{eqn:beta_OM}.
\item ML anisotropy profile: $\beta(r) = \beta_{\rm ML}(r, r_a)$, where $\beta_{\rm ML}$ is given in Equation~\ref{eqn:beta_ML}.
\item True anisotropy: measure $\beta(r_i)$ directly from the simulation in spherical bins.
\end{enumerate}
In the case of assuming constant anisotropy or an OM or ML anistropy profile, we simultaneously fit for the Burkert parameters $\{r_b, \rho_b\}$ and for the free parameter in the model for $\beta(r)$ (for constant anisotropy, $\beta_0$; for the OM and ML profiles, $r_a$) during $\chi^2$ minimization. In the case for which $\beta$ is measured directly from the simulation or is assumed to be 0 everywhere, we introduce no additional free parameters. Most observational studies of low-mass galaxies take one of the first 4 approaches, assuming isotropy, constant anisotropy, or a parameterized model for $\beta(r)$.

We use four different metrics to asses the accuracy of the Jeans model fits produced by each approach. The first is the minimum $\chi^2$ statistic given by Equation~\ref{eqn:chi_square}. This measures how accurately a given model can reproduce the known $v_{r, {\rm rms}}$ profile; in general, lower $\chi^2$ values indicate a better fit. Next, we measure $M_{\rm Jeans}/M_{\rm true}$, the ratio of the total mass inside $R_{90 \rm m}$ for the best-fit Jeans model to the true total mass inside $R_{90 \rm m}$. Next, we quantify the mean fractional error in the recovered density profile with the $\overline{\Delta \rho}$ statistic introduced in \citet{Li_2016}:
\begin{equation}
\label{eqn:delta_rho}
\overline{\Delta\rho}=\frac{1}{r}\int_{0}^{r}\frac{\left|\rho_{{\rm Jeans}}(r')-\rho_{{\rm true}}(r')\right|}{\rho_{{\rm true}}(r')}\,{\rm d}r' \, ,
\end{equation}
where $\rho_{{\rm Jeans}}(r)$ is the best-fit Burkert profile found from Jeans modeling and $\rho_{{\rm true}}(r)$ is the true density profile. $\overline{\Delta \rho}$ is positive-definite, with $\overline{\Delta \rho}=0$ corresponding to a perfect recovery of the true density profile and higher values of $\overline{\Delta \rho}$ indicating greater disagreement between the true and predicted profiles. In practice, we measure $\overline{\Delta \rho}$ at $r=R_{90 \rm m}$ and compute the integral using a trapezoidal approximation in 25 logarithmically space radius bins. Finally, we introduce an analogous statistic, $\overline{\Delta \beta}$, to quantify the mean error in the recovered anisotropy profile:
\begin{equation}
\label{eqn:delta_beta}
\overline{\Delta\beta}=\frac{1}{N}\sum_{i=1}^{N}\left|\beta_{{\rm Jeans}}(r_{i})-\beta_{{\rm true}}(r_{i})\right| \, .
\end{equation}
Here the sum is over all the bins in which $\beta$ values are required for Jeans modeling. We use the same 25 linearly spaced bins as we use in constructing the $v_{r,{\rm rms}}$ profile in Section~\ref{sec:modeling_methods}. A perfectly recovered anisotropy profile corresponds to $\overline{\Delta \beta}=0$, and larger errors in the predicted $\beta(r)$ profile will in general produce larger values of $\overline{\Delta \beta}$.

\begin{figure}
\includegraphics[width=\columnwidth]{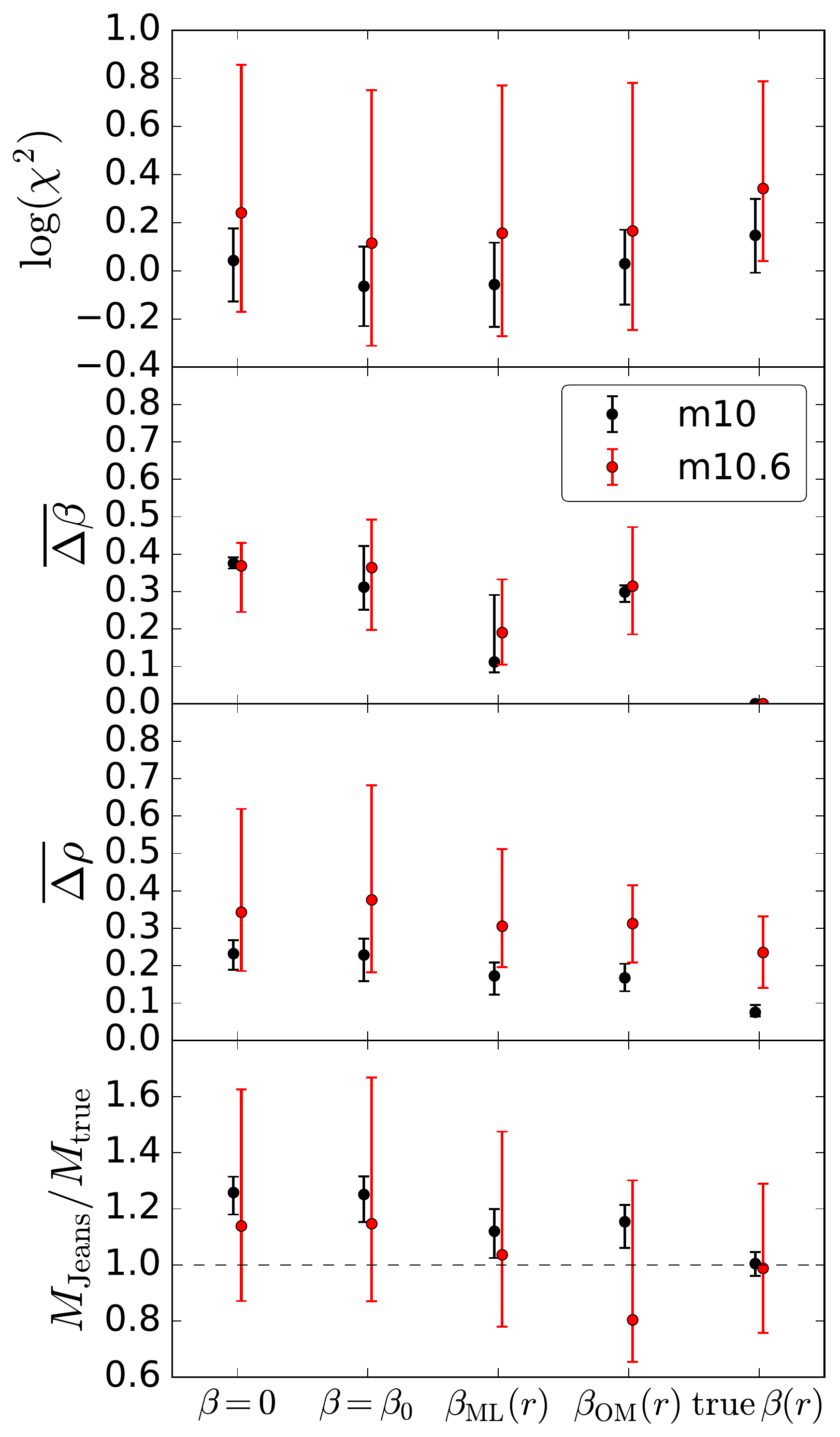}
\caption{
Results of Jeans modeling for two galaxies, using 5 different models for the velocity anisotropy, $\beta(r)$. Points show the median value of each statistic in the last 40 simulation snapshots since $z \approx 0.2$; error bars show the 68\% scatter across these snapshots.
\textbf{Top}: minimum $\chi^2$ value (Equation~\ref{eqn:chi_square}).
\textbf{2nd panel}: mean error in the predicted anisotropy profile (Equation~\ref{eqn:delta_beta}).
\textbf{3rd panel}: mean error in the density profile (Equation~\ref{eqn:delta_rho}).
\textbf{Bottom}: Ratio of $M_{\rm dyn}(R_{90 \rm m})$ predicted by Jeans modeling to the true total mass enclosed in $R_{90 \rm m}$. 
Problematically, Jeans modeling often prefers (through lower $\chi^2$ values) poor models for $\beta(r)$ that produce large $\overline{\Delta \beta}$ and $\overline{\Delta \rho}$.}
\label{fig:different_beta_guesses}
\end{figure}

Figure~\ref{fig:different_beta_guesses} shows the results of using each treatment of $\beta(r)$ to construct Jeans models for $\texttt{m10}$ and $\texttt{m10.6}$. For each treatment of $\beta(r)$ in a given snapshot, we follow the procedure described in Section~\ref{sec:modeling_methods} to find the anisotropy and Burkert profile parameters whose predicted $v_{r, {\rm rms}}$ profile best fits the true $v_{r, {\rm rms}}$ profile. (For the the ``true $\beta(r)$'' treatment, this is exactly the same procedure that was used in the last section.) We then calculate the minimum $\chi^2$ statistic and the value of $\overline{\Delta \beta}$, $\overline{\Delta\rho}$, and $M_{\rm Jeans}/M_{\rm true}$ for the best-fit parameters. To sample the galaxies' across several starburst cycles, we repeat this procedure for each of the last 40 simulation snapshots since $z\approx 0.2$. Points and error bars show the median and 68\% scatter in each of these statistics across the last 40 snapshots.

The top panel of Figure~\ref{fig:different_beta_guesses} shows minimum $\chi^2$ values. Across all models for $\beta(r)$, $\chi^2$ values are generally lower for \texttt{m10} than for \texttt{m10.6}. As in the previous sections, this is because \texttt{m10.6} undergoes repeated, strong potential fluctuations due to feedback-driven outflows. In contrast, \texttt{m10} undergoes only weak fluctuations that do not drive the galaxy far from equilibrium. Consistent with the results of the previous sections, the mean error in the predicted mass models as measured by $\overline{\Delta \rho}$ and $M_{\rm Jeans}/M_{\rm true}$ are higher for $\texttt{m10.6}$ than for \texttt{m10} for \textit{any} choice of anisotropy model.

As measured by both $\overline{\Delta \rho}$ (middle panel) and $M_{\rm Jeans}/M_{\rm true}$ (bottom panel), errors in the predicted mass profiles are smaller for both galaxies when the true $\beta(r)$ profile is used in Jeans modeling than when the anisotropy is not known and any of the parameterized models for $\beta(r)$ are used. This is not surprising, because the anisotropy profile encodes important information about the kinematic state of the galaxy, and the models we consider for $\beta(r)$ are necessarily approximations of the true profiles. The ML and OM anisotropy profiles produce a similar range of $\overline{\Delta \rho}$ and $M_{\rm Jeans}/M_{\rm true}$ values for both galaxies. The constant $\beta$ models lead to somewhat larger errors in the predicted density profiles and dynamical masses. This is also not surprising: as Figure~\ref{fig:median_anisotropy_profiles} shows, all of the galaxies in our sample have $\beta(r)$ profiles that increase with radius, from $\beta(r=0)\approx 0$ to $\beta(r=R_{90 \rm m}) \approx 0.6$. Constant anisotropy models cannot represent these profiles accurately, but the ML and OM profiles can provide a better fit. We therefore could reasonably expect the mass models recovered from the Jeans analysis to be more accurate for the ML and OM models than for constant anisotropy models.

More surprising is the relationship between the minimum $\chi^2$ values, which are a measure of how accurately the dynamical model could recover the input $v_{r, {\rm rms}}$ profile, and $\overline{\Delta \rho}$, which measures how accurately it recovered the true density profile. Although using the galaxies' true $\beta(r)$ profile leads to the most accurate mass profile and lowest $\overline{\Delta \rho}$ values, the corresponding $\chi^2$ values are actually \textit{higher} than those for many of the parameterized $\beta(r)$ models. In general, \textit{we find no correlation between how well a model for $\beta(r)$ recovers the input $v_{r,{\rm rms}}$ profile and how well it recovers the true mass or anisotropy profile.}

The poor correlation between the minimum $\chi^2$, $\overline{\Delta \rho}$, and $\overline{\Delta \beta}$ constitutes a serious concern for observational dynamical modeling studies, which generally have no a priori constraints on the form of $\beta(r)$ and thus use the minimum $\chi^2$ value as an indicator of how well the true anisotropy profile has been recovered. Indeed, a number of works \citep{vanderMarel_1994, Lokas_2001, Lokas_2002, Wilkinson_2004, Battaglia_2005, Mamon_2005, Battaglia_2008} explicitly use minimum $\chi^2$ values to constrain galaxies' anisotropy profiles. That is, they construct Jeans models using a variety of different profiles for $\beta(r)$, compute a minimum $\chi^2$ value for each profile, and conclude that the $\beta(r)$ profile that produces the smallest minimum $\chi^2$ value most closely matches the true anisotropy profile (and will most accurately recover the true dynamical mass profile). Other works \citep[e.g.,][]{Kleyna_2001, Gilmore_2007, Chen_2016} choose a particular form of $\beta(r)$ for their dynamical models on the grounds that it has been shown to accurately recover galaxies' observed $v_{r,{\rm rms}}$ profiles when combined with a suitable dynamical mass profile. But if, as Figure~\ref{fig:different_beta_guesses} suggests, there is not actually a direct correlation between how well different models can recover the input $v_{r, {\rm rms}}$ profile and how accurately they represent the true mass distribution and anisotropy, then $\beta(r)$ profiles inferred from Jeans modeling cannot be considered authoritative.

\begin{figure*}
\includegraphics[width=\textwidth]{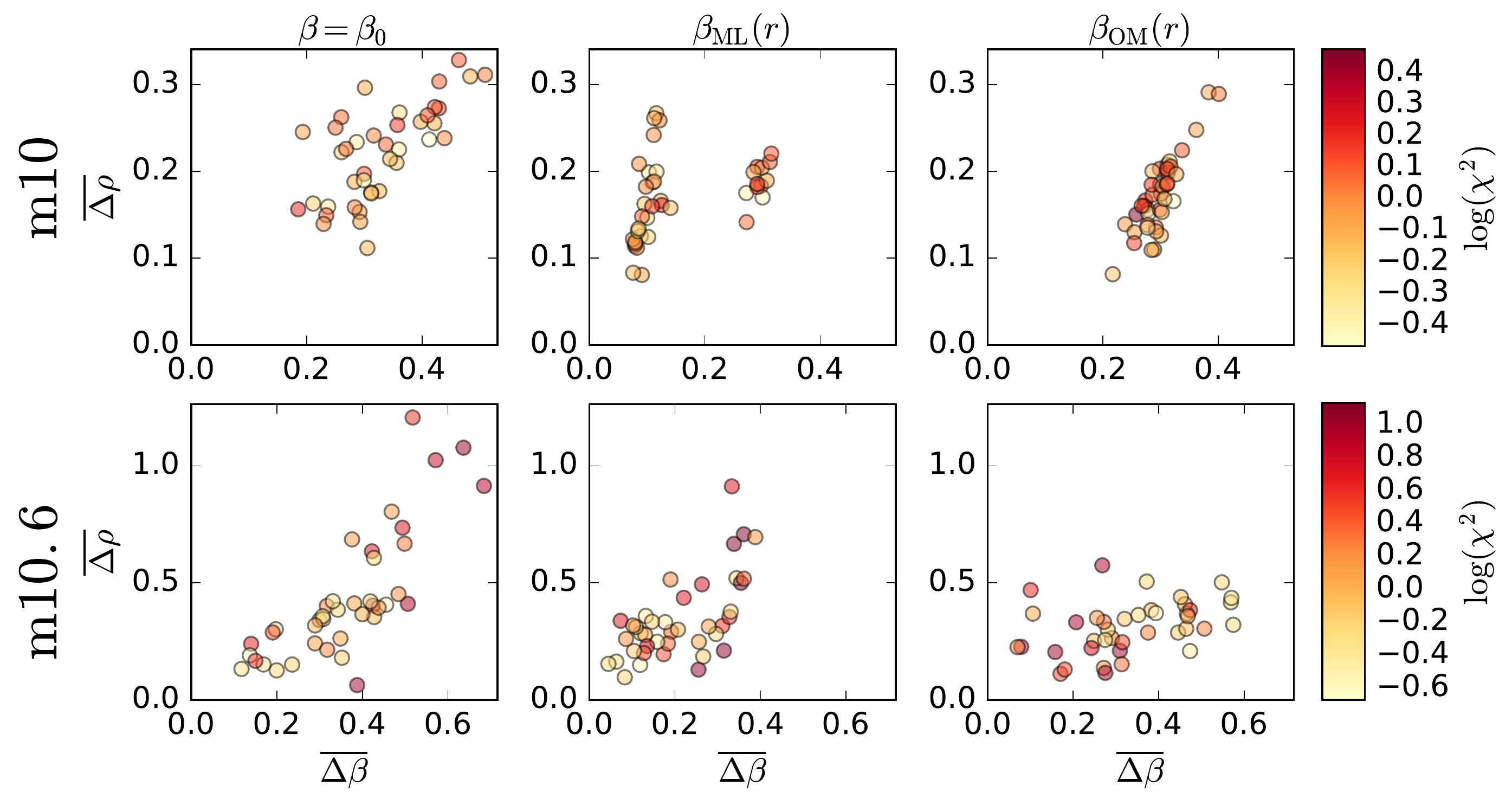}
\caption{
Results of Jeans modeling with unknown anisotropy in \texttt{m10} (top) and \texttt{m10.6} (bottom). We construct a Jeans model for each of the last 40 snapshots since $z \approx 0.2$ using three different functional forms of the anisotropy profile: constant anistropy (left), the ML anisotropy profile (Equation~\ref{eqn:beta_ML}; middle), and the OM anisotropy profile (Equation~\ref{eqn:beta_OM}; right). $\overline{\Delta \beta}$ (Equation~\ref{eqn:delta_beta}) quantifies how accurately the Jeans model recovers the true anisotropy profile; $\overline{\Delta \rho}$ (Equation~\ref{eqn:delta_rho}) measures how accurately it recovers the true density profile. Points are colored according to the minimum $\chi^2$ value (Equation~\ref{eqn:chi_square}), a measure of how accurately the input $v_{r, {\rm rms}}$ profile is recovered. Note that the top and bottom panels use different scales. In general, there is no significant correlation between $\chi^2$ values and $\overline{\Delta \beta}$ or $\overline{\Delta \rho}$. The ML and OM profiles lead to more accurate Jeans models (lower $\overline{\Delta \rho}$) than the constant anisotropy model, but they do not lead to lower $\chi^2$ values.}
\label{fig:delta_beta_vs_delta_rho}
\end{figure*}

We investigate the relation between $\overline{\Delta \beta}$, $\overline{\Delta \rho}$, and the minimum $\chi^2$ value further in Figure~\ref{fig:delta_beta_vs_delta_rho}, which shows the results of Jeans modeling for individual snapshots of \texttt{m10} and \texttt{m10.6} using three different one-parameter models of $\beta(r)$. For each form of $\beta(r)$, we calculate the best-fit Jeans model for each of the last 40 simulation snapshots since $z\approx 0.2$, always using the Burkert form of $M_{\rm dyn}(r)$. We then compute $\overline{\Delta \beta}$, $\overline{\Delta \rho}$, and the minimum $\chi^2$ value corresponding to each model. We plot $\overline{\Delta \rho}$ versus $\overline{\Delta \beta}$ and color points according to their minimum $\chi^2$ value.

In general, we find a positive correlation between $\overline{\Delta \beta}$ and $\overline{\Delta \rho}$: when the Jeans model accurately predicts the true anisotropy profile (lower $\overline{\Delta \beta}$), it is more likely to also predict the true density profile (lower $\overline{\Delta \rho}$). Conversely, larger errors in the predicted anisotropy profile often produce larger errors in the density profile. However, the relationship between $\overline{\Delta \beta}$ and $\overline{\Delta \rho}$ varies significantly between the two galaxies and between different models for $\beta(r)$. For example, there is a clear positive correlation between $\overline{\Delta \beta}$ and $\overline{\Delta \rho}$ for the OM profile in \texttt{m10}, but there is no clear correlation for the same profile in \texttt{m10.6}, or for the ML profile in \texttt{m10}.

These different trends can be understood as a result of differences both in the shape of the two galaxies' anisotropy profiles and in the scale of their potential fluctuations. 
Similar $\beta(r)$ and $v_{r, {\rm rms}}$ profiles across snapshots are not enough to guarantee that dynamical modeling will predict the same combination of $\beta(r)$ and $M_{\rm dyn}(r)$ profiles for all snapshots. A self-consistent Jeans model can explain a given flat $v_{r, {\rm rms}}$ profile either with a low dynamical mass combined with a positive $\beta(r)$ profile, \textit{or} with a high dynamical mass compensated for with tangential anisotropy (negative $\beta(r)$). This ``mass-anisotropy degeneracy'' presents a challenge for dynamical modeling even when galaxies are near equilibrium  \citep[see, for example,][]{Merritt_1987}. If $\beta$ is allowed to vary freely, even small differences in galaxies' $v_{r, {\rm rms}}$ profiles between different snapshots can thus cause significant scatter in the galaxies' inferred anisotropy and dynamical mass profiles. We explore this degeneracy further in Appendix~\ref{sec:mass_anisotropy_degeneracy}.

Because our galaxies all have positive true $\beta(r)$ profiles, dynamical models without radial anisotropy will tend to overestimate the dynamical mass, producing both larger $\overline{\Delta \rho}$ and $\overline{\Delta \beta}$ values. Because the OM and ML profiles are positive at all radii, they prevent Jeans modeling from converging on incorrect models in which tangential anisotropy drives down the predicted $v_{r, {\rm rms}}$ profile and allows for estimates of $M_{\rm dyn}(r)$ that are too high. The constant anisotropy model, on the other hand, sets fewer restrictions on $\beta(r)$ and is thus more susceptible to incorrect mass estimates stemming from the mass-anisotropy degeneracy. This gives rise to the points in the leftmost panels of Figure~\ref{fig:delta_beta_vs_delta_rho} with both $\overline{\Delta \beta}$ and $\overline{\Delta \rho}$ values that are larger than any of those for the ML and OM profiles. It also explains why the $\beta = 0$ and $\beta = \beta_0$ models in Figure~\ref{fig:different_beta_guesses} systematically overestimate the total mass (i.e., $M_{\rm Jeans}/M_{\rm true} > 1)$, while the other three treatments of $\beta(r)$ are converged on $M_{\rm Jeans}/M_{\rm true} \approx 1$.

The most successful Jeans model is, first and foremost, the one that results in the lowest values of $\overline{\Delta \rho}$ and $\overline{\Delta \beta}$ -- not the one that produces the tightest scaling between $\overline{\Delta \rho}$ and $\overline{\Delta \beta}$. With this in mind, the OM and ML models perform somewhat better in both galaxies than the constant anisotropy model, because our simulated galaxies have radially-biased anisotropy profiles.

While lower $\overline{\Delta \beta}$ values generally indicate a better recovery of the true $\beta(r)$ profile, it does not automatically follow that the $\beta(r)$ model that produces the lowest $\overline{\Delta \beta}$ values is the model that most closely resembles the true $\beta(r)$ profile. This is because there is no guarantee that the anisotropy profile predicted by Jeans modeling for a given model of $\beta(r)$ will match the profile obtained by directly fitting that model to the true anisotropy profile. This can be seen directly in the $\overline{\Delta \beta}$ values for the ML and OM models in \texttt{m10} (upper middle and upper right panels of Figure~\ref{fig:delta_beta_vs_delta_rho}.) The ML profile appears to perform slightly better than the OM profile, as the $\overline{\Delta \beta}$ values are somewhat smaller on average. However, when fit directly, the median anisotropy profile of \texttt{m10} is fit approximately equally well by the ML and OM profiles (Figure~\ref{fig:median_anisotropy_profiles}), with the OM profile actually producing a marginally better fit.

Just as in the comparison between Jeans fits across \textit{different} models of $\beta(r)$ in Figure~\ref{fig:different_beta_guesses}, Figure~\ref{fig:delta_beta_vs_delta_rho} shows that, even for a fixed model of $\beta(r)$, there is for the most part no correlation between the $\chi^2$ value and either $\overline{\Delta \beta}$ or $\overline{\Delta \rho}$. The exceptions are the constant-anisotropy and ML models in \texttt{m10.6}: in the lower left and middle panels of Figure~\ref{fig:delta_beta_vs_delta_rho}, snapshots with high $\overline{\Delta \beta}$ and $\overline{\Delta \rho}$ generally have higher minimum $\chi^2$ values than those with low $\overline{\Delta \beta}$ and $\overline{\Delta \rho}$. However, this correlation is weak. It appears to be primarily the result of the strong potential fluctuations experienced by \texttt{m10.6}. That is, the snapshots in which $v_{r, {\rm rms}}$ is poorly recovered (producing a high $\chi^2$ value) are primary those in which the galaxy is far from dynamical equilibrium (and thus has high $\left|\overline{v_r}\right|$).

This result raises the possibility of using the quality of Jeans model fits (as measured by the minimum $\chi^2$ value) as an indicator of galaxies' dynamical states. Galaxies that are currently far from equilibrium should produce larger $\chi^2$ values in Jeans modeling. One might therefore conclude that if Jeans modeling fails to accurately recover the input $v_{r, {\rm rms}}$ profile, the galaxy is out of equilibrium. However, this interpretation is probably unrealistic in practice. Large $\chi^2$ values also could indicate, for example, that a galaxy is not spherically symmetric, or that it has non-negligible net rotation. For the galaxies studied in this work, the scatter in $\chi^2$ values between different galaxies with similar masses is typically larger than the scatter between different snapshots of the same galaxy over the course of the burst/outflow cycle, so $\chi^2$ values would likely serve as a poor predictor of galaxies' dynamical states.

\section{Dynamical evolution of a gas-stripped galaxy}
\label{sec:dwarf_spheroidal}

The short-timescale fluctuations that we have explored occur in gas-rich, star-forming galaxies. However, almost all of the dwarf galaxies that are satellites within the virial radius of the MW or M31 contain little gas and have no recent star formation. We also seek to test the accuracy of Jeans modeling in such quiescent galaxies, which are more likely to be in dynamical equilibrium.

Many of the quiescent satellite dwarf galaxies in the Local Group show evidence of star formation within the last few Gyr \citep{Grebel_2004, Weisz_2015, Skillman_2016}, indicating that their star formation quenched semi-recently. Together with the fact that the fraction of quenched low-mass galaxies rises dramatically near massive halos as compared to in the field \citep{Geha_2012, Wetzel_2014, Weisz_2015}, this suggests that interactions with the host halo of the MW or M31 in the form of ram pressure and tidal stripping are required to remove the gas from low-mass galaxies \citep{Faber_1983, Grebel_2003, Mayer_2006, Toloba_2011}.

To isolate the effect of removing gas and turning off star formation, without the additional complications of tidal effects from falling into a massive host halo, we consider a simple ``toy model'' to simulate the removal of gas from a low-mass galaxy via ram-pressure stripping. 
This is not intended to represent a fully realistic ram-pressure stripping scenario; our goal is to determine specifically how the absence of gas (and gas outflows) affects the reliability of Jeans modeling. We will examine dwarf galaxies simulated self-consistently around a MW-mass host, as in \citet{Wetzel_2016}, in future work.

We construct this toy model as follows. We run the \texttt{m10.6} simulation normally until $z=0.1$. Then, from $z=0.1$ to $z=0$, we run two versions. The first, which we term the ``fiducial'' run, is allowed to evolve uninterrupted until $z=0$. In the second run, which we will call the ``gas-stripped'' run, we instantaneously impart a uniform velocity kick of $200\,{\rm km\,s^{-1}}$ to all gas particles, approximating the effect of rapid ram-pressure stripping on a dwarf galaxy orbiting through a MW-mass halo. Because this velocity kick exceeds the galaxy's escape velocity $(v_{\rm esc}\approx 160\,\rm km\,s^{-1})$, the gas is soon swept up in the Hubble flow, and the galaxy evolves passively, without gas or star formation until $z=0$. (We also tested removing all gas instantaneously from the galaxy, which leads to nearly identical results.)

Following the velocity kick, the gas dissipates within a few 100 Myr, and thereafter the galaxy evolves passively, without additional star formation. The potential at first expands rapidly when gas is removed, and the stellar distribution expands slightly, just as it does following normal feedback-driven outflows. However, these fluctuations soon die down, and the galaxy reaches equilibrium within a few 100 Myr. We note that gas removal in our toy model occurs likely more quickly than the inferred satellite quenching timescale of order a few Gyr \citep[e.g.,][]{Wetzel2015b, Fillingham2015}. We do not expect this to significantly affect our results, because we start our dynamical modeling only well after star formation has ceased and the galaxy has reached equilibrium.

\begin{figure}
\includegraphics[width=\columnwidth]{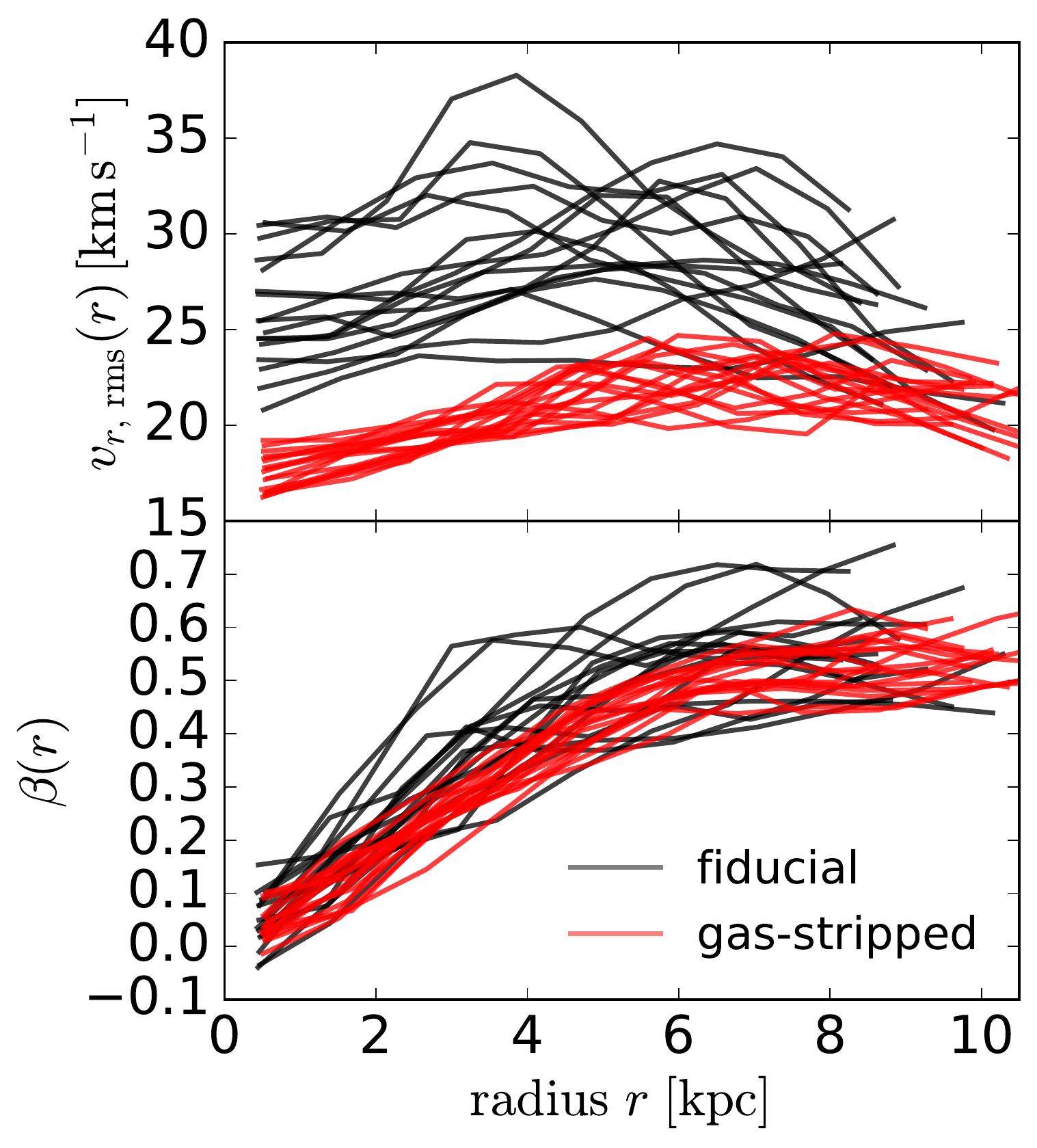}
\caption{
Profiles of rms radial velocity (top) and anisotropy (bottom) for the fiducial (gas-rich and star forming; black lines) run of \texttt{m10.6} and the re-run in which we remove gas at $z = 0.1$ via a $200\,{\rm km\,s^{-1}}$ velocity kick (red lines). Profiles are plotted out to $R_{90 \rm m}$ in each snapshot, which is generally larger in the gas-stripped run. Potential fluctuations in the gas-stripped run subside once gas is removed, leading to less scatter in the galaxy's $v_{r, {\rm rms}}$ and $\beta$ profiles across different snapshots.
}
\label{fig:m10.6_fiducial_vs_gas_stripped_profiles}
\end{figure}

Figure~\ref{fig:m10.6_fiducial_vs_gas_stripped_profiles} compares the $v_{r,{\rm rms}}(r)$ and $\beta(r)$ profiles for the fiducial and gas-stripped runs of \texttt{m10.6} between $z\approx 0.09$, when the gas-stripped run has settled back into equilibrium, and $z=0$. For each snapshot, we plot radial profiles out to $R_{90 \rm m}$, which is somewhat larger in the gas-stripped run ($R_{90 \rm m} \approx 10\,{\rm kpc}$) than in the fiducial run ($R_{90 \rm m} \approx 7-10\,{\rm kpc}$). There is less scatter across different snapshots in both the shape and overall normalization of $\beta(r)$ and especially $v_{r, {\rm rms}}(r)$ profiles in the gas-stripped run than in the gas-rich fiducial run. The gas-stripped $v_{r,{\rm rms}}$ profiles are systematically offset toward lower values than their counterparts in the fiducial run. The median $\beta(r)$ profiles of the fiducial and gas-stripped runs are quite similar, with marginally lower $\beta(r)$ values in the gas-stripped snapshots.

The distribution and kinematics of stars in the gas-stripped run are qualitatively similar to those in the most rarefied snapshots of the fiducial run, which also have the shallowest potential wells, lowest $v_{r, {\rm rms}}$ values, and largest $R_{90 \rm m}$ \citepalias{El-Badry_2016}. In this sense, complete gas removal can be viewed as a somewhat more extreme version of the post-starburst gas outflows that repeatedly expand the galactic potential in the fiducial gas-rich run. The key difference is that, unlike in outflow episodes in the fiducial run, gas in the gas-stripped run never reaccretes, so the potential never re-contracts, leaving the stellar distribution in its most diffuse configuration with no star formation and no potential fluctuations.

\begin{figure}
\includegraphics[width=\columnwidth]{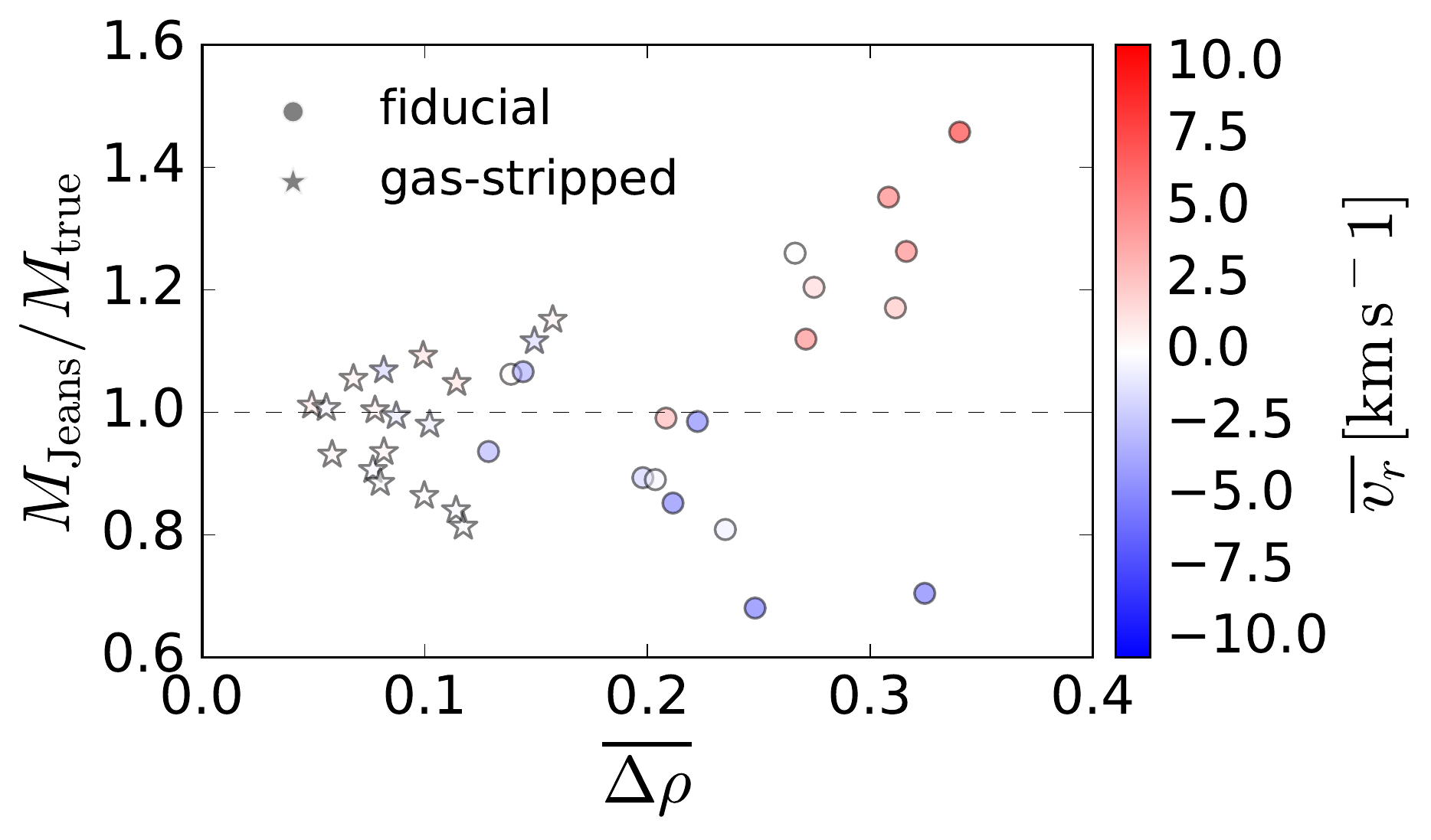}
\caption{
Accuracy of Jeans models fit to the last 18 snapshots of \texttt{m10.6} since $z \approx 0.1$. Round symbols show values for the fiducial run of the simulation; stars show values for the re-simulation in which gas is artificially stripped from the galaxy through a $200\,{\rm km\,s^{-1}}$ velocity kick at $z=0.1$. Points for both runs are color-coded according to $\overline{v_r}$, the means stellar radial velocity. As measured by both $\overline{\Delta \rho}$ and $M_{\rm Jeans}/M_{\rm true}$, Jeans modeling recovers the true dynamical mass $\sim 3 \times$ more accurately in the gas-stripped run.
}
\label{fig:m10.6_fiducial_vs_gas_stripped}
\end{figure}

Figure~\ref{fig:m10.6_fiducial_vs_gas_stripped} compares the accuracy of Jeans model fits carried out on the gas-stripped run of \texttt{m10.6} to those from the fiducial (gas-rich) simulation. We follow the same procedure in Section~\ref{sec:jeans_modeling_results} to build Jeans models for both the fiducial and gas-stripped runs. We measure the quality of the recovered density profile using the $\overline{\Delta \rho}$ and $M_{\rm Jeans}/M_{\rm true}$ statistics, and we color points for both runs according to $\overline{v_r}$ to show whether the galaxy is expanding or contracting in a given snapshot.

As expected, the mean radial velocities of the gas-stripped snapshots are all $\sim 0$: while the fiducial run continues to expand and contract, the gas-stripped run evolves passively and remains in dynamical equilibrium. As measured by both $\overline{\Delta \rho}$ and $M_{\rm Jeans}/M_{\rm true}$, the mass profiles predicted by Jeans modeling are significantly more accurate for the the gas-stripped galaxy. Across these 18 snapshots, the median error in the density profile for the fiducial run is $\langle\overline{\Delta\rho}\rangle \approx 0.25$, while for the gas-stripped run, $\langle\overline{\Delta\rho}\rangle \approx 0.08$.

Jeans modeling is thus significantly more reliable in the gas-stripped galaxy than in its gas-rich counterpart.
This implies that mass profiles inferred from dynamical modeling of dSph or dE galaxies are likely still reliable (though we have not accounted for tidal forces, which could complicate Jeans modeling further).

\section{Summary and Discussion}
\label{sec:summary_discussion}

\subsection{Summary}
\label{sec:Summary}

We investigated the effects of feedback-driven potential fluctuations on stellar kinematics in low-mass galaxies using a suite of cosmological zoom-in baryonic simulations from the FIRE project. We explored how accurately dynamical modeling can recover the mass profiles of simulated galaxies, both when the stellar velocity anisotropy is unknown, as is the case for typical dynamical studies of nearby galaxies, and in the idealized scenario in which the anisotropy is known exactly.
Our main results are as follows.

\begin{enumerate}[leftmargin=*]

\item \textit{Potential fluctuations drive stellar kinematics}: Stellar feedback-driven potential fluctuations cause the stellar kinematics of our low-mass galaxies to fluctuate significantly on short timescales. The $v_{r, {\rm rms}}(r)$ profiles of our galaxies -- the principal ``input'' data through which Jeans modeling can constrain galaxy masses -- can change by as much as a factor of two between periods of maximum outflow and maximum inflow, corresponding to timescales of only a few 100 Myr (see Figure~\ref{fig:anisotropy_age_profiles}).

\item \textit{Stellar kinematics strongly correlate with star formation rate}: The time-evolution of the sSFR and $\sigma_{\rm los}$ in our simulations are remarkably similar (Figure~\ref{fig:m10.6_sfr_sigma_time}). Both quantities are regulated by gas outflows, but there is a $\sim 50$ Myr offset between them, because the sSFR rapidly declines during a gas outflow, but stars do not change their kinematics until enough gas is displaced to change the overall potential. We predict a positive correlation between sSFR and $\sigma_{\rm los}$ for observed galaxies at fixed mass, which should be strongest when the observed sSFR is an average over the past $\sim 100$ Myr. (Figure~\ref{fig:m10.6_sfr_vs_sigma_offset}). This prediction provides a clear test of the role of stellar feedback in regulating stellar (and dark-matter) densities within low-mass galaxies, as predicted by many theoretical works. Preliminary comparisons of SFRs and dispersions measured from stacked SDSS spectra in other works appear to confirm this trend.

\item \textit{Potential fluctuations undermine the accuracy of dynamical models}: Jeans modeling treats galaxies as virialized systems in dynamical equilibrium. Fluctuations in the gravitational potential -- and the resulting time-varying stellar kinematics -- thus can non-trivially bias dynamical mass estimates, even when the anisotropy is known exactly. Jeans modeling systemically overpredicts galaxy masses during periods of net outflow/expansion ($\overline{v_r} > 0$) and underpredicts them during periods of inflow/contraction ($\overline{v_r} < 0$) (see Figure~\ref{fig:jeans_error_vs_vr_time}). Errors in Jeans dynamical mass estimates are largest for galaxies in the mass regime in which potential fluctuations are most energetic ($M_{\rm star}\approx 10^{8-9.5} M_{\odot}$; see Figure~\ref{fig:all_err_vs_mass}). Dynamical mass estimates from Jeans modeling are typically biased by $\sim 20\%$ in this mass regime, with errors of nearly a factor of 2 during the largest non-equilibrium fluctuations.

\item \textit{Unknown anisotropy can introduce significant biases}:
When the stellar velocity anisotropy, $\beta(r)$, is unknown -- as is typical in observations -- Jeans modeling is susceptible to converging on joint mass + anisotropy models that overestimate galaxies' total mass and underestimate their radial anisotropy. Errors in mass modeling can be mitigated by using more realistic models for $\beta(r)$, in particular, radially increasing and positive-definite anisotropy models -- such as the Osipkov-Merritt (OM) and Mamon-{\L}okas (ML) models -- perform better than isotropic or constant-anisotropy models. However, Jeans modeling cannot accurately constrain the shape of the anisotropy profile, and worse, Jeans modeling frequently favors models for $\beta(r)$ that are poor fits to the true anisotropy (see Figure~\ref{fig:different_beta_guesses}).
Equally troubling, there is no significant correlation between how accurately a particular Jeans model recovers the input $v_{r, {\rm rms}}$ profile (as measured by a $\chi^2$ value), and how accurately it predicts the true mass and anisotropy profile (see Figure~\ref{fig:delta_beta_vs_delta_rho}). This is a serious concern for the validity of constraints on the anisotropies and mass profiles of galaxies derived from dynamical modeling, because observational studies frequently choose between models for $\beta(r)$ by comparing how well they recover the input $v_{r, {\rm rms}}$ profile.

\item \textit{Jeans modeling is more reliable in galaxies without gas}: To verify that errors in Jeans modeling primarily result from feedback-driven potential fluctuations, we re-simulated the late-time evolution evolution of one of our galaxies, removing its gas through a late-time velocity kick designed to roughly approximate ram-pressure stripping of a galaxy falling into a MW-mass halo. Potential fluctuations soon died down after gas was removed and star formation ceased (see Figure~\ref{fig:m10.6_fiducial_vs_gas_stripped_profiles}). Thereafter, Jeans modeling could typically recover the galaxy's mass to within 10\%, $\sim 3 \times$ more accurate than in the fiducial (gas-rich) simulation (see Figure~\ref{fig:m10.6_fiducial_vs_gas_stripped}).

\end{enumerate}

\subsection{Comparison with previous works}

\citet{Li_2016} recently investigated the accuracy of Jeans modeling by using Jeans models to recover the dynamical masses of galaxies in the Illustris simulation \citep{Vogelsberger_2014}. They modeled the anisotropy as a constant and used the same $\overline{\Delta \rho}$ statistic as this work to quantify the mean error in the density profile predicted by Jeans modeling. They found that at $z=0$, the density profiles of 68\% of their galaxies could be recovered to an accuracy of $\overline{\Delta \rho} < 0.26$, and almost all of their models yielded $\overline{\Delta \rho} < 1$. They also investigated how accurately Jeans modeling could recover the total mass within the stellar radius, finding typical mass errors of $\sim 10\%$. These numbers are broadly consistent with our results (compare to Figure~\ref{fig:different_beta_guesses}, Figure~\ref{fig:gNFW_vs_Burkert}, and Figure~\ref{fig:all_err_vs_mass}). It is important to note, however, that their galaxy sample was quite different from ours. First, most of their galaxies are more massive and are supported by rotation; for this reason, they use axisymmetric models with an additional term in the Jeans equation for streaming motions. More importantly, because of the Illustris simulation's lower resolution and different treatment of feedback, their low-mass galaxies do not undergo the strong feedback-driven potential fluctuations which are the primary drivers of errors in Jeans modeling in our analysis.

Using idealized high-resolution simulations of an isolated low-mass $(M_{200} \approx 10^9\,M_{\odot})$ galaxy, \citet{Read_2016b} examined whether stellar feedback-driven outflows could bias dynamical mass estimates obtained from both Jeans modeling of stars and gas rotation curve fitting. They found that stellar feedback adds significant disordered motion to the gas in starburst and post-starburst galaxies, and as a result, rotation curve fitting produces dynamical mass estimates that are systemically too large. However, the potential fluctuations introduced by these outflows were not large enough to introduce systematic errors in dynamical mass estimates from Jeans modeling. This result is not inconsistent with our findings, because in our simulations feedback-driven outflows only become efficient at altering the gravitational potential at higher masses ($M_{200} \gtrsim 10^{10} M_{\odot}$ and $M_{\rm star} \gtrsim 10^{7} M_{\odot}$; see \citetalias{El-Badry_2016}). Nevertheless, there is a clear need to investigate further whether other models for low-mass galaxies produce similar potential fluctuations as the FIRE model. For example, some feedback prescriptions do not significantly generate cores in dwarf galaxies' inner density profiles \citep[e.g.,][]{Fattahi_2016b}, while others create large cores even when only a small fraction of available feedback energy is coupled to the ISM \citep{Maxwell_2015}.

\subsection{Looking forward: implications of proper motion studies}
\label{sec:observations}

At present, dynamical studies of nearby low-mass galaxies are severely limited by the available kinematic data. Resolved stellar kinematics cannot be reliably measured for galaxies at distances greater than $1-2\,{\rm Mpc}$, so current dynamical studies of galaxies outside the Local Group are based on kinematic properties of all stars integrated along the line-of-sight. Even in the nearest dwarf spheroidal galaxies, where it is possible to obtain resolved kinematics for individual stars \citep[e.g.,][]{Walker_2009}, only line-of-sight velocities are measured, making it impossible to distinguish between radial and tangential velocity components, and thus, to measure the velocity anisotropy.

The quality and volume of kinematics data of nearby dwarf galaxies will increase dramatically in the next decade with the introduction of next generation observatories both on the ground (TMT and GMT) and in space (Gaia, JWST, and WFIRST). These facilities will allow us to supplement existing stellar line-of-sight velocity data with plane-of-the-sky velocities calculated from proper motions.

The power of proper motions for dynamical modeling has been demonstrated compellingly in studies of the MW's open clusters \citep{Leonard_1989} and globular clusters \citep{Anderson_2010, vanderMarel_2010}. Crucially, proper motions provide an excellent probe of the stellar velocity anisotropy, putting stronger constraints on $\beta(r)$ than those that can be obtained from line-of-sight velocities alone.\footnote{
In principle, proper motions can be combined with line-of-sight velocities to yield full 3-dimensional stellar kinematics, from which $\beta(r)$ can be computed directly. However, one still can calculate $\beta(r)$ with high fidelity in the absence of $v_{\rm los}$ measurements \citep{Leonard_1989}, which is important because Gaia's flux limit for astrometry is expected to be much dimmer than the corresponding limit for spectroscopy \citep{An_2012}.}
For example, using HST proper motions over a 4-year baseline, \citet{vanderMarel_2010} performed a detailed dynamical analysis of the globular cluster Omega Cen. At a distance of $D \approx 5\,{\rm kpc}$, their astrometric resolution of $\sim 100\,{\rm \mu as\,yr^{-1}}$ translated to a velocity precision of $\sim 2\,{\rm km\,s^{-1}}$ in each coordinate. Using spherical Jeans modeling, they constrained the cluster's central density with an uncertainty of only $\sim 0.02\,{\rm dex}$.

By the end of its 5-year mission, Gaia \citep{Lindegren_2016} is expected to measure proper motions for the brightest stars that can be individually resolved within dwarf galaxies in the Local Group at an accuracy of $\sim 30 - 60\,{\rm \mu as\,yr^{-1}}$ \citep{deBruijne_2014, Evslin_2015}.
For a dwarf galaxy at a distance of 80 kpc, this translates to a projected velocity precision of $\sim 12 - 22 \,{\rm km\,s^{-1}}$.
Furthermore, \citet{Evslin_2015} found that, if Gaia data are combined with TMT observations in 2022, this could provide proper motions for hundreds of stars in the Sculptor dwarf spheroidal galaxy $(D\approx 80\,{\rm kpc})$ with a precision of $\sim 5\,{\rm km\,s^{-1}}$.
Finally, combining existing and upcoming HST proper-motion imaging with follow-up observations from next-generation missions, such as JWST and WFIRST, is expected to improve the precision of these measurements even further \citep{Kallivayalil_2015}. Optimistically, we can expect resolved stellar kinematics of nearby dwarf galaxies to enable dynamical modeling with a precision comparable to that currently feasible in globular clusters.

We are thus on the brink of a new era of precision astrometry that will allow us to test our theoretical predictions. As discussed in Section~\ref{sec:different_beta_models}, ignorance of galaxies' true anisotropy profiles can cause significant errors in Jeans modeling, and attempts to constrain anisotropy profiles using dynamical modeling of $v_{r, {\rm rms}}$ profiles alone are subject to important biases and uncertainties. Direct observational constraints on true $\beta(r)$ profiles will greatly improve the reliability of dynamical mass profiles derived from Jeans modeling, though significant uncertainties in inferred mass profiles will remain for gas-rich star-forming galaxies, which are more likely to experience potential fluctuations. Beyond their utility for dynamical modeling studies, measurements of galaxies' anisotropy profiles are also valuable for discriminating different evolutionary histories \citep{vanAlbada_1982, Londrillo_1991, Dekel_2005, Onorbe_2007, Wu_2014, Rottgers_2014}

Resolved stellar kinematics of dwarf galaxies in the Local Group also will enable direct measurements of stellar radial streaming velocities, $\overline{v_r}$. Unlike currently available line-of-sight velocities, these radial velocity measurements are sensitive to the expansion and contraction of the stellar distribution. They will thus test directly whether low-mass star-forming galaxies are undergoing potential fluctuations as predicted by our model.

\acknowledgments{
We thank the referee for helpful comments, Justin Read for advice on Jeans modeling, and Andrew Hearin for suggesting comparisons with observations. We also thank Dan Weisz, Ryan Trainor, Chung-Pei Ma, Susan Kassin, and Jenny Greene for productive discussions.
K.E. gratefully acknowledges support from the Caltech SURF program, a Berkeley graduate fellowship, and a Hellman award for graduate study.
A.R.W. was supported by the Moore Center for Theoretical Cosmology and Physics at Caltech via a Moore Prize Fellowship, and by Carnegie Observatories via a Carnegie Fellowship in Theoretical Astrophysics.
M.G. was supported by a fellowship from the John S. Guggenheim Memorial Foundation.
E.Q. was supported by NASA ATP grant 12-ATP-120183, a Simons Investigator award from the Simons Foundation, and the David and Lucile Packard Foundation.
P.F.H. was supported by an Alfred P. Sloan Research Fellowship, NASA ATP Grant NNX14AH35G, and NSF Collaborative Research Grant \#1411920 and CAREER grant \#1455342.
D.K. and T.K.C. were supported by NSF grant AST-1412153 and funds from the University of California San Diego. D.K. was additionally supported by the Cottrell Scholar Award.
C.-A.F.-G.  was supported by the NSF through grants AST-1412836 and AST-1517491, and by NASA through grant NNX15AB22G.
We ran numerical calculations on the Caltech compute cluster ``Zwicky'' (NSF MRI award \#PHY-0960291) and allocation TG-AST120025 granted by the Extreme Science and Engineering Discovery Environment (XSEDE) supported by NSF.
}

\bibliographystyle{yahapj}

\appendix
\section{Jeans Model Fitting}
\label{sec:Jeans_Model_Explanation}

\subsection{Mass models}
\label{sec:mass_models}

We experimented with using two different models for the total matter distribution of our galaxies during Jeans modeling. The first is the two-parameter Burkert profile \citep{Burkert_1995}, given by Equation~\ref{eqn:Burkert_rho}, which forces a core in the density profile, that is, ${\rm d}\ln\rho/{\rm d}\ln r={\rm 0}$ as $r\to 0$. The corresponding dynamical mass profile $M_{\rm dyn}(r)$ is
\begin{equation}
\label{eqn:Burkert_Mr}
M_{{\rm Burkert}}(r)=\pi \rho_{b}r_{b}^{3}\left[\ln\left(\frac{(r_{b}^{2}+r^{2})(r_{b}+r)^{2}}{r_{b}^{4}}\right)-2\tan^{-1}\left(\frac{r}{r_{b}}\right)\right] \, .
\end{equation}
We also tried fitting a ``generalized NFW'' profile \citep[gNFW;][]{Zhao_1996, Wyithe_2001}, given by
\begin{equation}
\label{eqn:gNFW_rho}
\rho_{\rm gNFW}(r)=\frac{\rho_{s}}{(r/r_{s})^{\gamma}\left[1+(r/r_{s})\right]^{3-\gamma}} \, .
\end{equation}
The gNFW density profile scales as $\rho \propto r^{-\gamma}$ at small radius and $\rho \propto r^{-3}$ at large radius; the transition between these regimes occurs at $r \sim r_s$. The standard NFW profile \citep{Navarro_1997} is a special case of the gNFW profile corresponding to $\gamma = 1$, while setting $\gamma=0$ produces a cored profile similar to the Burkert model; a number of authors \citep{Kravtsov_1998, Moore_1998, Ghigna_2000, Klypin_2001, Power_2003} have proposed their own ``universal'' profiles corresponding to a $\gamma$ parameter in the range $0 \leq \gamma \leq 3/2$. The gNFW dynamical mass profile is given by
\begin{equation}
\label{eqn:m_dyn_gNFW}
\begin{split}
M_{{\rm gNFW}}(r)=4\pi\rho_{s}r_{s}^{3}\frac{(r/r_{s})^{\xi}}{\xi}\times {}_{2}F_{1}\left(\xi,\xi;\xi+1;-r/r_{s}\right),
\end{split}
\end{equation}
where $\xi=3-\gamma$ and $_{2}F_{1}\left(a,b;c;z\right)$ is Gauss' hypergeometric function \citep{Abramowitz_1965}. Note that Equation~\ref{eqn:m_dyn_gNFW} diverges for finite values of $r$ when $\gamma\geq3$, so physical profiles must have $\gamma < 3$.

Some Jeans modeling studies explicitly express the dynamical mass as the sum of luminous and dark contribution to the total mass, that is, $M_{{\rm dyn}}(r) = M_{{\rm stars}}(r) + M_{{\rm dark}}(r)$. In this case, $M_{\rm stars}(r)$ typically is expressed as a stellar mass-to-light ratio $\Upsilon_{\rm star}$ times a functional fit to the deprojected light profile, and $M_{\rm dark}(r)$ either represents the total nonluminous mass in one term or explicitly separates the mass contributions of gas and dark matter. We opt to express the total dynamical mass as a single profile, as this decreases the degeneracy in our model and makes it more straightforward to disentangle the effects of potential fluctuations from other issues, such as the significant changes in $\Upsilon_{\rm star}$ over the course of galaxy's burst cycles.

\subsection{Example Jeans model fits}
\label{sec:example_fits}

\begin{figure*}
\includegraphics[width = \textwidth]{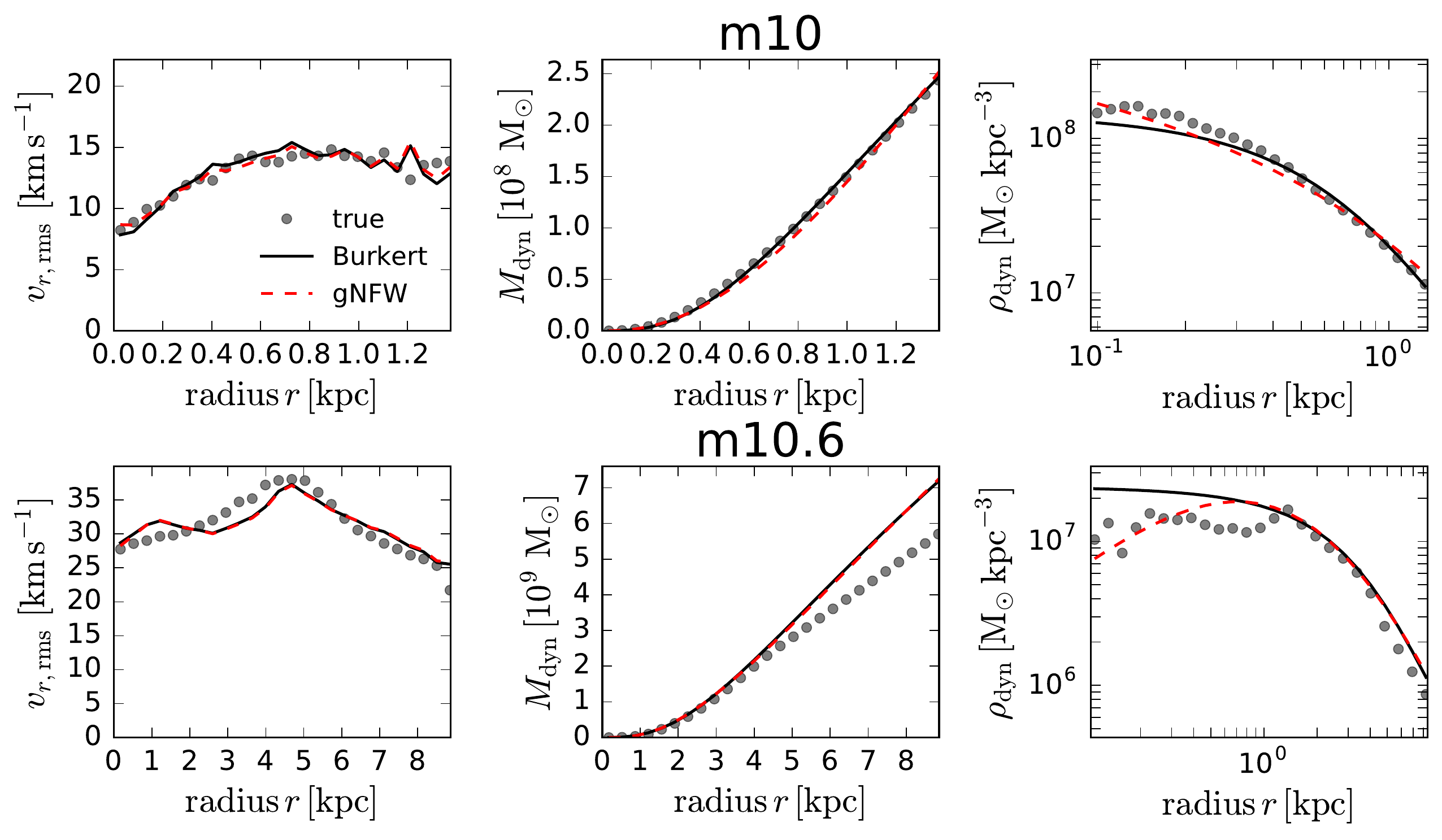}
\caption{
Best-fit Burkert (black) and generalized NFW (gNFW; red) profiles from Jeans modeling. Top (bottom) panels show fits to \texttt{m10} (\texttt{m10.6}).
\textbf{Left}: gray points show $\overline{v_r}$ profiles values in linearly spaced spherical shells between $r=0$ and $r=R_{90 \rm m}$. Black (red) line shows the $\chi^2$ minimizing $v_{r, {\rm rms}}$ profile predicted by Equation~\ref{eqn:vr_square} assuming a Burkert (gNFW) form for $M_{\rm dyn}(r)$.
\textbf{Middle}: best-fit Burkert and gNFW $M_{\rm dyn}(r)$ profiles corresponding to the lines in the left panel. True $M_{\rm dyn}$ points are shown in black for comparison.
\textbf{Right}: density profiles corresponding to the $M_{\rm dyn}(r)$ profiles in the middle panel. While the best-fit gNFW and Burkert profiles produce essentially identical $M_{\rm dyn}(r)$ and $v_{r, {\rm rms}}$ profiles, the density profiles differ.
}
\label{fig:two_galaxies_comparision}
\end{figure*}

Figure~\ref{fig:two_galaxies_comparision} illustrates the Jeans modeling procedure, showing both the Burkert and gNFW fits to the $z=0$ snapshots of two of our simulated galaxies. The left panels show stellar radial velocities. Gray points show $v_{r,{\rm rms},i}$ values, that is, the true, measured, rms velocities. Overplotted lines show the best-fit $v_{r,{\rm rms}}(r_i)$ profiles computed using Equation~\ref{eqn:vr_square} for both the Burkert (black) and gNFW (red) forms of $M_{\rm dyn}(r)$, with the best-fit model parameters found through $\chi^2$ minimization as described in the next section. Here $\beta(r)$ is measured directly from the simulation for simplicity; we discuss models with unknown anisotropy in Section~\ref{sec:mass_anisotropy_degeneracy}.

The center panels show dynamical mass profiles. Black and red lines compare the best-fit Burkert and gNFW $M_{\rm dyn}(r)$ profiles associated with the corresponding $v_{r,{\rm rms}}$ profiles in the left panel. Gray points show the true total mass enclosed within each shell. Finally, the right panels shown the mean density $\rho_{{\rm dyn}}=({\rm d}M_{{\rm dyn}}/{\rm d}r)/(4\pi r^{2})$ in logarithmically spaced bins. Black and red lines show densities for the same best-fit Burkert and gNFW profiles in the other two panels, and gray points represent the true density calculated in each spherical shell.

In \texttt{m10}, the best-fit Burkert and gNFW profiles both predict a $v_{r,{\rm rms}}$ profile in excellent agreement with the true profile, suggesting that the galaxy is approximately in dynamical equilibrium. Consistent with this interpretation, both best-fit $M_{\rm dyn}(r)$ profiles recover the true dynamical mass profile with high fidelity, and the Burkert and gNFW mass profiles agree well with each other.

Nevertheless, we find visible differences between the best-fit gNFW and Burkert density profiles. Despite the similarity of their cumulative $M_{\rm dyn}(r)$ profiles, the two density profiles disagree by $\approx 50\%$ near the galactic center.

This disagreement highlights a central weakness of Jeans modeling: qualitatively different density profiles can have similar total enclosed mass profiles at large radius, and because it is $M_{\rm dyn}(r)$ rather than $\rho_{\rm dyn}(r)$ that appears explicitly in the Jeans equation, Jeans modeling is less sensitive to the shape of galaxies' density profiles. Our Jeans analysis in particular is not optimized to recover the slope of the density profile at small radius: because linearly spaced $v_{r, {\rm rms}}$ measurements are used during $\chi^2$ minimization, the relatively few points at small radius are not weighted heavily in determining the best-fit model parameters.

In \texttt{m10.6}, the best-fit $v_{r, {\rm rms}}$ predictions differ by $\lesssim 10\%$ from the true rms radial velocity profile, though the predictions of the Burkert and gNFW profiles agree excellently with each other. The recovered dynamical mass profiles are also less accurate than in \texttt{m10}, though there is again good agreement between the gNFW and Burkert fits. The best-fit density profiles are nearly identical at large radii but exhibit qualitatively different behavior at $r \ll 1$; neither the Burkert nor the gNFW model recovers the true form of $\rho_{\rm dyn}(r)$ accurately.

In this snapshot, \texttt{m10.6} is likely not in full dynamical equilibrium, for two reasons.
First, no self-consistent set of orbits can produce the true $v_{r,{\rm rms}}$ profile given the galaxy's $\beta(r)$ and $n(r)$ profile, at least not as permitted by the Burkert and gNFW models.
Second, the model parameters that produce the $v_{r,{\rm rms}}$ profile that agrees best with the true values fail to accurately recover the true form of $M_{\rm dyn}(r)$.

\subsection{Chi-Square Minimization}
\label{sec:chi_square_minization}

We find the global minimum in the $\chi^2$ function using an approach combining brute force search with a well-studied optimization algorithm. We begin by laying down a coarse grid spanning all plausible regions of parameter space and find the gridpoint ${\mathbf p_0}$ at which the $\chi^2$ value is smallest.\footnote{
For the two-parameter Burkert profile, we use a $50\times50$ grid in $\{r_b,\rho_b\}$ space; for the three-parameter gNFW profile, we use a $20\times20\times20$ grid in $\{\gamma, r_s,\rho_s\}$ space. We have verified that increasing or decreasing the grid resolution by a factor of 10 does not change the global minimum on which the optimizer converges whatsoever.}
We then employ the Nelder-Mead downhill simplex method \citep[NMS;][]{Nelder_1965} to find the global minimum of the $\chi^2$ function, using ${\mathbf p_0}$ as the initial guess. The NMS algorithm uses a geometric object called a ``simplex'' which moves through parameter space and adapts to the local topology using a small number of allowed transformations until it contracts into a local minimum. It is used elsewhere in the astronomical literature for parameter-fitting \citep[e.g.,][]{Kallrath_1987, Gray_2001, Prvsa_2005} and is well-suited for optimization problems such as this one, in which the gradient of the function being minimized cannot be calculated explicitly. The initial brute-force search step is not always necessary -- our $\chi^2$ functions are generally well-behaved and have only a single, global, minimum -- but it is a useful precaution to ensure that the algorithm begins searching near the minimum so it does not converge on a non-stationary point \citep{Wright_1996, McKinnon_2006}.

\begin{figure*}
\includegraphics[width=\textwidth]{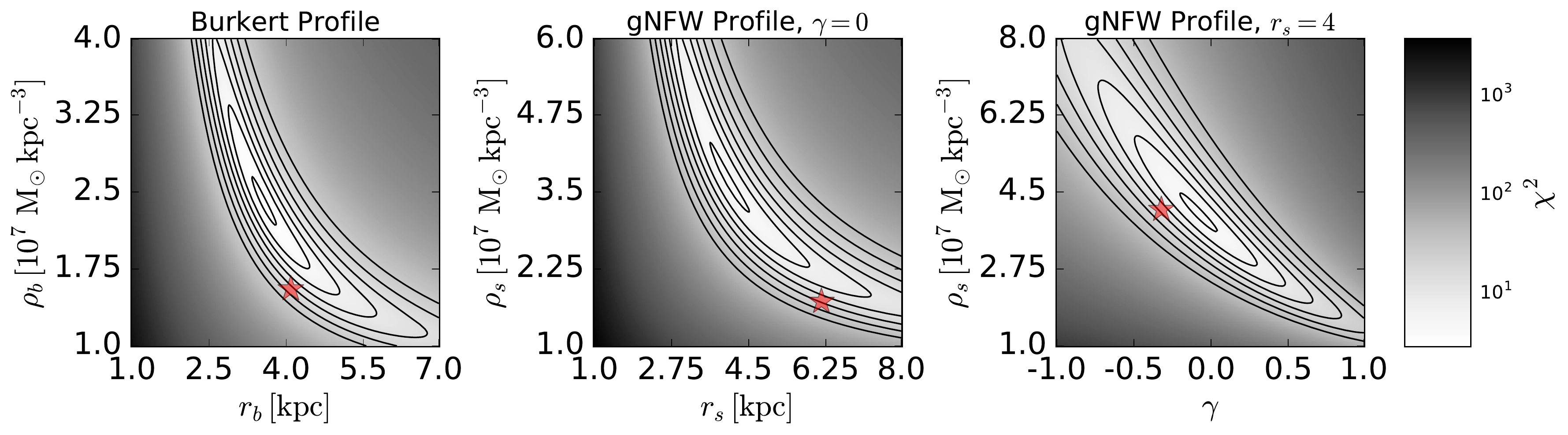}
\caption{
$\chi^2$ functions for Burkert (left) and generalized NFW (gNFW; middle, right) profile fits to \texttt{m10.6} at $z=0$. The color of each pixel shows the $\chi^2$ value (Equation~\ref{eqn:chi_square}) corresponding to the $v_{r, {\rm rms}}$ profile predicted by equation~\ref{eqn:vr_square}. Contours are logarithmically spaced between $1.05 \chi_{\rm min}^2$ and $10 \chi_{\rm min}^2$ in each panel. Because the gNFW profile has three free parameters, its full $\chi^2$ function cannot be easily visualized; instead, we inspect its projections in 2D parameter space. In the middle panels, we fix the parameter $\gamma=0$ to facilitate comparison with the Burkert profile, and in the right panel, we fix $r_s = 4$. These projections highlight the degeneracies between the different model parameters. Red stars show the ``true'' best-fit values obtained by directly fitting the density profile.
}
\label{fig:chi_square_plot}
\end{figure*}

To assess the degeneracy between the different parameters of our dynamical models, we can visually inspect contours of the $\chi^2$ function in high-probability regions of parameter space. Figure~\ref{fig:chi_square_plot} shows the $\chi^2$ function for Burkert and gNFW profile fits to the $z=0$ snapshot of \texttt{m10.6}. The Burkert profile (left panel) shows degeneracy between $r_b$ and $\rho_b$, in that a low $\chi^2$ value can be attained for either large $r_b$ and low $\rho_b$ or for small $r_b$ and high $\rho_b$, because the two combinations of parameters produce similar total mass profiles. The $r_s$ and $\rho_s$ parameters of the gNFW profile show comparable degeneracy when $\gamma$ is held fixed (middle panel), because the gNFW profile with $\gamma=0$ has a similar (but not identical) shape to the Burkert profile. Finally, holding $r_s$ fixed (right panel) reveals significant degeneracy between $\gamma$ and $\rho_s$. Cored profiles (low $\gamma$) with high $\rho_s$ produce similar $M_{\rm dyn}(r)$ profiles to cuspy profiles (high $\gamma$) with lower $\rho_s$. This degeneracy highlights the difficulty of obtaining tight constraints on the central density profiles of low-mass galaxies from Jeans modeling \citep{Walker_2009, Adams_2014, Zhu_2016}.

In general, we find that degeneracies between our model parameters increase as we increase the number of parameters. Especially for the gNFW profile, similar predicted $v_{r, {\rm rms}}$ profiles and $\chi^2$ values can be produced by a variety of different combinations of $\{\gamma, r_s, \rho_s\}$. Adding additional free parameters to the $\beta(r)$ profile further exacerbates the degeneracy. This is a generic problem that arises in fitting functions with many free parameters to data \citep{Klypin_2001} and is not a shortcoming of our particular procedure for Jeans modeling. By experimenting with several different parameter-fitting methods, including both using different optimization algorithms to find the $\chi^2$ minimum and using MCMC sampling methods to explore the multidimensional posterior,\footnote{
With flat priors, this is for practical purposes equivalent to plotting projections of the $\chi^2$ function.} we verified that there is little danger of our procedure converging on a false local minimum, even when there is high degeneracy. The real cost of higher degeneracy is that the model parameters returned by Jeans modeling become more poorly constrained, so that slight changes in the input $v_{r, {\rm rms}}$ data can produce significantly different best-fit parameters.

\subsection{Burkert vs gNFW models}
\label{sec:burkert_vs_gNFW}

We next compare the performance of the Burkert and gNFW models in recovering galaxies' mass and density profiles. We use three of the metrics introduced in Section~\ref{sec:different_beta_models} to asses the accuracy of our Jeans model fits: the minimum $\chi^2$ value, the ratio of the total enclosed mass inside $R_{90 \rm m}$ for the best-fit Jeans model to the true enclosed mass, $M_{\rm Jeans}/M_{\rm true}$, and $\overline{\Delta \rho}$, the mean error in the density profile, as defined in Equation~\ref{eqn:delta_rho}.

\begin{figure}
\includegraphics[width=\columnwidth]{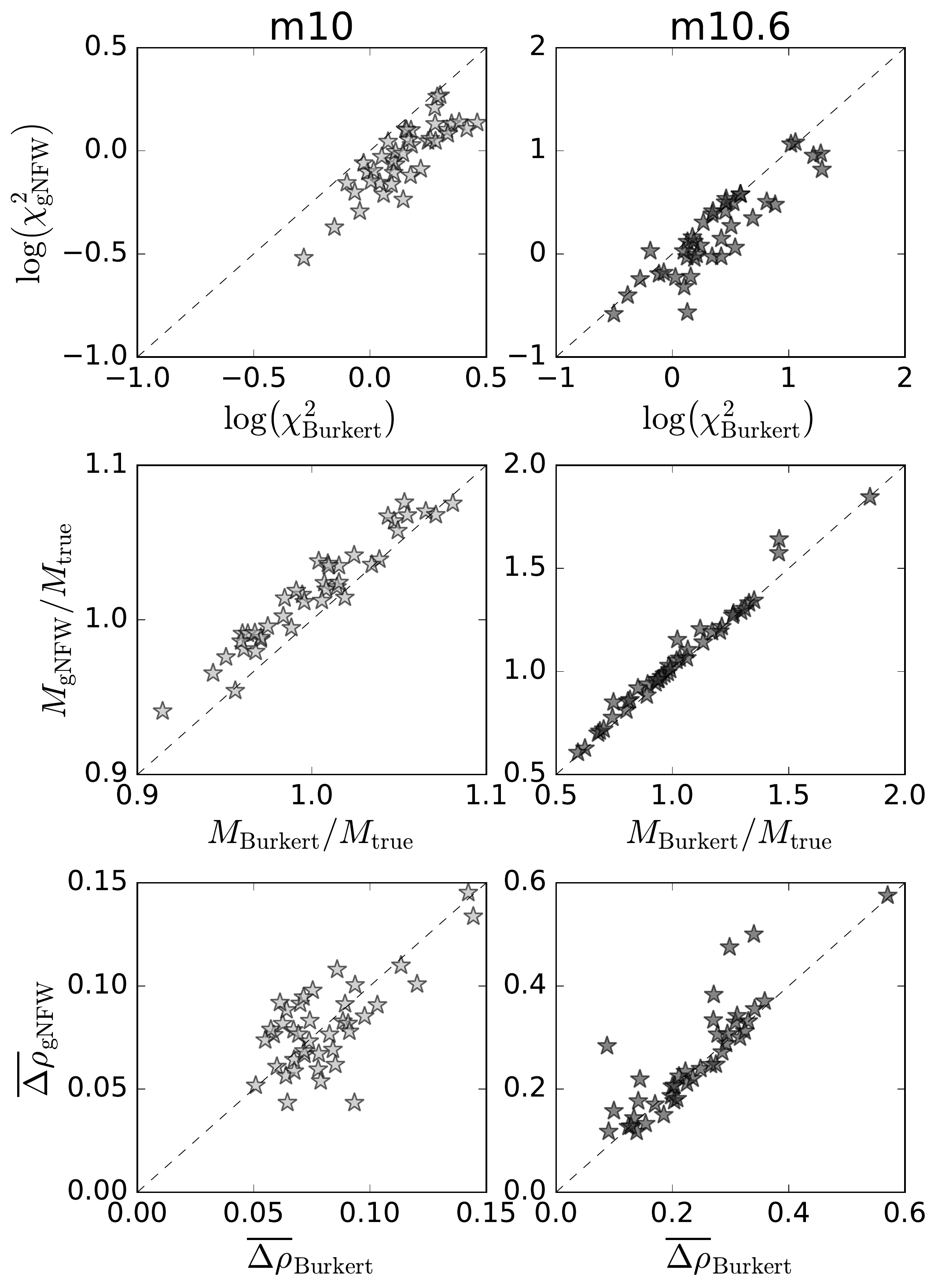}
\caption{
Comparison of Burkert and generalized NFW (gNFW) fits obtained from Jeans modeling in \texttt{m10} and \texttt{m10.6} in the last 40 snapshots since $z\approx 0.2$.
\textbf{Top}: minimum $\chi^2$ value obtained from Equation~\ref{eqn:chi_square}. \textbf{Middle}: Ratio of $M_{\rm dyn}(R_{90 \rm m})$ enclosed by best-fit Burkert and gNFW profiles obtained through Jeans modeling, to the true total mass enclosed in $R_{90 \rm m}$. \textbf{Bottom}: mean error in the density profile, as defined in Equation~\ref{eqn:delta_rho}. Dashed lines show the one-to-one relation. Note the different scales on the left and right-hand panels; because  \texttt{m10.6} has stronger potential fluctuations, errors in its Jeans modeling are generally larger.
}
\label{fig:gNFW_vs_Burkert}
\end{figure}

Figure~\ref{fig:gNFW_vs_Burkert} compares the values of these statistics for the last 40 snapshot since $z\approx 0.2$ of \texttt{m10} (left) and \texttt{m10.6} (right). The top panel compares minimum $\chi^2$ values. In both galaxies, the gNFW profile almost always produces lower minimum $\chi^2$ values than the Burkert profile. This is not surprising, because the gNFW profile has an extra free parameter and thus can produce a wider variety of $v_{r,{\rm rms}}$ profile shapes. Both profiles generally produce lower $\chi^2$ values in \texttt{m10} than in \texttt{m10.6}, primarily because \texttt{m10} undergoes smaller potential fluctuations and is closer to dynamical equilibrium. 

The middle panel compares errors in the total dynamical mass, $M_{\rm dyn}(R_{90 \rm m})$, for the two profiles. For both \texttt{m10} and \texttt{m10.6}, the Burkert and gNFW models usually agree within a few percent. The difference between the masses recovered using the two different profiles is typically less than the difference between the mass recovered using either profile and the true mass, indicating that mass errors are driven more by systematics than by the choice of profile. We note a slight systematic offset between the mass estimates obtained from the two profiles: the gNFW profile fits typically have dynamical masses a few percent higher than those from the Burkert profiles. This offset is more noticeable in \texttt{m10}, because the scatter and average mass errors are smaller, but it is true for both galaxies. The Burkert mass estimates are centered around $M_{\rm Jeans}/M_{\rm true} \approx 1$, whereas the gNFW model overestimates the true mass, on average.

Finally, Figure~\ref{fig:gNFW_vs_Burkert} (bottom) compares the mean error in the Burkert and gNFW density profiles, as given by Equation~\ref{eqn:delta_rho}. In most snapshots, the Burkert and gNFW models produce similar $\overline{\Delta \rho}$ values, with mean density errors of approximately $\sim 10\%$ in \texttt{m10} and $\sim 25\%$ in \texttt{m10.6}. However, for a few snapshots of \texttt{m10.6}, the error is significantly larger for the gNFW profile than for the Burkert profile (these snapshots all have high $|\overline{v_r}|$ values, indicating they are far from dynamical equilibrium). Here, Jeans modeling predicts a cusped ($\gamma \sim 1$) gNFW profile, while the true galaxy retains a core.

This suggests that $\gamma$ is especially poorly constrained by Jeans modeling when the potential is fluctuating. We emphasize, however, that a poorly constrained central density profile is a generic problem for Jeans modeling studies, even when galaxies are in equilibrium. First, $\gamma$ is highly degenerate with $\rho_s$ and $\beta$ \citep[see also][]{Zhu_2016}. Second, because $\gamma$ only describes the density profile at $r \ll r_s$, it is more difficult to constrain when one is fitting to the entire $v_{r,{\rm rms}}$ profile. Because $v_{r,{\rm rms}}$ is sampled in linearly spaced bins out to $r=R_{90 \rm m}$, most of the ``input'' $v_{r,{\rm rms}}$ data are at $r \gg r_s$. Only the central few $v_{r, {\rm rms}}$ points constrain $\gamma$, and their contribution to the $\chi^2$ statistic is easily outweighed by the data at larger radius. Some of our gNFW Jeans models fail pathologically at small radius (predicting, for example, $\gamma = -3$). These models produce $v_{r, {\rm rms}}$ and $M_{\rm dyn}(r)$ profiles that are identical to those predicted by the true density profile except at $r \ll r_s$.

Better constraints on $\gamma$ can be obtained either by fitting only to the central $v_{r, {\rm rms}}$ profile (that is, discarding points at $r \gg r_s$ when calculating the $\chi^2$ statistic) or by sampling $v_{r, {\rm rms}}$ in logarithmic radial bins, so that there are more data points at small radii. Not surprisingly, this comes with the tradeoff of poorer constraints on the density profile at large radii and less accurate total mass predictions. Because our work focuses primarily on studying how potential fluctuations affect Jeans modeling estimates of the total mass, we use linearly spaced bins out to $R_{90 \rm m}$.\footnote{
Alternately \citep[e.g.,][]{Li_2016}, one can place a prior on $\gamma$, forcing e.g.  $-1<\gamma<1$ or forcing $\gamma$ to take on one of a few discrete values. This prevents catastrophic failures in the Jeans gNFW models, but we find that constraints on $\gamma$ are still poor when using only a few $v_{r,{\rm rms}}$ points at small radii.}

The Burkert model has less freedom in the shape of the density profile at small radii, allowing less opportunity for catastrophic failure. Of course, the trade-off is that the profile cannot fit galaxies with steep central cusps. Because all of the low-mass galaxies in our simulations develop cores to some degree \citep{Chan_2015}, the Burkert Jeans models are generally more well-behaved at small radii than the gNFW models. We therefore use the Burkert profile exclusively, but we emphasize that this choice is motivated by a priori knowledge of the true density profiles.

\subsection{Choice of Bin Spacing and Radial Fitting Region}

\label{sec:bin_spacing}
\begin{table}[tbp]
\centering
\caption{
Absolute fractional error in the total dynamical mass predicted by Jeans modeling, $\Delta M = \left|M_{\rm Jeans}-M_{\rm true}\right|/M_{\rm true}$, for two of our simulations. $\Delta M_{50}$ and $\Delta M_{90}$ are the fractional mass errors within $R_{\rm 50m}$ and $R_{\rm 90m}$, respectively. We report median values for different fitting regions and bin spacings, all for the last 40 snapshots since $z \approx 0.2$.}
\label{tab:bin_spacing}
\begin{tabular}{p{1.4cm} | p{0.8cm}| p{0.7cm} | p{0.7cm} | p{0.7cm} | p{0.7cm}}
bin spacing & $R_{\rm max}$  & $\Delta M_{50}$, \texttt{m10} & $\Delta M_{50}$, \texttt{m10.6} & $\Delta M_{90}$, \texttt{m10} & $\Delta M_{90}$, \texttt{m10.6}\\ 
\hline
linear &  $R_{90 \rm m}$ & 0.07 & 0.14 & 0.03 & 0.19 \\
linear &  $R_{50 \rm m}$ & 0.07 & 0.16 & 0.30 & 0.28 \\
log    &  $R_{90 \rm m}$ & 0.05 & 0.13 & 0.08 & 0.21 \\
log    &  $R_{50 \rm m}$ & 0.07 & 0.16 & 0.32 & 0.28 \\ \hline
\end{tabular}
\end{table}

Throughout our primary analysis, we use linearly spaced measurements of $v_{r, {\rm rms}}$ between $r=0$ and $r=R_{90 \rm m}$ as our inputs in Jeans modeling. However, it is not always possible to measure $v_{r,{\rm rms}}$ out to such a large radius. Observational Jeans modeling studies can typically measure stellar kinematics out to $(1.5 - 2.5)R_{e}$ for integrated light studies \citep{Adams_2014, Bundy_2015} and out to more than $5R_{e}$ for studies of nearby dwarf spheroidal galaxies with resolved stellar kinematics \citep{Walker_2009}. For comparison, $R_{90m}$ in \texttt{m10.6} varies between $(2 - 4)R_{e}$ \citepalias{El-Badry_2016}.

As discussed in the previous section, the region in which $v_{r,{\rm rms}}$ is sampled can have nontrivial effects on the predicted mass profile. For example, using logarithmically spaced measurements of $v_{r, {\rm rms}}$ or sampling $v_{r, {\rm rms}}$ primarily at small radius will cause the Jeans modeling procedure to prioritize recovery of the $v_{r, {\rm rms}}$ profile at small radius over large radius, potentially leading to less accurate constraints on the total mass.

To assess the effects of varying the sampling of $v_{r, {\rm rms}}$ measurements to which our Jeans models are fit, we experimented with using logarithmic bins and with bins extending only out to $r=R_{50 \rm m}$, the radius enclosing 50\% of the stellar mass, rather than $R_{90 \rm m}$. Table~\ref{tab:bin_spacing} presents the median fractional mass errors over the last 40 snapshots of \texttt{m10} and \texttt{m10.6} for each of these binning schemes. For each galaxy and each binning scheme, we measure both $\Delta M_{90}$, the fractional mass error inside $R_{90m}$ and $\Delta M_{50}$, the fractional error inside $R_{50m}$.

Errors in the total mass within $R_{90m}$ are minimized when $v_{r, \rm rms}$ is measured in linearly spaced bins extending out to $R_{90 \rm m}$. Using logarithmic bins out $R_{90 \rm m}$ produces slightly larger mass errors, while restricting the fit to $v_{r, {\rm rms}}$ measurements inside $R_{50 \rm m}$ leads to significantly larger mass errors. In fact, when $v_{r, \rm rms}$ data are only available within $R_{50m}$, $\Delta M_{90}$ is comparable in \texttt{m10} and \texttt{m10.6}, suggesting that the error due to incomplete radial coverage dominates over errors arising from potential fluctuations.

This is not surprising: when $v_{r, {\rm rms}}$ data are only available within $R_{50 \rm m}$, Jeans modeling cannot constrain the mass profile between $R_{50 \rm m}$ and $R_{90 \rm m}$, and thus, extrapolations in $M_{\rm dyn}(r)$ to larger radius are based entirely on the form of the mass profile at small radius. 

On the other hand, errors in the total mass within $R_{50m}$ remain reasonable small even when kinematic data are only available with $R_{50m}$, and are comparable to $\Delta M_{90}$ when data are available with $R_{90m}$. That is, Jeans modeling can recover with reasonable accuracy the dynamical mass within the region where kinematic data are available. It becomes significantly less accurate when one attempts to measure the dynamical mass beyond the maximum radius where $v_{r, \rm rms}$  measurements are available, as there is no guarantee that the extrapolated mass profile will provide a good fit at larger radius.

We stress that, once a fitting region and binning scheme are chosen, our results are well converged with bin size: increasing the number of linearly spaced bins between $r=0$ and $r=R_{90 \rm m}$ by a factor of two causes sub percent-level changes in the predicted dynamical masses.

\subsection{Effects of unknown anisotropy} 
\label{sec:mass_anisotropy_degeneracy}

Thus far, we have limited our discussion to the simplified case in which the true form of $\beta(r)$ is known a priori. This is valid for systems for which one can obtain 3D stellar kinematics via proper motions (see Section~\ref{sec:observations}). However, measuring $\beta$ directly in nearby low-mass galaxies is not yet feasible, so one generally uses a parameterized model for $\beta(r)$. Here, we show how each of our three parameterized models for $\beta(r)$ affects our Jeans model fit for the total density profile.

During $\chi^2$ minimization, we simultaneously fit for $\{r_b,\rho_b\}$ and the free parameter characterizing the anisotropy profile -- $\beta_0$ for constant anisotropy, and the anisotropy radius, $r_a$, for the ML and OM profiles. We carry out $\chi^2$ minimization as described in Section~\ref{sec:chi_square_minization}, with the additional free parameter from the anisotropy increasing the dimension of the parameter space from 2 to 3.

\begin{figure*}
\includegraphics[width=\textwidth]{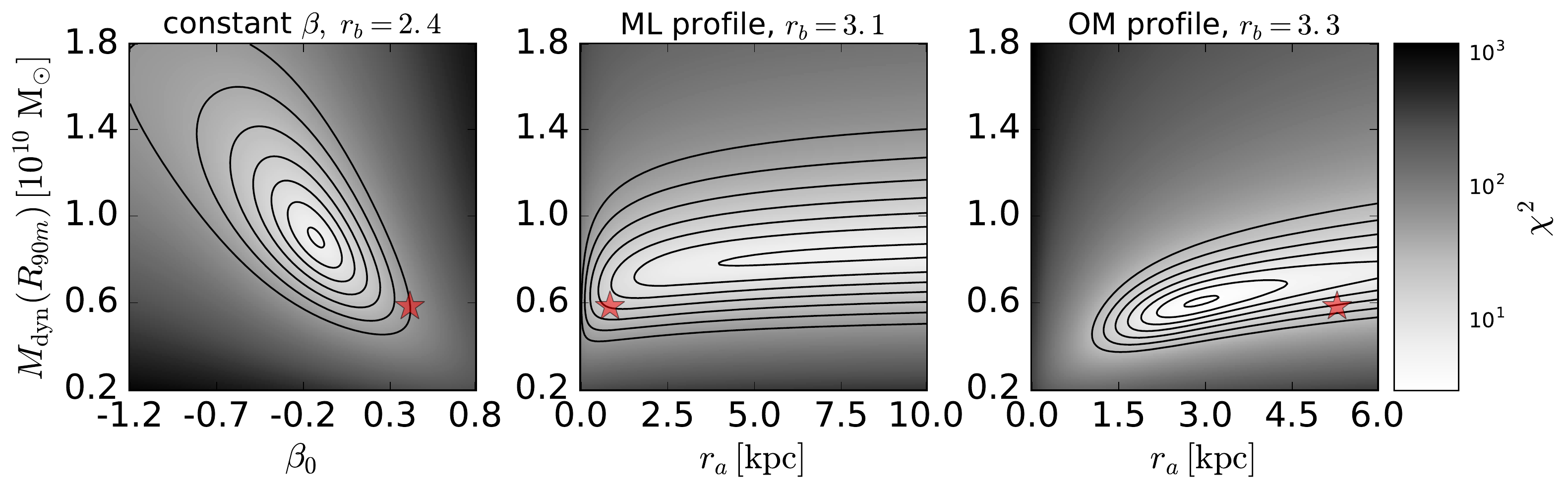}
\caption{
$\chi^2$ values for Jeans model fits to \texttt{m10.6} at $z = 0$. Colors and contours as in Figure~\ref{fig:chi_square_plot}. Each panel uses a different model for $\beta(r)$ and a Burkert model for the gravitational potential.
\textbf{Left}: constant anisotropy, that is, $\beta(r)=\beta_0$.
\textbf{Middle}: ML anisotropy profile, that is, $\beta(r)$ = $\beta_{\rm ML}(r, r_a)$, where $\beta_{\rm ML}(r, r_a)$ is defined in Equation~\ref{eqn:beta_ML}.
\textbf{Right}: OM anisotropy profile, that is, $\beta(r) = \beta_{\rm OM}(r, r_a)$, where $\beta_{\rm OM}(r, r_a)$ is defined in Equation~\ref{eqn:beta_OM}. Each panel shows a slice of parameter space with $r_b$ held fixed at the value corresponding to the global $\chi^2$ minimum for that model. Red symbols show the true enclosed mass and best-fit $\beta(r)$ model parameters, which we compute directly from the simulation.
}
\label{fig:beta_model_chi2s}
\end{figure*}

Figure~\ref{fig:beta_model_chi2s} shows $\chi^2$ values for fixed-$r_b$ slices of parameter space for each of the models for $\beta(r)$ that we tested in Section~\ref{sec:different_beta_models}. In each panel, $r_b$ is set to the value corresponding to the global $\chi^2$ minimum. To allow for straightforward comparison between the dynamical masses predicted by the different models for $\beta(r)$, which all favor different values of $r_b$, we plot total dynamical masses rather than $\rho_b$ on the y-axis. At fixed $r_b$, these relate to $\rho_b$ via a multiplicative constant (see Equation~\ref{eqn:Burkert_Mr}).

\begin{figure}
\includegraphics[width=\columnwidth]{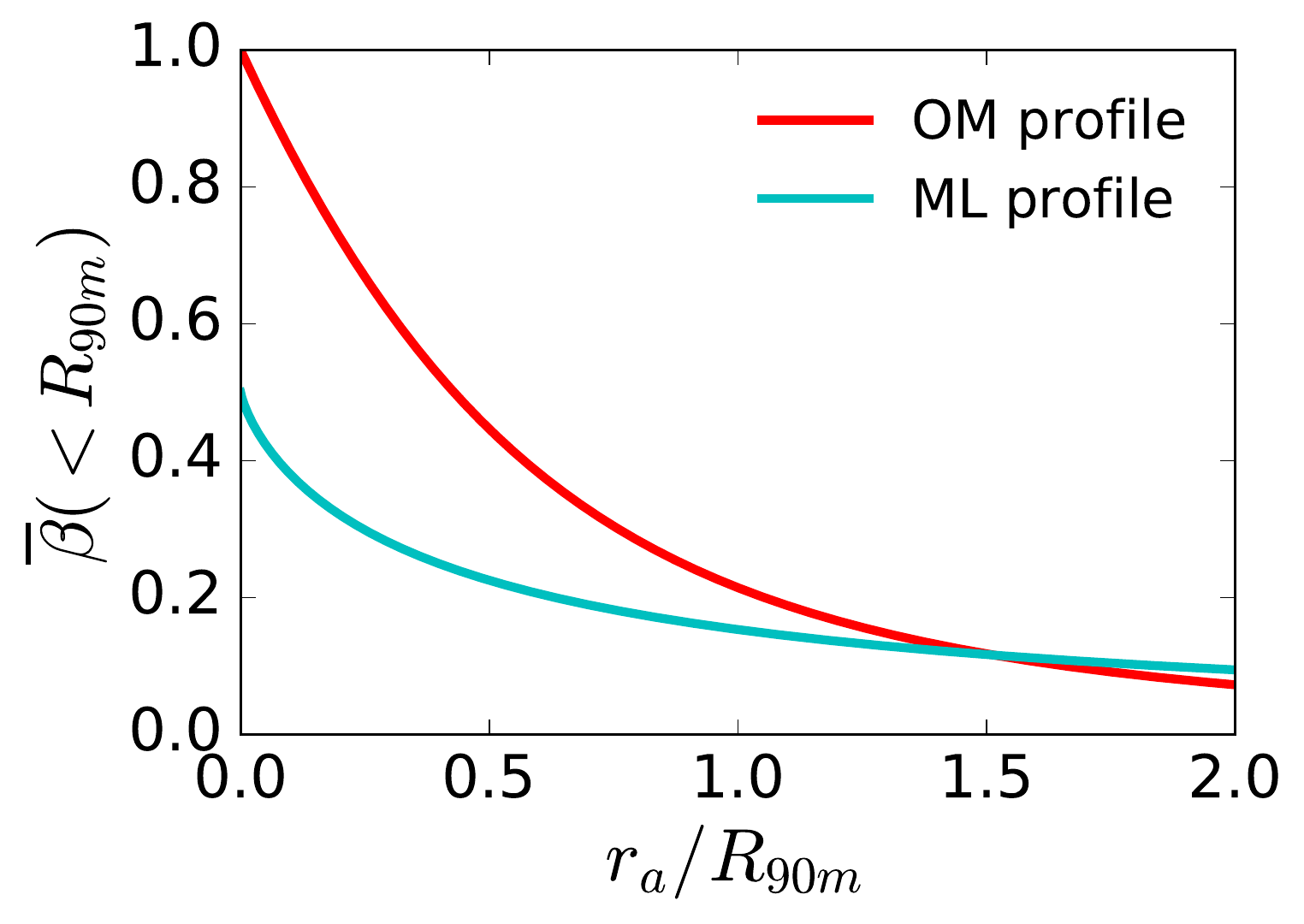}
\caption{Mean anisotropy of the ML and OM models between $r = 0$ and $r = R_{90 \rm m}$.}
\label{fig:OM_vs_ML_mean_beta}
\end{figure}

All three forms of $\beta(r)$ show some degeneracy between mass and anisotropy: models with radial orbits (high $\beta$) and low $M_{\rm dyn}$ predict similar $v_{r, {\rm rms}}$ profiles and $\chi^2$ values to those with more tangential orbits (low $\beta$) and high $M_{\rm dyn}$.\footnote{
If this is not apparent for the ML and OM profiles, note that larger $r_a$ values for these profiles produce lower overall $\beta$ values, because $r_a$ is approximately the radius at which the profile increases from $\beta = 0$ to $\beta = 0.5$ (for the ML profile) or $\beta = 1$ (for the OM profile). Figure~\ref{fig:OM_vs_ML_mean_beta} shows this explicitly.}
This degeneracy is well-studied in the literature \citep[e.g.,][]{Merritt_1987}, and breaking it is among the primary incentives for measuring $\beta(r)$ through resolved proper motions studies. In part because of this degeneracy, the $\chi^2$ minimum does not coincide with the true $M_{\rm dyn}$ and the best-fit anisotropy values recovered by direct fitting, which Figure~\ref{fig:beta_model_chi2s} shows in red symbols. However, the mass-anisotropy degeneracy is manifest differently for different models of $\beta(r)$, as the three models' differently shaped $\chi^2$ contours show. The largest error in $M_{\rm dyn}$ occurs for the constant anisotropy model. Here, Jeans modeling converges on a combination of high $M_{\rm dyn}$ and low $\beta$, which produces a similar $v_{r, {\rm rms}}$ profile to the true combination of lower $M_{\rm dyn}$ and higher $\beta$.

The different constraints on $\beta(r)$ that different anisotropy profiles provide was highlighted by \citet{Mamon_2013}, who used a dynamical modeling framework similar to ours to investigate how accurately Jeans modeling could recover the anisotropy profiles of dark matter halos from dissipationless cosmological simulations. They tried a variety of different models for $\beta(r)$, including constant anisotropy and the ML profile. As in our simulations, the true anisotropy profiles of their halos generally increased with radius and were reasonably well-fit by the ML model. Despite this, they found that using the ML profile in Jeans modeling provided poor constraints on the anisotropy profile: even when the true $\beta(r)$ closely resembled an ML profile, their dynamical modeling procedure could not reliably recover the true $\beta(r)$ and was unable to distinguish between models with different anisotropy radii $r_a$.

In Figure~\ref{fig:beta_model_chi2s}, the shape of the contours of the $\chi^2$ function in $r_a-M_{\rm dyn}$ parameter space illustrates the poor constraints on $r_a$. For the ML profile (and, to a lesser extent, the OM profile), $\chi^2$ minimization is unlikely to strongly constrain $r_a$, because similar $\chi^2$ values are produced by a wide range of $r_a$ values (that is, the $\chi^2$ contours are almost horizontal). This occurs for two reasons.
First, adjusting $r_a$ in the ML and OM profiles changes the total anisotropy less than adjusting $\beta_0$ in the constant-anisotropy models; that is, the ML and OM profiles have a smaller dynamic range of allowed $\beta$ values. Figure~\ref{fig:OM_vs_ML_mean_beta} demonstrates this explicitly, by showing the mean anisotropy between $r = 0$ and $R_{90 \rm m}$, $\overline{\beta}(< R_{90 \rm m})=\frac{1}{R_{90 \rm m}}\int_{0}^{R_{90 \rm m}}\beta\left(r\right)\,{\rm d}r$, of the ML and OM anisotropy functions for different values of $r_a$. While varying $\beta_0$ for the constant-anisotropy model allows $\overline{\beta}$ to take on any value in $(-\infty,1]$, varying $r_a$ in the ML and OM profiles only allows it to take values in the intervals $[0, 0.5]$ and $[0, 1]$, respectively. These profiles have less freedom -- and less opportunity for failure -- in $\beta(r)$. The range of allowed $M_{\rm dyn}$ values is thus also smaller.

Second, Equation~\ref{eqn:int_factor} shows that it is the behavior of $\beta(r)$ at small radius that most significantly affects the $v_{r, {\rm rms}}$ profile predicted by Jeans modeling; at larger radii, the contribution from $\beta(r)$ falls off as $1/r$. Both the OM and ML profiles force $\beta \to 0$ as $r \to 0$, so there is less freedom in $\beta$, especially at small radii. On the other hand, the constant anisotropy model allows $\beta(r=0)$ to vary over the full range of $(-\infty, 1]$, and as such, different values of $\beta_0$ produce $v_{r, {\rm rms}}$ profiles that are more distinct from one another than the range of profiles predicted by different choices of $r_a$ in the ML and OM models.

The OM and ML models thus have less relative predictive power for constraining the shape of the anisotropy profile. On the other hand, in many dynamical modeling studies, the dynamical mass profile is of primary interest, and the anisotropy profile is viewed as a nuisance parameter. For such works, the ML and OM profiles provide somewhat more accurate measurements of $M_{\rm dyn}$ than the constant-anisotropy profile, which tends to systemically overestimate the total mass (see Section~\ref{sec:different_beta_models}).

\end{document}